\documentclass[12pt]{article}

\begin{document}

\title{
Hamiltonian Methods for the Study of Polarized Proton Beam
	Dynamics in Accelerators and Storage Rings\\}

\author{V.V. Balandin and N.I. Golubeva \\
\\
\small\it
Institute for Nuclear Research of RAS, \\
\small\it
60th October Anniversary Pr., 7a, Moscow 117 312, Russia
}
\date{}
\maketitle

\begin{abstract}
The equations of classical spin-orbit motion can be extended 
to a {\bf Hamiltonian system} in 9-dimensional phase space by 
introducing a {\bf coupled spin-orbit Poisson bracket} (\ref{fom2}) 
and a {\bf Hamiltonian function} (\ref{fom3}).
After this extension and by establishing connections
between initial and extended systems
it becomes possible to apply
the {\bf methods of the theory of Hamiltonian systems} to the study of
polarized particle beam dynamics in circular accelerators and storage
rings. Some of those methods have been implemented in the
{\bf computer code FORGET-ME-NOT} \cite{start1,start2}.
\end{abstract}

\clearpage

\tableofcontents

\clearpage

\section{Introduction}

\hspace*{0.5cm}
We began writing this paper 
in the summer of 1994
year following the suggestion of D.Barber to summarize our
results of the investigation of the dynamics of unpolarized and 
polarized proton  beams, which we obtained during 1991-1993  
and which were only partly available and only in the very 
compressed form of conference proceedings.

After starting this work it became clear that 
it would be better to separate the available material
into several papers, and for the first one we 
chose those results which
can be incorporated together with help of 
the research technique which we call {\bf Hamiltonian 
extension of the equations of classical spin-orbit motion}.
Such important topics as the computation and analysis of
the one-turn Taylor maps for spin and orbit motion,
spin dependent invariant functions and their connection 
with the Derbenev-Kondratenko vector, the normal form algorithm
for equations of spin motion in the $\mbox{SU}(2)$ representation
and, of course, many more practical subjects related to
the question how to preserve the polarization of a particle
beam during acceleration 
will not appear in the present paper and will
be published later.

For various reasons the publication   
has been prolonged for almost three years, and now we
are glad to be free from this, but, at the same time,
we are not too enthusiastic about the necessity to
write up the remaining unpublished results.

\subsection{The Classical Equations of Spin-Orbit Motion}

\hspace*{0.5cm}
The quasi-classical description of the motion of a relativistic
nonradiating point particle with spin in accelerators and storage
rings includes the equations of orbit motion which we write
in the Hamiltonian form
\begin{eqnarray}
\frac{d \vec{q}}{dt}\;=\;  \frac{\partial H_{orbt}}{\partial \vec{p}},
\hspace{1.5cm}
\frac{d \vec{p}}{dt}\;=\;-\:\frac{\partial H_{orbt}}{\partial \vec{q}}
\label{vk1}
\end{eqnarray}
and the Thomas-BMT equation \cite{tom,BMT} for the classical
spin vector $\;\vec{s}$
\begin{eqnarray}
\frac{d \vec{s}}{dt}\;=\;\left[\:\vec{W} \times \vec{s}\;\right] 
\label{vk2}
\end{eqnarray}
Here
\begin{eqnarray}
H_{orbt}\;=\; c\:\sqrt{{\vec{\pi}\:}^2 \:+\: m_0^2\: c^2}\;+\;e\:\Phi 
\nonumber
\end{eqnarray}
\begin{eqnarray}
\vec{W}\; =\; -\frac{e}{m_0 \gamma c}
\left(
\left(1 + \gamma G \right) \vec{\cal B}\: -\:
\frac{G \left(\vec{\pi} \cdot \vec{\cal B}\:\right) \vec{\pi}}
{m_0^2c^2(1+\gamma)} \:-\:
\frac{1}{m_0 c} \left(G + \frac{1}{1+\gamma}\right)
\left[\vec{\pi} \times \vec{\cal E}\:\right]
\right) 
\nonumber
\end{eqnarray}
and $\:t\:$ is the time. The vectors
$\:\vec{q}\:=\:(q_1, q_2, q_3)$ 
and $\:\vec{p}\:=\:(p_1, p_2, p_3)$ 
are canonical position and momentum variables,
and $\;\vec{s}\:=\:(s_1, s_2, s_3)\;$ is the classical 
spin vector of length $\:\hbar /2\:$.
The parameters
$\:e\:$ and $\:m_0\:$ are the charge and the rest mass of the particle,
$\:c\:$ is the velocity of light, $\:G = (g-2)/2\:$ which quantifies 
the anomalous spin $\:g\:$ factor, $\:\gamma\:$ is the Lorentz factor, 
$\:\vec{\pi}\:$ is kinetic momentum vector, $\:\vec{\cal E}\:$ and 
$\:\vec{\cal B}\:$ are the electric and magnetic fields, 
and $\:\vec{A}\:$ and $\:\Phi\:$ are the vector and scalar potentials.
\begin{eqnarray}
\vec{\cal B} \;=\;  \mbox{curl}_{\:\vec{q}} \; \vec{A}
\label{mf_ga}
\end{eqnarray}
\begin{eqnarray}
\vec{\cal E} \;=\; -\:\mbox{grad}_{\:\vec{q}} \; \Phi \;-\; 
\frac{1}{c}\: \frac{\partial \vec{A}}{\partial t}  
\label{ef_ga}
\end{eqnarray}
\begin{eqnarray}
\vec{\pi} \;=\; \vec{p} \;-\; \frac{e}{c}\: \vec{A}  
\nonumber
\end{eqnarray}
\begin{eqnarray}
\gamma \;=\; 
\frac{H_{orbt} - e \Phi}{m_0 c^2} \;=\;
\sqrt{1 \:+\: \left(\frac{\vec{\pi}}{m_0 c}\right)^2}
\label{gamama}
\end{eqnarray}
Later on we will refer to the system (\ref{vk1})-(\ref{vk2}) as the
{\bf triangular system} (the equations of spin motion contain the 
orbital variables but the evolution of orbital variables does not 
depend on the spin degree of freedom).

{\bf Remark:} In the Hamiltonian picture of orbital motion we 
cannot define the Lorentz factor (\ref{gamama}), in the usual way, 
in terms of 
the absolute value $\:v\:$ of the particle velocity $\:\vec{v}$
\begin{eqnarray}
\gamma \;=\; \frac{1}{\sqrt{1 \:-\: \left(v/c\right)^2}},  
\hspace{1.0cm}
v^2 \;=\; \vec{v} \cdot \vec{v}
\nonumber
\end{eqnarray}
but instead need to express it through canonical variables.

\section{The Hamiltonian Extension of the Equations 
of Classical Spin-Orbit Motion}

\hspace*{0.5cm}

The Poisson bracket lies at
the basis of the 
modern point of view of the Hamiltonian formalism. The idea of
axiomatic introduction of the bracket 
very likely 
belongs to Dirac. In this section we
recall the necessary definitions from the theory of Hamiltonian systems and
introduce coupled spin-orbit Poisson brackets which give us the canonical
extension of the equations of classical spin-orbit motion.

\subsection{Poisson Bracket}

\hspace*{0.5cm}
Let $\:M\:$ be a finite-dimensional manifold $\:$(phase space)$\:$ 
and $\;C^{\infty}(M)\;$
be the linear space of smooth functions 
$\;f : M \rightarrow R\;$. The binary
operation
\begin{eqnarray}
\{*,\:* \} \: : \: C^{\infty}(M) \times C^{\infty}(M) 
\;\rightarrow\;C^{\infty}(M)
\nonumber
\end{eqnarray}
called the {\bf Poisson bracket} satisfies the relations
\begin{eqnarray}
\begin{array}{ll}
\\
{\bf bilinearity}: & \{\lambda f \:+\: \mu g,\; h \} 
\;=\; \lambda \:\{f,\; h \} \;+\; \mu \:\{g, \;h \} 
\nonumber \\
\\
{\bf antisymmetry}: & \{f, \;g \} \;=\; -\{g,\; f \} 
\nonumber \\
\\
{\bf Leibnitz} \hspace{0.2cm} {\bf rule}: & 
\{f, \;g \cdot h \} \;=\;
\{f,\; g \} \cdot h \;+\; g \cdot \{f,\; h \} 
\nonumber \\
\\
{\bf Jacobi} \hspace{0.2cm} {\bf identity}: & 
\{ \{f, \;g \},\; h \} \;+\;
\{ \{h,\; f \},\; g \} \;+\; \{ \{g,\; h \},\; f \} \;=\; 0 \\
\end{array}
\nonumber
\end{eqnarray}
where $\;\lambda,\;\mu\;$ are arbitrary constants.

Let $\;\vec{z}\;$ be local coordinates on a 
manifold $\;M\:$. From the bilinearity and
Leibnitz rule it follows that for any fixed function 
$\;h\;$, the linear operator
\begin{eqnarray}
\{h, \;* \}\: :\: C^{\infty}(M) \;\rightarrow\; C^{\infty}(M)  
\nonumber
\end{eqnarray}
is a differentiation and hence may be represented in the form
(see, for example \cite{nemstep,olver})
\begin{eqnarray}
\{h,\; * \} \;=\; A_i^h (\vec{z}) \; \frac{\partial \:*}{\partial z_i}
\label{d1}
\end{eqnarray}
(here and further on the summation over repeated indices is implied). 
From (\ref{d1}) one finds by direct substitution of functions 
$\;z_i\;$ that 
\begin{eqnarray}
A_{\:i}^{\:h} (\vec{z}\,) \;=\; \{h,\; z_i\}  
\label{d2}
\end{eqnarray}
Comparing (\ref{d1}) and (\ref{d2}) we obtain
\begin{eqnarray}
\{h, \;* \}\; =\; \{h, \;z_i\} \; \frac{\partial \:* }{\partial z_i}
\label{d3}
\end{eqnarray}
Consider the Poisson bracket $\;\{f,\; g\}\;$. 
By successively applying (\ref{d3})
to the expressions $\;\{f,\: * \}\;$ and $\;\{*,\: z_j\}\;$ we obtain
\begin{eqnarray}
\{f,\;g\}\; =\; \{f,\; z_j\} \; \frac{\partial g}{\partial z_j} 
\;=\; \{z_i,\;z_j\} \; 
\frac{\partial f}{\partial z_i} \:
\frac{\partial g}{\partial z_j}
\label{d4}
\end{eqnarray}
Now we introduce the skew-symmetric matrix 
$\;\hat{J} \:=\: (\{z_i,\; z_j\})\;$ 
and represent the Poisson bracket (\ref{d4}) in the form
\begin{eqnarray}
\{f,\;g\} \;=\; \mbox{grad}_{\:\vec{z}} \; f \cdot \hat{J}
\hspace{0.15cm}\mbox{grad}_{\:\vec{z}} \; g
\label{d5}
\end{eqnarray}
Thus in a fixed local coordinate system $\;\vec{z}\;$, 
the Poisson bracket is
completely defined if we know the values 
$\;{\hat{J}}_{\:ij}\; =\; \{z_i, \;z_j\}\;$ as
functions of $\;\vec{z}\:$. 

The matrix $\:\hat{J}\:$ is called {\bf the
structural matrix} of a Poisson bracket.

\subsection{Hamiltonian Dynamical Systems}

\hspace*{0.5cm}
Again let $\;\vec{z}\;$ be local coordinates on a manifold $\;M\,$. 
{\bf Hamiltonian systems} by definition have the form:
\begin{eqnarray}
\frac{d \vec{z}}{d \tau} \;=\; \{\vec{z},\: H\}  
\label{fom1}
\end{eqnarray}
where $\;H\:=\:H(\tau,\: \vec{z})\;$ is an arbitrary function 
(possibly depending explicitly on $\;\tau$\footnote{Until 
now the equations of classical spin-orbit 
motion (\ref{vk1})-(\ref{vk2}) are written using the time $t$ as
independent variable, but later on we will change this independent
variable to be a path length along the design orbit $z$. This is the 
reason for us, here and further on, to formulate general results, 
which are not sensitive to the specific form of the variables and 
Hamiltonians used, we denote the independent variable as $\tau$.}), 
called the {\bf Hamiltonian}. Using the representation of Poisson 
bracket (\ref{d5}) we can rewrite equations (\ref{fom1}) as
\begin{eqnarray}
\frac{d \vec{z}}{d \tau} \;=\; \hat{J}(\vec{z}\,) \hspace{0.15cm} 
\mbox{grad}_{\:\vec{z}} \; H
\label{gom2}
\end{eqnarray}

Let $\;h \:=\: h(\vec{z}\,)\;$ be an arbitrary function and
$\;\vec{z}(\tau)\;$ 
be the solution of the system (\ref{fom1}) with initial condition
$\;\vec{z}({\tau}_0) \:=\: {\vec{z}}_0\;$. It is clear that the 
derivative 
of the function $\;h(\tau)\:=\: h(\vec{z}(\tau))\;$ 
for the Hamiltonian system (\ref{fom1}) has the form
\begin{eqnarray}
\frac{d h}{d \tau} \;=\; \{h,\: H\}  
\label{gom3}
\end{eqnarray}
From (\ref{gom3}) we can obtain the expansion of $\;h(\tau)\;$ with 
respect to a small $\;\triangle \tau \;=\; \tau \:-\: {\tau}_0$
\begin{eqnarray}
h(\tau) \;=\; h({\tau}_0) \:+\: \triangle \tau \,
\{h,\; H\}({\tau}_0) 
\:+\: O(\triangle {\tau}^2)  
\label{gom4}
\end{eqnarray}
Writing $\;f(\tau)\;$ and $\;g(\tau)\;$ instead $\;h(\tau)\;$ 
in (\ref{gom4}) 
and calculating their Poisson brackets with respect to the variables
$\;{\vec{z}}_0\,$, we see that
\begin{eqnarray}
\{f,\; g \}(\tau) \;=\; \{f, \;g \}({\tau}_0) \;+ 
\nonumber 
\end{eqnarray}
\begin{eqnarray}
 +\; \triangle \tau\: (\{f,\;\{g,\; H\}\}({\tau}_0) \;+\; 
\{ \{f,\;H\}, \;g\}({\tau}_0)) \;+\;
O(\triangle {\tau}^2)   
\label{gom5}
\end{eqnarray}
Applying Jacobi identity to the terms of order 
$\;\triangle \tau\;$ in (\ref{gom5}) one finds
\begin{eqnarray}
\{f, \;g \}(\tau) \;=\;
 \{f,\; g \}({\tau}_0) \;+\; \triangle \tau \:\{\{f,\; g\},\;H\} 
({\tau}_0) \;+\; O(\triangle {\tau}^2)  
\nonumber
\end{eqnarray}
or, equivalently
\begin{eqnarray}
\frac{\{f, \;g \}(\tau)\: -\: \{f,\; g\}({\tau}_0)}
{\triangle \tau} \;=\; 
\{\{f,\; g\},\; H\}({\tau}_0)\; +\; O(\triangle \tau) 
\label{gom6}
\end{eqnarray}
The formula (\ref{gom6}) in the limit when 
$\;\triangle \tau \:\rightarrow \:0\;$
gives us the fundamental property of the solution of 
the Hamiltonian system:

{\bf The flow of a Hamiltonian system preserves the Poisson 
bracket:}\footnote{We use the term $''$preserve$''$ to
mean that the algebraic form is not changed.}
\begin{eqnarray}
\frac{d}{d \tau}\:\{f, \;g \} \;=\; \{\{f,\; g\},\; H\}  
\label{gom7}
\end{eqnarray}

\subsection{Hamiltonian Extension of the  Equations
of Classical Spin-Orbit Motion}

\hspace*{0.5cm}
We now introduce the Poisson bracket\footnote{Note that 
when applied to the spin variables only, the 
Poisson bracket (\ref{fom2}) gives the usual result 
$\{\,s_i,\,s_j\,\}\:=\:\epsilon_{ijk}\,s_k$.}
\begin{eqnarray}
\{f(\vec z),\; g(\vec z)\} \;=\; 
f_{\:\vec q} \:\cdot\: g_{\:\vec p}\; -\; 
f_{\:\vec p}\: \cdot\: g_{\:\vec q}\; +\; 
\left[\,f_{\:\vec s} \:\times\: g_{\:\vec s}\:\right] \cdot \vec s  
\label{fom2}
\end{eqnarray}
in the 9-dimensional phase space $\;\vec{z}\: =\: (\vec{x},\:\vec{s})\;$ 
of 6 orbital variables
$\;\vec{x} \:=\: (\vec{q},\:\vec{p})\;$ and 3 spin 
variables $\;\vec{s}\;$ 
and consider a Hamiltonian system of ordinary differential equations
\begin{eqnarray}
\frac{d \vec{z}}{d t} \;=\; \{\vec{z},\; H\}  
\label{sm1}
\end{eqnarray}
with the Hamiltonian function
\begin{eqnarray}
H \;=\; H_{orbt}(t, \vec{x}) \;+\; \vec{W}(t, \vec{x}) \cdot \vec{s}
\label{fom3}
\end{eqnarray}
In the variables $\;\vec{q}$, $\;\vec{p}\:$ and $\;\vec{s}\;$ the system 
(\ref{sm1}) can be written as
\begin{eqnarray}
\frac{d \vec{q}}{d t} \;=\; 
\frac{\partial H_{orbt}}{\partial \vec{p}} \;+\; 
\frac{\partial \left(\vec{W} \cdot \vec{s} \:\right)}
{\partial \vec{p}}
\label{f31}
\end{eqnarray}
\begin{eqnarray}
\frac{d \vec{p}}{d t} \;=\; 
-\,\frac{\partial H_{orbt}}{\partial \vec{q}}\; -\; 
\frac{\partial \left(\vec{W} \cdot \vec{s}\:\right)}
{\partial \vec{q}}
\label{f32}
\end{eqnarray}
\begin{eqnarray}
\frac{d \vec{s}}{d t} \;=\; 
\left[\vec{W} \times \vec{s}\:\right]  
\label{f33}
\end{eqnarray}
and we will understand the equations (\ref{f31})-(\ref{f33}) as the
{\bf Hamiltonian extension of the equations of classical spin-orbit 
motion} (\ref{vk1})-(\ref{vk2}).

Note that the matrix $\:\hat{J}(\vec{z})\:$
for the spin-orbit Poisson bracket (\ref{fom2}) has the form
\begin{eqnarray}
\hat{J}(\vec{z}) \;=\; \left(
\begin{array}{rrrrrrrrr}
0 & 0 & 0 & 1 & 0 & 0 & 0 & 0 & 0 \\
0 & 0 & 0 & 0 & 1 & 0 & 0 & 0 & 0 \\
0 & 0 & 0 & 0 & 0 & 1 & 0 & 0 & 0 \\
-1 & 0 & 0 & 0 & 0 & 0 & 0 & 0 & 0 \\
0 & -1 & 0 & 0 & 0 & 0 & 0 & 0 & 0 \\
0 & 0 & -1 & 0 & 0 & 0 & 0 & 0 & 0 \\
0 & 0 & 0 & 0 & 0 & 0 & 0 & s_3 & -s_2 \\
0 & 0 & 0 & 0 & 0 & 0 & -s_3 & 0 & s_1 \\
0 & 0 & 0 & 0 & 0 & 0 & s_2 & -s_1 & 0
\end{array}
\right)  \label{jm}
\end{eqnarray}
and is not a constant matrix (in contrast to the case of classical
Poisson brackets), but depends on spin variables.
The structural matrix (\ref{jm}) can also be written in the more 
compact block diagonal form
\begin{eqnarray}
\hat{J}(\vec{z}) \;=\; \mbox{diag}\, (J, \hspace{0.1cm} J_s(\vec{s}\,))  
\nonumber
\end{eqnarray}
where a $\,6 \times 6\,$ constant matrix $\:J\:$ is the symplectic unit
\cite{guil} and
\begin{eqnarray}
J_s(\vec{s}\:) \;=\; \left(
\begin{array}{rrr}
0 & s_3 & -s_2 \\
-s_3 & 0 & s_1 \\
s_2 & -s_1 & 0
\end{array}
\right)  
\nonumber
\end{eqnarray}

\subsection{Connection between the Triangular System 
and its Hamiltonian Extension}

\hspace*{0.5cm}
We now wish to point out some ways for
establishing the connections between properties and solutions
of the system (\ref{f31})-(\ref{f33}) and the initial triangular
system (\ref{vk1})-(\ref{vk2})
({\bf truncation procedures}).
We will do this {\bf without ascribing any physical sense to the spin
dependent members in
the right sides of equations (\ref{f31}), (\ref{f32})}
\footnote{
Note that some authors (see, for example \cite{derb,russm})
ascribe the spin dependent members on the right sides
of equations (\ref{f31})-(\ref{f32}) to a
quasi-classical effect of the spin on the orbit motion, 
so that in that case
no truncation procedures are needed.}.

Because the statements, which will be listed below, are connected
not with the specific form of the spin-orbit Hamiltonian, but only 
with properties of the spin-orbit Poisson bracket,
let us consider an arbitrary smooth Hamiltonian function
possibly depending nonlinearly
on the variables $\;\vec{s}\:$ 
\begin{eqnarray}
H\;=\;H(\tau,\, \vec{z})\;=\;H(\tau,\, \vec{x}, \,\vec{s}\,)
\label{generham}
\end{eqnarray}
and introduce, in correspondence 
to the Hamiltonian system
\begin{eqnarray}
\frac{d \vec{x}}{d \tau} \;=\; J \hspace{0.15cm} 
\mbox{grad}_{\:\vec{x}}\; H,
\hspace{1.0cm}
\frac{d \vec{s}}{d \tau} \;=\; J_s(\vec{s}\,) \hspace{0.15cm} 
\mbox{grad}_{\:\vec{s}} \; H
\label{genernon}
\end{eqnarray}
the triangular truncated system defined as
\begin{eqnarray}
\frac{d \vec{x}}{d \tau} \;=\; J \cdot 
\left.\left(
\mbox{grad}_{\:\vec{x}} \; H\right)\right|_{\vec{s}=\vec{0}},
\hspace{0.7cm}
\frac{d \vec{s}}{d \tau} \;=\; J_s(\vec{s}) \cdot 
\left.\left(
\mbox{grad}_{\:\vec{s}} \;H\right)\right|_{\vec{s}=\vec{0}}
\label{generlin}
\end{eqnarray}

{\bf a)} If 
\begin{eqnarray}
\vec{z}(\tau,\,\tau_0,\,\vec{z}_0) \;=\;
(\vec{x}(\tau,\,\tau_0,\,\vec{x}_0,\,\vec{s}_0), \;
\vec{s}(\tau,\,\tau_0,\,\vec{x}_0,\,\vec{s}_0)) 
\nonumber
\end{eqnarray}
is the solution of
(\ref{genernon}) which passes
through the point 
$\;\vec{z}_0 \:=\: (\vec{x}_0,\;\vec{s}_0)\;$ when
$\;\tau \:=\: \tau_0\,$, then
\begin{eqnarray}
\vec{z}_{*}(\tau,\,\tau_0,\,\vec{z}_0) \;=\; 
(\vec{x}_{*}(\tau,\,\tau_0,\,\vec{x}_0),\;
\vec{s}_{*}(\tau,\,\tau_0,\,\vec{x}_0,\,\vec{s}_0))
\nonumber
\end{eqnarray}
where
\begin{eqnarray}
\vec{x}_{*}(\tau,\,\tau_0,\,\vec{x}_0) \;=\; 
\vec{x}(\tau,\,\tau_0,\,\vec{x}_0, \,\vec{0})
\hspace{0.45cm} \mbox{and} \hspace{0.45cm}
\vec{s}_{*}(\tau,\,\tau_0,\,\vec{x}_0,\,\vec{s}_0) \;=\;
\left. \frac{\partial \vec{s}}{\partial\vec{s}_0}
\right|_{\vec{s}_0 = \vec{0}} \cdot \vec{s}_0
\nonumber
\end{eqnarray}
gives us the solution of (\ref{generlin}).

{\bf b)} If the system (\ref{genernon}) admits
an invariant function $\;V(\tau,\, \vec{z}\,)\;$ which can be 
represented in the form
\begin{eqnarray}
V(\tau, \,\vec{z}\,) \;=\; V_m(\tau, \,\vec{z}\,)\; +\; 
V_{>m}(\tau,\,\vec{z}\,)  
\nonumber
\end{eqnarray}
where $\:V_m\:$ is a homogeneous polynomial of degree 
$\:m\:$ in variables $\:\vec{s}\:$, and
\begin{eqnarray}
\lim \limits_{|\vec{s}| \rightarrow 0} 
\;\frac{V_{>m}}{|\vec{s}\,|^m} \;=\; 0
\nonumber
\end{eqnarray}
then $\;V_m(\tau,\,\vec{z}\,)\;$ is a first integral of the system
(\ref{generlin}).

{\bf c)} If $\; \vec{x}(\tau,\,\tau_0,\,\vec{x}_0)\;
\stackrel{{\rm
def}}{=}\;\vec{\phi}(\tau,\,\tau_0,\,\vec{x}_0)\;$
is a solution of the first of the equations (\ref{generlin}),
then the system (\ref{generlin}) can be written as a family of
Hamiltonian systems of the type (\ref{genernon})
depending on parameters $\;(\tau_0,\,\vec{x}_0)\;$
with the Hamiltonian function
\begin{eqnarray}
\vec{W}\left(\tau, \,\vec{\phi}(\tau,\,\tau_0,\,\vec{x}_0)\right)\cdot
\vec{s}
\hspace{0.6cm} 
\mbox{where}
\hspace{0.6cm}
\left.\vec{W}(\tau, \,\vec{x}\,) \;=\; 
\left(\mbox{grad}_{\:\vec{s}} \; H\right)\right|_{\vec{s}=\vec{0}}
\nonumber
\end{eqnarray}

\section{Hamiltonian Methods 
for the Extended System}

\subsection{Degenerate Poisson Brackets and Reduction 
of the Order of a Hamiltonian System}

\hspace*{0.5cm}
If there are nontrivial functions ({\bf Casimir functions})
$f_l(\vec{z}\,)$ (maybe given locally on the manifold $M$) such that
\begin{eqnarray}
\{f_l,\; h \}\; =\; 0  
\label{k1}
\end{eqnarray}
for any function $h(\vec{z}\,)$ then the matrix
$\:\hat{J}(\vec{z}\,)\:$ is
degenerate and this Poisson bracket is said to be {\bf degenerate}. (For a
degenerate matrix $\:\hat{J}(\vec{z}\,)\:$ of constant rank, the functions
$\;f_l\;$
in (\ref{k1}) locally always exist.)

If all (at least all functionally independent) Casimir functions 
$\;f_l\;$ have
been found, then from properties (\ref{gom3}) and (\ref{k1}) 
it follows that
for an arbitrary Hamiltonian the trajectories of the system 
(\ref{fom1}) $\;\vec z(\tau )\;$ lie on the intersecting level surfaces
\begin{eqnarray}
f_l(\vec z\,)\;=\;c_l\;=\;const
\hspace{0.4cm}
(l\;=\;1,2,\ldots ,m)  
\label{k2}
\end{eqnarray}
where the Poisson bracket no longer remains degenerate.

Locally any Poisson bracket of constant rank can be brought 
into the form ({\bf Darboux theorem})
\begin{eqnarray}
\hat{J} \;=\; \left(
\begin{array}{ccc}
O_{nn} & I_n & O_{nm} \\
-I_n & O_{nn} & O_{nm} \\
O_{mn} & O_{mn} & O_{mm}
\end{array}
\right)  \label{jmnm}
\end{eqnarray}
where $\;O_{kl}\;$ is a $\:k \times l\:$ zero matrix and 
$\;I_n\;$ is a $\:n\times n\:$
identity matrix, and $\;\mbox{dim}\, M \:=\: 2n + m\,$. 
Thus in local coordinates we
obtain the classical Hamiltonian system with $\,n\,$ degrees of freedom
depending on $\,m\,$ parameters ($c_1, c_2, \ldots , c_m$ in (\ref{k2})).

\subsection{Degenerate Poisson Brackets for Global Variables, 
or Local Darboux Coordinates ?}

\hspace*{0.5cm}
The spin-orbit Poisson bracket (\ref{fom2}) is degenerate. It has the
nontrivial Casimir function $\:f_1 \:=\: {\mid \vec{s} \mid}^{2}\:$ 
and on the level
surface $\:f_1 \:=\: const\: > \:0\:$ its rank is constant 
and is equal to 8. This means
that we can decrease the dimensions of the system (\ref{genernon}) by
introducing Darboux coordinates. Thus we obtain the 
{\bf classical Hamiltonian system with 4 degrees of freedom
depending on one parameter} $\:{\mid \vec{s} \mid}^{2}\:$.
It is clear that the Darboux coordinates are not
unique and may be introduced in various ways.
We consider only one typical example.

Let $\:\vec{i}$, $\:\vec{j}$, $\:\vec{k}\:$ be an 
arbitrary orthogonal system of
unit vectors in three dimensional space 
$\:R^3\:$ satisfying the condition
\begin{eqnarray}
\vec{i} \cdot \left[\,\vec{j} \times \vec{k}\:\right] \;=\; 1  
\nonumber
\end{eqnarray}
We introduce three new spin variables $\:\psi$, $\:J$, $\:I\:$  
by the equations
\begin{eqnarray}
\left\{
\begin{array}{l}
\vec{s} \cdot \vec{i}\; =\; J \\
\\
\vec{s} \cdot \vec{j} \;=\; \sqrt{I - J^2} \:\cos (\psi) \\
\\
\vec{s} \cdot \vec{k} \;=\; \sqrt{I - J^2} \:\sin (\psi)
\end{array}
\right.  
\label{f1233}
\end{eqnarray}
or, equivalently
\begin{eqnarray}
\vec{s}(\psi,\, J,\, I) \;=\; J \cdot \vec{i}\: +\: 
\sqrt{I - J^2} \,\left( \cos (\psi) \cdot \vec{j}\: +\: 
\sin (\psi) \cdot\vec{k}\,\right)
\nonumber
\end{eqnarray}
Here $\:J\:$ is the projection of the spin vector on the 
$\:\vec{i}$-axis,
$\:\psi\:$
is the polar angle in the transverse plane and 
$\:I \:=\: {\mid \vec{s} \mid}^2\,$.
In the new variables the spin part of motion equations (\ref{genernon})
becomes
\begin{eqnarray}
\dot{\psi} \;=\; H_J, \hspace{1.0cm} 
\dot{J} \;=\; -\, H_{\psi}, 
\hspace{1.0cm}
\dot{I}\;=\; 0  
\label{ggg}
\end{eqnarray}
where the Hamiltonian (\ref{generham}) takes on the form
\begin{eqnarray}
H(\tau,\, \vec{x},\, \psi,\, J,\, I) \;=\; 
H(\tau,\, \vec{x},\, \vec{s}(\psi,\, J, \,I)) 
\nonumber
\end{eqnarray}
Unfortunately, when 
\begin{eqnarray}
{\left( (\mbox{grad}_{\:\vec{s}}\: H)
 \cdot \vec{j} \,\right)}^2 \;+\; {\left( (\mbox{grad}_{\:\vec{s}}\: H)
 \cdot \vec{k}\,
\right)}^2 \;\neq\; 0  
\label{generham2}
\end{eqnarray}
this coordinate system cannot be extended onto the whole sphere 
$I = const > 0$, since it has a singularity for $\:I - J^2 = 0$. This
means we need to
have
a whole atlas of local coordinates systems (at least two local coordinate
systems defined by different vectors 
$\:\vec{i}_1$, $\:\vec{j}_1$, $\:\vec{k}_1\:$
and $\:\vec{i}_2$, $\:\vec{j}_2$, $\:\vec{k}_2\:$ 
in (\ref{f1233})) for a complete
description of spin motion on the sphere in the electric and magnetic fields
depending on time and position of the particle. We will have the same
difficulties with any other Darboux coordinates because they are defined by
the topological properties of the sphere.
So the {\bf way pointed
by the Darboux theorem does not look like the most natural or straightforward
approach to the problem of investigation of polarized beam dynamics} and we
prefer to study the equations of spin-orbit motion using initial global
variables $\:\vec x$, $\:\vec s\:$ and the Poisson bracket (\ref{fom2})
(degenerate).

{\bf Remark:} Note that in Darboux coordinates considered here 
the Hamiltonian (\ref{fom3}) linear with respect to spin variables  
takes on the form
\begin{eqnarray}
H \:=\: H_{orbt} \:+\: \left( \vec{W} \cdot \vec{i}\, \right) J 
\:+\: 
\sqrt{I- J^2}
\left( \left( \vec{W} \cdot \vec{j} \,\right) \cos \left( \psi \right) 
\:+\:
\left( \vec{W} \cdot \vec{k}\, \right) \sin \left( \psi \right) \right)
\nonumber
\end{eqnarray}
and the condition (\ref{generham2}) now reads as
\begin{eqnarray}
{\left( \vec{W} \cdot \vec{j}\, \right)}^2 \:+\: 
{\left( \vec{W} \cdot\vec{k}\,\right)}^2\; \neq\; 0
\nonumber
\end{eqnarray}

\subsection{The Properties of Solutions of the Extended Equations 
of Spin-Orbit Motion Independent of the Specific Choice 
of the Hamiltonian}

\hspace*{0.5cm}
Which properties of solutions are independent of the specific choice 
of the Hamiltonian function in the equations of motion (\ref{fom1})? 
These are, for example, properties connected with the existence of 
Casimir functions and with the preservation of the Poisson bracket 
along the trajectories of the Hamiltonian system.

The spin-orbit Poisson bracket (\ref{fom2}) has the Casimir function 
${\mid \vec{s} \mid}^{2}$. This means that the length of the vector 
$\,\vec{s}(\tau)\,$ is preserved during the motion, i.e.
\begin{eqnarray}
\mid \vec{s}(\tau)\mid \;\equiv\; \mid \vec{s}(\tau_0) \mid
\nonumber
\end{eqnarray}

Let 
$\;\vec{z}(\tau)\: =\: \vec{\phi} (\tau,\, \tau_{0},\, \vec{z}_{0})$
be the solution of the system (\ref{genernon}),
where $\;\vec{\phi}(\tau_0,\, \tau_0,\,\vec{z}_0)\:=\:\vec{z}_0$.
Using the Taylor series expansion of the function 
$\;\vec{\phi}(\tau, \,\tau_{0},\,\vec{z}_{0})\;$ with respect 
to spin variables we obtain:
\begin{eqnarray}
\left\{
\begin{array}{ll}
\vec{x}(\tau) \;=\; \vec{F}(\tau,\, \tau_{0},\, \vec{x}_{0}) 
\:+\: O(\mid \vec{s}_{0}\mid) & \\
&\\
\vec{s}(\tau) \;=\; A(\tau,\, \tau_{0},\, \vec{x}_{0}) \cdot 
\vec{s}_{0}\:+\: O(\mid\vec{s}_{0}\mid ^{2} ) &
\end{array}
\right.  
\label{mm5}
\end{eqnarray}
where $A(\tau, \,\tau_{0}, \,\vec{x}_{0})$ is a $3 \times 3$ matrix.
The map (\ref{mm5}) preserves the
Poisson bracket (\ref{fom2}). Using this property we find:

{\bf a)} The Jacobian matrix of the vector-function 
$\vec{F}(\tau,\, \tau_{0},\, \vec{x}_{0})$ is symplectic:
\begin{eqnarray}
\left( {\frac{\partial \vec{F} }{\partial \vec{x}_{0}}} \right) ^{\top} J
\left( {\frac{\partial \vec{F} }{\partial \vec{x}_{0}}} \right) \;=\; J
\nonumber
\end{eqnarray}
Here the $\,6 \times 6\,$ matrix $\:J\:$ is the symplectic unit
\cite{guil} and the
symbol '$\top$' indicates transpose of a matrix.

{\bf b)} Every element of the matrix 
$\:A(\tau, \,\tau_{0}, \,\vec{x}_{0})\:$ is equal to
its own cofactor. For a $3 \times 3$ real nonsingular matrix 
this means that
$\:A(\tau,\, \tau_{0},\, \vec{x}_{0})\:$ is an orthogonal matrix and 
$\:\mbox{det}\,A(\tau,\,\tau_{0},\, \vec{x}_{0}) \:=\: 1\:$, i.e.
\begin{eqnarray}
A(\tau,\, \tau_{0}, \,\vec{x}_{0}) \;\in\; \mbox{SO} (3)  
\nonumber
\end{eqnarray}

We have briefly discussed the properties connected with the existence of
Casimir functions and with the preservation of the Poisson bracket, but
without doubt, Liouville's theorem on the conservation of volume is
one of the most popular properties of classical Hamilton systems (at least in
the accelerator physics) which is independent of the choice of a specific
Hamiltonian. Strictly speaking, this is a statement about the existence of
an integral invariant 
($\mbox{dim}\:\vec z$-dimensional) of the density 
$\:g(\vec z\,)\:\equiv\:1$. 

It is well known \cite{NS} that the system 
of ordinary differential equations
\begin{eqnarray}
{\frac{d \vec{z} }{d \tau}} \;=\; \vec{f} (\tau, \vec{z}\,)  
\nonumber
\end{eqnarray}
with continuously differentiable right hand side
admits a
non-negative continuously differentiable
function $\:g(\vec{z}\,)\:$ (in particular the function 
$\:g(\vec{z}\,)\: \equiv \:1$) which serves as
the density appearing in an integral invariant if and only if
\begin{eqnarray}
\mbox{div}_{\:\vec{z}} \; 
\left(g(\vec{z\,})\: \vec{f}(\tau, \,\vec{z}\,)\right) \;=\; 0  
\label{div}
\end{eqnarray}
For the Hamiltonian system (\ref{gom2}) the equality (\ref{div}) 
has the form
\begin{eqnarray}
\mbox{div}_{\:\vec{z}} \; (g(\vec{z}\,) \; \hat{J}(\vec{z}\,) 
\; \mbox{grad}_{\:\vec{z}} \: H(\tau,\vec{z}\,)) \;=\; 0
\label{divgrad}
\end{eqnarray}
It follows from (\ref{divgrad}) that the Hamilton system (\ref{gom2}) will
preserve the phase space volume for a given Hamiltonian 
function $\:H\:$ if and only if
\begin{eqnarray}
\mbox{div}_{\:\vec{z}} \; (\hat{J} (\vec{z}\,) \;
\mbox{grad}_{\:\vec{z}}\: H) \;=\;
{\frac{{\partial \hat{J}_{ij}} }{{\partial z_i}}} \cdot
{\frac{{\partial H} }{{\partial z_j}}} \;=\;
\sum \limits_{j} \left( \sum \limits_{i} \frac{\partial\hat{J}_{ij}}
{\partial z_i} \right) \frac{\partial H}{\partial z_j} \;=\; 0
\label{intinvc}
\end{eqnarray}

{\bf Example}: Defining in the two-dimensional Euclidean space 
the Poisson bracket by the equality
\begin{eqnarray}
\{ z_1, \;z_2 \} \;=\; z_1
\nonumber
\end{eqnarray}
we will have the Hamiltonian system
\begin{eqnarray}
{\frac{d z_1 }{d \tau}} \;=\; z_1 \: 
{\frac{{\partial H} }{{\partial z_2}}}, 
\hspace{1.0cm} 
{\frac{d z_2 }{d \tau}} \;=\; - z_1 \: 
{\frac{{\partial H} } {{\partial z_1}}}
\label{ex1}
\end{eqnarray}
Suppose $\;H \:=\: a \cdot z_2\,$. 
Then a solution of the system (\ref{ex1}) is
given by means of the formulae
\begin{eqnarray}
z_2(\tau) \;=\; z_2(0), 
\hspace{1.0cm} 
z_1(\tau) \;=\; z_1(0) \cdot e^{a \tau}
\nonumber
\end{eqnarray}
Consequently, an image of the single square
\begin{eqnarray}
0 \:\leq \:z_1(0)\: \leq \: 1, 
\hspace{1.0cm} 
0\: \leq \:z_2(0)\: \leq \: 1
\nonumber
\end{eqnarray}
will have the area $\;e^{a \tau}\;$ after a displacement along
trajectories.
The area is preserved for $\:a\,=\,0\,$, and 
$\:\longrightarrow 0\:$ for $\:a \,<\, 0\,$, and 
$\:\longrightarrow \infty\:$ for $\:a > 0$.

Examining the condition (\ref{intinvc}) we find that 
the Hamiltonian system
will
conserve phase space volume independently from the choice of Hamiltonian
function if and only if
\begin{eqnarray}
\left|\sum\limits_i{\frac{\partial \hat J_{ij}}{{\partial z_i}}}\right| 
\;=\;0, 
\hspace{1.0cm}
j\:=\:1,\ldots ,\mbox{dim}\:\vec z  
\label{intinv2}
\end{eqnarray}
For the spin-orbit Poisson bracket all 
the values (\ref{intinv2})
vanish and consequently, the phase volume in 
the 9-dimensional space
is preserved.

Moreover the phase volume is the integral invariant and in the reduced
8-dimensional phase space, this is a direct product of the two-dimensional
sphere of a fixed radius $\rho >0$ and the 6-dimensional space of orbital
variables (level surface of the Casimir function 
$\:{\mid \vec{s} \mid}^{2}\:$).
To proof this fact consider an arbitrary sufficiently small
domain $\:D\:$ and introduce the spherical coordinates
\begin{eqnarray}
\left\{
\begin{array}{l}
\vec s\cdot \vec i\;=\;\rho \cos \theta \sin \varphi \\
\\
\vec s\cdot \vec j\;=\;\rho \cos \theta \cos \varphi \\
\\
\vec s\cdot \vec k\;=\;\rho \sin \theta
\end{array}
\right.  
\label{intinv3}
\end{eqnarray}
where $\;-{\frac \pi 2}\leq \theta \leq {\frac \pi 2}$,
$\;0\leq\varphi \leq 2\pi\;$ and where the unit vectors $\:\vec{i}$, 
$\:\vec{j}\:$ and
$\:\vec{k}\:$, satisfying the condition 
$\;\vec i\cdot \left[\vec j \times \vec k\:\right]\:=\:1\;$
and forming an orthogonal basis, are chosen
in such a way that the coordinate transformation (\ref{intinv3}) is
nonsingular within some open set which includes the domain $\:D$. In new
variables the equations of motion (\ref{genernon}) have the form
\begin{eqnarray}
{\frac{dx}{d\tau }}\;=\;J \:
\mbox{grad}_{\:\vec x} \: H,
\hspace{1.0cm}
{\frac{d\rho} {d\tau }}\;=\;0
\label{ex2}
\end{eqnarray}
\begin{eqnarray}
{\frac{{d\theta }}{{d\tau }}}\;=\;{\frac 1{{\rho \cos \theta }}}
{\frac{{\partial H}}{{\partial \varphi }}},
\hspace{1.0cm}
{\frac{{d\varphi }}{{d\tau }}}\;=\; -{\frac 1{{\rho \cos \theta }}}
{\frac{{\partial H}}{{\partial \theta }}}
\label{ex3}
\end{eqnarray}
If we neglect the equation
\begin{eqnarray}
{\frac{{d\rho }}{{d\tau }}}\;=\;0  
\nonumber
\end{eqnarray}
it is easy to check that the remaining system of eight equations has the
integral invariant with the density
\begin{eqnarray}
g\;=\;\rho ^2\cos \theta   
\label{ex4}
\end{eqnarray}
but the density (\ref{ex4}) is just the density of a volume in the
8-dimensional phase space considered.

{\bf Remark:} One can check that the  
properties listed above are correct not only for 
the extended but also for the
original triangular system (in fact, this was the 
criterion for their selection). For the triangular system
these properties can be regarded
as properties independent of the choice of 
external electromagnetic field.

\section{Canonical Transformations and 
the Introduction of Machine Coordinates
for Circular Accelerators}

\hspace*{0.5cm}
In the theory of circular accelerators it is useful to describe the
spin-orbit motion in terms of a curvilinear coordinate system
associated with the design orbit. 
In the previous sections we have introduced Hamiltonian extension of 
the equations of classical spin-orbit motion. Hamiltonian
systems have a very special form, and the special form is not preserved by
an
arbitrary change of variables. In this section we describe the
transformations of phase space that are canonical with respect to 
Poisson bracket (\ref{fom2}) and preserve that special form, and which
allow us to make coordinate transformations using the
Hamiltonian function (\ref{fom3}) directly instead of
the equations of motion.

If we linearize the resulting Hamiltonian equations with respect to
spin variables and then neglect the effect of spin on the orbit
motion ({\bf triangular truncation procedure}) we obtain the
transformed version of the triangular system too.

\subsection{Canonical Transformations of Phase Space}

\hspace*{0.5cm}
As was already noted, in a fixed local coordinate system $\:\vec z\:$ the
Poisson bracket is completely defined if we know the values of the elements
of the skew-symmetric matrix $\;{\hat J}\:=\:(\{z_i,\:z_j\})\:$ as
functions of $\:\vec z $. Before introducing the canonical
transformations we will discuss
the converse problem. Under what conditions on the skew-symmetric 
$\:k\times k\:$
matrix $\:\tilde J(\vec z\,)\:$ will the binary operation
\begin{eqnarray}
F_{\tilde J}(f,\,g)\;=\;\mbox{grad}_{\:\vec{z}} \: f \cdot 
\tilde J \;\mbox{grad}_{\:\vec{z}} \: g
\nonumber
\end{eqnarray}
be a Poisson bracket? This operation is automatically bilinear and
antisymmetric and satisfies the Leibnitz rule. Hence the operation 
$\,F_{\tilde J}(*,\,*)\,$ will be the Poisson bracket if and only if it
satisfies
the Jacobi identity. The Jacobi identity written in terms of elements of
the matrix $\,\tilde J\,$ has the form (see, for example 
\cite{nemstep,olver})
\begin{eqnarray}
{\tilde J}_{ml} \cdot {\frac{\partial {\tilde J}_{ij}}{{\partial y_l}}}
\:+\:
{\tilde J}_{jl} \cdot {\frac{\partial {\tilde J}_{mi}}{{\partial y_l}}} 
\:+\:
{\tilde J}_{il}\cdot {\frac{\partial {\tilde J}_{jm}}{{\partial y_l}}}
\:=\:0, 
\hspace{0.5cm}i,j,m\:=\:1,\ldots ,k  
\label{jac}
\end{eqnarray}

Returning to the main theme,
consider a coordinate transformation from old variables $\:\vec{z}\:$ to
new
variables $\:\vec{y}\:$ in the Hamiltonian system of ordinary differential
equations (\ref{gom2})
\begin{eqnarray}
\vec{z} \;=\; \vec{\varphi} (\vec{y}\,)  
\label{*}
\end{eqnarray}
with a nondegenerate Jacobian matrix 
($\mbox{dim}\,\vec{z} \:=\: \mbox{dim}\,\vec{y} \:=\:
\mbox{dim}\,M\: =\: k$)
\begin{eqnarray}
\left( {\frac{\partial \vec{\varphi} }{\partial \vec{y}}} \right) 
\;=\; \left(
\begin{array}{ccc}
{\frac{\partial {\varphi}_1 }{\partial y_1}} & \cdots & 
{\frac{\partial {\varphi}_1 }{\partial y_k}} \\
\vdots &  & \vdots \\
{\frac{\partial {\varphi}_k }{\partial y_1}} & \cdots & 
{\frac{\partial {\varphi}_k }{\partial y_k}}
\end{array}
\right)  
\nonumber
\end{eqnarray}
If we take into account the connection between the gradients of
the Hamiltonian
function $\:H\:$ in old and new variables
\begin{eqnarray}
\left. 
\left({\frac{\partial \vec{\varphi}}{\partial \vec{y}}}\right)^{- \top}
\mbox{grad}_{\:\vec{y}} \; H(\tau,\, \vec{\varphi}(\vec{y}\,))
\;=\; \mbox{grad}_{\:\vec{z}} \; H(\tau, \,\vec{z}\,)
\right|_{\vec{z} = \vec{\varphi}(\vec{y}\,)}
\label{gradc}
\end{eqnarray}
we easily obtain from (\ref{gom2}) the differential equations 
for $\:\vec{y}$
\begin{eqnarray}
{\frac{{d \vec{y}} }{{d \tau}}} \;=\; 
\left( {\frac{ \partial \vec{\varphi} }{\partial \vec{y}}} \right)^{-1} 
\hat{J} (\vec{\varphi}(\vec{y}\,))\; 
\left({\frac{\partial \vec{\varphi} }{\partial \vec{y}}} \right)^{- \top} 
\mbox{grad}_{\:\vec{y}} \; H(\tau,\, \vec{\varphi}(\vec{y}\,))  
\label{nbh}
\end{eqnarray}
We now introduce the function $\;\tilde{H}(\tau,\, \vec{y}\,) \:=\: 
H(\tau,\,\vec{\varphi}(\vec{y}\,))\;$ and the skew-symmetric matrix
\begin{eqnarray}
\tilde{J}(\vec{y}\,) \;=\; 
\left({\frac{\partial \vec{\varphi}}{\partial \vec{y}}}\right)^{-1} 
\hat{J} (\vec{\varphi}(\vec{y}\,)) 
\left({\frac{\partial \vec{\varphi}}{\partial \vec{y}}}\right)^{-\top}  
\nonumber
\end{eqnarray}
and rewrite the system (\ref{nbh}) in the form
\begin{eqnarray}
{\frac{{d \vec{y}} }{{d \tau}}} \;=\; \tilde{J}(\vec{y}\,)
\;\mbox{grad}_{\:\vec{y}}
\; \tilde{H}  
\label{ghhg}
\end{eqnarray}
Equation (\ref{ghhg}) resembles a Hamiltonian system with a new
Hamiltonian function $\tilde{H}(\tau, \vec{y})$. But in fact 
it will actually only be a Hamiltonian system
(independently from the choice of specific Hamiltonian) if
the matrix $\tilde{J}$ satisfies the Jacobi identity (\ref{jac}). This
condition looks very complicated, but there are two important situations
when it becomes trivial:

{\bf 1. The matrix $\;\tilde{J}(\vec{y}\,)\;$ is constant} 
(independent from $\:\vec{y}\:$).

{\bf 2. The matrix $\;\tilde{J}(\vec{y}\,)\;$ is equal to the 
matrix $\;\hat{J}(\vec{y}\,)$}.

{\bf Example 1:} For the Darboux coordinates (\ref{f1233}) the 
matrix $\tilde{J}$ has the form (\ref{jmnm}) and hence is constant.

{\bf Example 2:} Introduce the new spin-orbit variables by the equations
\begin{eqnarray}
\left\{
\begin{array}{l}
\vec{x}_{old} \;=\; \vec{F}(\vec{x}_{new}) \\
\\
\vec{s}_{old} \;=\; \vec{s}_{new}
\end{array}
\right.  
\label{tex}
\end{eqnarray}
If the Jacobian matrix of the vector-function $\vec{F}$ is symplectic then
the matrix $\tilde{J}$ is equal to the matrix $\hat{J}$ (\ref{jm}).

Now we are ready to introduce the necessary definitions.

{\bf Definition 1:} {\sl A coordinate transformation 
$\vec{z} = \vec{\varphi}(\vec{y}\,)$ is a 
{\bf Poisson transformation} of phase space if the
matrix $\;\tilde{J}(\vec{y}\,)\;$ satisfies the Jacobi identity
(\ref{jac}).}

{\bf Definition 2:} {\sl A coordinate transformation 
$\vec{z} = \vec{\varphi}(\vec{y}\,)$ is a {\bf Canonical transformation}
of phase space if
the matrix $\;\tilde{J}(\vec{y}\,)\;$ is equal to the matrix
$\;\hat{J}(\vec{y}\,)$.}

The Poisson transformations preserve the Hamiltonian form of 
the initial system,
but their subset, canonical transformations, has additional helpful
properties. Rewrite the condition 
$\;\tilde{J}(\vec{y}\,) \:=\: \hat{J}(\vec{y}\,)\;$ in
the form
\begin{eqnarray}
\hat{J}(\vec{\varphi}(\vec{y}\,)) \;=\; 
\left({\frac{\partial \vec{\varphi}}{\partial \vec{y}}}\right)
\:\hat{J}(\vec{y}\,)\: 
\left({\frac{\partial \vec{\varphi}}{\partial \vec{y}}}\right)^{\top}
\label{lgtl}
\end{eqnarray}
The equality (\ref{lgtl}) means that the image of the Poisson brackets of
coordinates functions is equal to the Poisson brackets of the images. Thus a
canonical transformation is an $''$isometry$''$ of phase space, and we can
find
the Hamiltonian equations in new variables by just making the coordinate
transformation in the Hamiltonian function (as in the usual way).

{\bf Example 3:} The solution 
$\vec{z}(\tau)\, =\, \vec{\phi} (\tau, \,{\tau}_{0},\,\vec{z}_{0})$ 
of the Hamiltonian system (\ref{fom1}) for any fixed 
$\:\tau \,\geq\, \tau_0\:$ is a canonical transformation of phase space:
\begin{eqnarray}
\hat{J}(\vec{\phi}(\tau, \,{\tau}_{0}, \,\vec{z}_{0})) \;=\;
\left(
{\frac{\partial \vec{\phi} }{\partial \vec{z}_{0}}} \right)\:
\hat{J}(\vec{z}_{0}) \:
\left({\frac{\partial \vec{\phi}}{\partial \vec{z}_{0}}}\right)^{\top}
\label{lgtl1}
\end{eqnarray}

{\bf Remark 1}: The condition for the map (\ref{*}) to be
symplectic is often used not in the form following 
from (\ref{lgtl})
\begin{eqnarray}
\left(\frac{\partial \vec{\varphi}}{\partial \vec{y}}\right) 
\;J\;
\left(\frac{\partial \vec{\varphi}}{\partial \vec{y}}\right)^{\top}
\;= \;J  
\label{lgtl_001}
\end{eqnarray}
where the matrix $\:J\:$ is the symplectic unit, but in the form
\begin{eqnarray}
\left(\frac{\partial \vec{\varphi}}{\partial \vec{y}}\right)^{\top}
\;J\;
\left(\frac{\partial \vec{\varphi}}{\partial \vec{y}}\right)
\;= \;J  
\label{lgtl_002}
\end{eqnarray}
It can be easily shown that since $\:J^2\, =\, -I\:$
these two
conditions are equivalent, but it is not 
true in general that (\ref{lgtl_001}) and (\ref{lgtl_002})
are equivalent  
even for a
constant nondegenerate matrix $\:\hat{J}\:$ in (\ref{lgtl}). 
To illustrate this let us introduce 
a Poisson bracket 
in four-dimensional 
Euclidean space by means of the structural matrix
\begin{eqnarray}
\hat{J} \;=\;
\left(
\begin{array}{rrrr}
 0 & 1 &  0 & 0 \\
-1 & 0 &  0 & 0 \\
 0 & 0 &  0 & a \\
 0 & 0 & -a & 0 
\end{array}
\right), 
\hspace{1cm} 
a \;=\; const
\nonumber
\end{eqnarray}
and consider the map  
\begin{eqnarray}
z_1 \;=\; y_1, 
\hspace{0.5cm}
z_2 \;=\; y_2 - y_4,
\hspace{0.5cm}
z_3 \;=\; a \: y_1 + y_3,
\hspace{0.5cm}
z_4 \;=\; y_4
\nonumber
\end{eqnarray}
which is generated as a displacement 
along trajectories of the Hamiltonian system 
with $\:H = z_1 \cdot z_4\:$ for the time $\:\tau = 1$.
Calculating the Jacobian matrix of this map we can check that
(\ref{lgtl_001}) is satisfied 
but that (\ref{lgtl_002}) gives  
\begin{eqnarray}
\left(\frac{\partial \vec{z}}{\partial \vec{y}}\right)^{\top}
\hat{J}\;
\left(\frac{\partial \vec{z}}{\partial \vec{y}}\right)
\;= \;\hat{J} \:+\:
\left(
\begin{array}{cccc}
0     & 0 & 0 & a^2-1 \\
0     & 0 & 0 & 0     \\
0     & 0 & 0 & 0     \\
1-a^2 & 0 & 0 & 0 
\end{array}
\right) \;\neq\; \hat{J}
\hspace{0.5cm}
\mbox{if}
\hspace{0.5cm}
a^2 \:\neq\: 1
\nonumber
\end{eqnarray}

{\bf Remark 2}: Let a Poisson transformation satisfy the equality 
\begin{eqnarray}
\hat{J} (\vec{\varphi}(\vec{y}\,)) \;=\; 
c \; 
\left({\frac{{\partial \vec{\varphi}}}{{\partial \vec{y}}}} \right) 
\:\hat{J}(\vec{y}\,)\: 
\left( {\frac{{\partial \vec{\varphi}}}
{{\partial\vec{y}}}}\right)^{\top}
\label{retbac}
\end{eqnarray}
for some nonzero constant $\,c$.
If the relation (\ref{retbac}) holds, then
in the new local coordinates the structural matrix of the 
Poisson bracket is determined by the matrix 
$\;\tilde{J}(\vec y\,) = c \: \hat{J}(\vec{y}\,)\:$.
If $\:c \neq 1\:$ this transformation is not 
canonical. Sometimes for $\:c \neq 1\:$ it is useful to
introduce for the transformed system a {\bf new Poisson bracket}, 
determined by the matrix 
$\;\tilde{J}(\vec y\,) \,=\,\hat{J}(\vec{y}\,)\,$. 
Defining the new Hamiltonian by means of the rule
\begin{eqnarray}
\tilde{H}(\tau,\,\vec{y}\,) \;=\; 
c \cdot H(\tau,\,\vec{\varphi}(\vec{y}\,))  
\nonumber
\end{eqnarray}
we can consider the transformed system as being Hamiltonian with the
$''$same$''$ Poisson bracket. For instance, in accelerator physics this
method
is used in studying betatron oscillations when the transverse
momenta are normalized by the value of the kinetic momentum of a 
reference particle, which is assumed be a constant.

\subsection{Nonautonomous Canonical Transformations}

\hspace*{0.5cm}
Except for transformations of the type (\ref{*}), in this paper more
general transformations of variables are used
\begin{eqnarray}
\vec{z} \;=\; \vec{\varphi}(\tau,\, \vec{y}\,)
\label{ct1}
\end{eqnarray}
depending on $\tau$ as a parameter. Then we need to answer the
question: $''$In what case 
is the transformation (\ref{ct1}) canonical and how
is the new Hamiltonian to be calculated?$''$

{\bf Definition 3:} {\sl A coordinate transformation 
$\vec{z}=\vec{\varphi}(\tau, \vec{y}\,)$ is a 
{\bf nonautonomous canonical transformation}
if for any fixed $\:\tau\:$
the condition (\ref{lgtl}) holds and there is a differentiable
function 
$\:\hat{F}(\tau,\,\vec{y}\,)\:$ such that
\begin{eqnarray}
\hat{J}(\vec{y}\,) \cdot \mbox{grad}_{\:\vec{y}\,}\; \hat{F} 
\;=\; 
-\left({\frac{{\partial \vec{\varphi}}}{{\partial \vec{y}}}}\right)^{-1} 
\cdot {\frac{{\partial\vec{\varphi}} }{{\partial \tau}}}  
\label{ct2}
\end{eqnarray}
Then the new Hamiltonian is given by the formula
\begin{eqnarray}
\tilde{H}(\tau,\, \vec{y}\,) \;=\; 
H(\tau,\, \vec{\varphi}(\tau,\,\vec{y}\,)) 
\:+\: \hat{F}(\tau,\, \vec{y}\,)  
\nonumber
\end{eqnarray}
}

{\bf Example 4:} Defining in the two-dimensional Euclidean 
space the Poisson
bracket by the equality
\begin{eqnarray}
\{z_1, \;z_2\} \;=\; z_1 \:+\: z_2  
\nonumber
\end{eqnarray}
it easy to check that the coordinate transformation
\begin{eqnarray}
\left\{
\begin{array}{l}
z_1 \;=\; (1+b)\, y_1 \:+\: b\, y_2 \:+\: c \\
\\
z_2 \;=\; -b\, y_1 \:+\: (1-b)\, y_2 \:-\: c
\end{array}
\right.  
\label{ddhh6}
\end{eqnarray}
is canonical independently from the choice of constants $b$ and $c$. 
Letting now $b$ and $c$ be functions of $\tau$ we find that for the set
$\:y_1 + y_2 \,\neq\, 0\:$ the equation (\ref{ct2}) has the solution
\begin{eqnarray}
\hat{F}(\tau,\, y_1,\, y_2) \;=\; - (y_1 + y_2)\, \frac{d b}{d \tau} 
\;-\;\mbox{sign}(y_1 + y_2)\, \ln |y_1 + y_2|\, \frac{d c}{d \tau}
\nonumber
\end{eqnarray}
and that this solution is unique up to an additive arbitrary
function
of the variable $\tau$. This means that if
\begin{eqnarray}
\frac{d c}{d \tau} \;\neq\; 0  
\nonumber
\end{eqnarray}
we cannot consider the transformation (\ref{ddhh6}) to be a nonautonomous
canonical transformation because the equation (\ref{ct2}) does not have
differentiable
solutions (at least in the neighbourhood of the set $y_1 + y_2 = 0$).
So for general Poisson brackets the solvability of the equation
(\ref{ct2}) does not follow from satisfying the condition (\ref{lgtl})
for all values of $\tau$.

We will not set ourselves
the target of studying the general properties of
nonautonomous canonical transformations, 
but instead we
consider some examples directly
connected with the purpose of this paper.

{\bf Example 5:} Let 
$\vec{z}(\tau) = \vec{\phi}(\tau, \tau_0, \vec{z_0})$
be a solution of a canonical system with the Hamiltonian 
$\;\tilde{F}(\tau, \,\vec{z}\,)\;$, so that
\begin{eqnarray}
\left.{\frac{{\partial \vec{\phi}} }{{\partial \tau}}} \;=\; 
\hat{J} (\vec{\phi}\,)
\cdot \mbox{grad}_{\:\vec{z}} \; \tilde{F}(\tau,\, \vec{z}\,)
\right|_{\vec{z} = \vec{\phi}}
\label{ct3}
\end{eqnarray}
Taking into account (\ref{gradc}) and (\ref{lgtl1}) one obtains 
from (\ref{ct3})
\begin{eqnarray}
\left({\frac{{\partial \vec{\phi}}}{{\partial z_0}}} \right)^{-1} 
\frac{\partial \vec{\phi}}{\partial \tau} \;=\; 
\hat{J} (\vec{z}_0) \cdot 
\mbox{grad}_{\:\vec{z}_0} \; 
\tilde{F}(\tau,\, \vec{\phi}(\tau,\, \tau_0,\, \vec{z_0}))
\label{ct4}
\end{eqnarray}
From (\ref{ct4}) and (\ref{lgtl1}) it follows that a 
coordinate
transformation
\begin{eqnarray}
\vec{z} \;=\; \vec{\phi} (\tau,\, \tau_0,\, \vec{y}\,)  
\nonumber
\end{eqnarray}
is nonautonomous canonical and the new Hamiltonian is given by
\begin{eqnarray}
\tilde{H}(\tau, \,\tau_0,\, \vec{y}\,)\; =\; 
H(\tau,\, \vec{\phi}(\tau,\, \tau_0, \,\vec{y}\,)) \:-\: 
\tilde{F} (\tau,\, \vec{\phi}(\tau,\, \tau_0,\, \vec{y}\,))  
\label{ct5}
\end{eqnarray}
As a particular case, the formula (\ref{ct5}) contains the Hamiltonian
version of the method of variation of constants when 
$H = H_1 + H_2$ and the function $\:\tilde{F}\:$ is chosen to be 
$\:H_1\:$.

{\bf Example 6:} Generalizing example 2 consider the transformation:
\begin{eqnarray}
\left \{
\begin{array}{l}
\vec{x}_{old} \;=\; \vec{F} (\tau, \,\vec{x}_{new}) \\
\\
\vec{s}_{old}\; =\; \vec{s}_{new}
\end{array}
\right.  
\nonumber
\end{eqnarray}
With the assumption that the Jacobian matrix of the vector function 
$\vec{F}(\tau, \vec{x})$ with respect to the variables $\:\vec{x}\:$ 
is symplectic for all
values of $\:\tau\,$, the equation (\ref{ct2}) is reduced to
\begin{eqnarray}
\mbox{grad}_{\:\vec{x}} \;\hat{F}\; =\; 
\left( {\frac{{\partial \vec{F}} }{{\partial \vec{x}}}} \right)^{\top} 
J \hspace{0.15cm} 
{\frac{{\partial \vec{F}} }{{\partial \tau}}}
\label{ct6}
\end{eqnarray}
The same symplecticity condition allows us to show that the matrix
\begin{eqnarray}
{\frac{{\partial} }{{\partial \vec{x}}}} 
\left( 
\left( {\frac{{\partial \vec{F}} }{{\partial \vec{x}}}} \right)^{\top} 
J \hspace{0.15cm} 
{\frac{{\partial \vec{F}} }{{\partial \tau}}} 
\right)  
\nonumber
\end{eqnarray}
is symmetric and, consequently, that the equation (\ref{ct6}) has a
solution
(defined up to some additive Casimir function).

Omitting a proof, we also point out that if only orbital variables are
transformed, then we can use the classical technique of generating
functions.

{\bf Example 7:} The linear transformation of spin variables
\begin{eqnarray}
\left \{
\begin{array}{l}
\vec{x}_{old} \;=\; \vec{x}_{new} \\
\\
\vec{s}_{old} \;=\; A(\tau) \; \vec{s}_{new}
\end{array}
\right.
\label{isco}
\end{eqnarray}
will satisfy (\ref{lgtl}) if and only if 
$A \in \mbox{SO}(3)$ for all
values of $\tau$. The condition (\ref{ct2}) now becomes: 
\begin{eqnarray}
\left(
\begin{array}{rrr}
0 & \frac{\partial \hat{F}}{\partial s_3} &
-\frac{\partial \hat{F}}{\partial s_2} \\
& & \\
-\frac{\partial \hat{F}}{\partial s_3} & 0 &
\frac{\partial \hat{F}}{\partial s_1} \\
& & \\
\frac{\partial \hat{F}}{\partial s_2} & -
\frac{\partial \hat{F}}{\partial s_1} & 0
\end{array}
\right)  \;=\; A^{\top} \;\frac{d A}{d \tau}
\label{gytra}
\end{eqnarray}
where the
function
$\hat{F}$
does not depend on $\vec{x}$.

Taking the derivative with respect to $\:\tau\:$ in the identity
\begin{eqnarray}
 A^{\top}(\tau)\: A(\tau)\; =\; I
\nonumber
\end{eqnarray}
we find that the matrix
\begin{eqnarray}
 A^{\top} \:\frac{d A}{d \tau}
\nonumber
\end{eqnarray}
is skewsymmetric, and hence that (\ref{gytra}) has a solution 
which can
be
expressed as follows
\begin{eqnarray}
\hat{F} \;=\;
\left( A^{\top} \frac{d A}{d \tau}\right)_{23} \cdot s_1 \:-\:
\left( A^{\top} \frac{d A}{d \tau}\right)_{13} \cdot s_2 \:+\:
\left( A^{\top} \frac{d A}{d \tau}\right)_{12} \cdot s_3
\nonumber
\end{eqnarray}
So (\ref{isco}) will be a nonautonomous canonical transformation for 
an arbitrary differentiable matrix $A \in \mbox{SO}(3)$.

\subsection{The Coordinate Frame Connected 
with the Closed Design Orbit}

\hspace*{0.5cm}
In this section the words $''$the closed design orbit$''$ mean some
suitable
closed curve which has a continuous unit tangent vector. Let the closed
design orbit be described by the vector $\:\vec{r}_0(z)\,$, where $\:z\:$
is the
length along this curve. Supplement the unit tangent vector
\begin{eqnarray}
\vec{T} \;=\; \frac{d \vec{r}_0}{d z}  
\nonumber
\end{eqnarray}
with two unit 
vectors $\vec{N}$
and $\vec{B}$ satisfying the conditions
\begin{eqnarray}
\vec{B}\; =\; \left[ \vec{T} \times \vec{N}\,\right], 
\hspace{0.5cm}
\vec{N} \;=\; \left[ \vec{B} \times \vec{T}\,\right], 
\hspace{0.5cm} 
\vec{T} \;=\; \left[ \vec{N} \times \vec{B}\,\right]
\nonumber
\end{eqnarray}
or, equivalently
\begin{eqnarray}
\left[ \vec{T} \times \vec{N}\,\right] \cdot \vec{B} \;=\; 1
\nonumber
\end{eqnarray}
The triplet $\:\vec{T}\:$, $\:\vec{N}\:$, $\:\vec{B}\:$ thus forms
an orthogonal right handed coordinate system. 
We assume that the evolution of this coordinate system as
the variable $z$  changes is described by a periodic solution of the
system of ordinary differential equations of the Fresnet type
\begin{eqnarray}
{\frac{d \vec{T} }{d z}} & = & -h \vec{N} - \alpha \vec{B}  
\nonumber \\
{\frac{d \vec{N} }{d z}} & = & +h \vec{T} + \mbox{\ae} \vec{B}  
\nonumber \\
{\frac{d \vec{B} }{d z}} & = & +\alpha \vec{T} - \mbox{\ae} \vec{N}  
\nonumber
\end{eqnarray}
We will find the transformation from the old spin-orbit coordinates to the
new spin-orbit coordinates connected with the vectors $\vec{T}$, $\vec{N}$
, $\vec{B}$ as a composition of two successive transformations: the first
one
changes the orbital variables and second one changes the spin variables and
the longitudinal momentum.

\subsubsection{Transformation of Orbital Variables}

\hspace*{0.5cm}
In the new coordinate system an arbitrary orbit-vector $\;\vec{r}\;$
lying in a sufficiently small neighbourhood of the closed design orbit can
be
written in the form
\begin{eqnarray}
\vec{r} \;=\; \vec{r}_0(z) \:+\: x\, \vec{N} \:+\: y\, \vec{B}  
\nonumber
\end{eqnarray}
We will take the parameters
$\;z,\, x,\, y\;$ to be the new orbital
variables. The
transition from the old coordinates to the new coordinates is made, as
usual, with the help of the generating function depending on the new
position and old momentum variables
\begin{eqnarray}
F(\vec{r},\, \vec{p}\,) \;=\; 
-\left(\vec{r}_0(z) \:+\: x \,\vec{N} \:+\: y\,\vec{B}\,\right) 
\cdot \vec{p}
\nonumber
\end{eqnarray}
The new momenta are given by the equations
\begin{eqnarray}
P_z & = & -{\frac{{\partial F} }{{\partial z}}} \; = \; \vec{p} \cdot
\left((1+hx+\alpha y)\,\vec{T} \:+\:\mbox{\ae}\, 
\left(x\,\vec{B} \:-\: y\,\vec{N}\,\right)\right)  
\label{tre1} \\
\nonumber
\\
P_x & = & -{\frac{{\partial F} }{{\partial x}}} \; = \; \vec{p} \cdot
\vec{N}
\label{tre2} \\
\nonumber
\\
P_y & = & -{\frac{{\partial F} }{{\partial y}}} \; = \; \vec{p} \cdot
\vec{B}
\label{tre3}
\end{eqnarray}
This transformation is a canonical transformation 
with respect to the spin-orbit
Poisson bracket.

\subsubsection{Transformation of Spin Variables 
and Longitudinal Momentum}

\hspace*{0.5cm}
The new spin variables are introduced via the equation
\begin{eqnarray}
\vec{s}_{old} \;=\; C(z)\: \vec{s}_{new}  
\label{tspn}
\end{eqnarray}
where 
$\;C(z)\:=\:\left(\vec{N}(z),\, \vec{B}(z),\, \vec{T}(z)\,\right)$
is a $\:3 \times 3\:$ matrix.

Since the matrix $C$ depends on the variable $z$ the coordinate
transformation
(\ref{tspn}) is not canonical. To make it canonical we change the
longitudinal momentum too:
\begin{eqnarray}
P_z^{old} \;=\; P_z^{new} \:-\: \alpha\, s_x \:+\: h\, s_y 
\:-\: \mbox{\ae}\,s_z 
\label{ts}
\end{eqnarray}
Here $\;s_x,\, s_y,\, s_z\;$ are the components of the spin vector
$\;\vec{s}_{new}\,$.

The coordinate transformation (\ref{tspn}), (\ref{ts}) satisfies the
condition (\ref{lgtl}) and hence is canonical. Since the old Hamiltonian
contains all vectors as scalar and cross products then the new Hamiltonian
will have the same form as the old one if we imagine that all vectors in
(\ref{fom3}) are written in terms of the projections on the unit vectors 
$\:\vec{N}$, $\:\vec{B}$, $\:\vec{T}\:$ and we take into account the
formulae
\begin{eqnarray}
p_{\vec{N}} \;=\; \vec{p} \cdot \vec{N} \;=\; P_x, 
\hspace{1.0cm}
p_{\vec{B}} \;=\; \vec{p} \cdot \vec{B} \;=\; P_y  
\nonumber
\end{eqnarray}
\begin{eqnarray}
p_{\vec{T}} \;=\; \vec{p} \cdot \vec{T} \;=\; 
\frac{1}{1+hx+\alpha y}
\left(P_z \,-\, \mbox{\ae}\, (x P_y \,-\, y P_x) \,-\, 
\alpha \,s_x \,+\, h\, s_y
\,-\, \mbox{\ae}\, s_z\right)
\nonumber
\end{eqnarray}
which we can easily obtain from (\ref{tre1})-(\ref{tre3}) and (\ref{ts}).
Here we have reverted to using the symbol $\;P_{z}\;$ instead of
$\;P_{z}^{new}$.

\subsection{Change of Independent Variable in 
Nonautonomous Hamiltonian Equations}

\hspace*{0.5cm}
Let the right hand part of the first equation 
of the Hamiltonian system (\ref{fom1}) satisfy the condition
\begin{eqnarray}
\frac{d z_1}{d \tau} \;=\; \{ z_1,\: H(\tau,\, \vec{z}\,)\}\; \neq\; 0
\label{cvar}
\end{eqnarray}
This means that the variable $\;z_1(\tau)\;$ changes monotonically with 
changing $\:\tau\:$ (strictly increasing or strictly decreasing) 
so that one
can introduce it as new independent variable. In many cases of practical
importance the new $''$time$''$ scale connected with $\:z_1\:$ gives us
certain
advantages and we wish to discuss the procedure of its introduction for 
Hamiltonian systems.

For convenience we introduce following notation
\begin{eqnarray}
\vec{z}  
\;\stackrel{{\rm def}}{=}\;  
(q,\, p,\, y_1,\, y_2,\, \ldots ,\, y_{k-2}) 
\;=\; (q,\, p,\,\vec{y}\,)
\label{convarq}
\end{eqnarray}
for the components of the vector $\:\vec z\:$
and assume that the matrix $\hat{J}$ for the Hamiltonian equations 
(\ref{fom1}) has the form
\begin{eqnarray}
\hat{J}(\vec{z}\,) \;=\; \left(
\begin{array}{rcccc}
0 & 1 & 0 & \ldots & 0 \\
-1 & 0 & 0 & \ldots & 0 \\
0 & 0 &  &  &  \\
\vdots & \vdots &  & \bar{J}(\vec{y}) &  \\
0 & 0 &  &  &
\end{array}
\right)  
\label{newj}
\end{eqnarray}
This is not a restriction, because on the one hand the spin-orbit Poisson
bracket (\ref{fom2}) has the necessary form and on the other hand any matrix
$\hat{J}$ can be brought into this form if the condition (\ref{cvar}) holds.
With the new notations the system (\ref{fom1}) becomes
\begin{eqnarray}
\frac{d q}{d \tau} =  \frac{\partial H}{\partial p}, 
\hspace{0.3cm} 
\frac{d p}{d \tau} = -\frac{\partial H}{\partial q}, 
\hspace{0.3cm}
\frac{d y_i}{d \tau} = \bar{J}_{ij}(\vec{y}\,) 
\frac{\partial H}{\partial y_j}, 
\hspace{0.3cm} i = 1, \ldots, k-2
\label{se1}
\end{eqnarray}
and the condition (\ref{cvar}) now reads as
\begin{eqnarray}
\frac{\partial H}{\partial p} \;\neq\; 0  
\label{cvar1}
\end{eqnarray}

We will interpret the procedure of changing the independent variable as
a procedure of the reduction of an autonomous Hamiltonian system to a
family of nonautonomous Hamiltonian equations of smaller dimension
defined on the level surfaces of the initial Hamiltonian function. With this
aim in mind we introduce two additional canonical variables 
($E,\upsilon$) and
the new Hamiltonian
\begin{eqnarray}
{\cal H}(\upsilon,\, E, \,\vec{z}\,) \;=\; H(\upsilon, \,\vec{z}\,) 
\;-\; E
\nonumber
\end{eqnarray}
to obtain an autonomous Hamiltonian system in a higher dimensional phase
space
\begin{eqnarray}
\frac{d q}{d \tau}\;=\; \frac{\partial {\cal H}}{\partial p}, 
\hspace{1.0cm} 
\frac{d p}{d \tau}\;=\;-\frac{\partial {\cal H}}{\partial q}  
\label{dfer1}
\end{eqnarray}

\begin{eqnarray}
\frac{d y_i}{d \tau}\;=\;\bar{J}_{ij}(\vec{y}\,)\: 
\frac{\partial {\cal H}}{\partial y_j}, 
\hspace{0.7cm} 
i \;=\; 1, \ldots, k-2  
\label{dfer2}
\end{eqnarray}

\begin{eqnarray}
\frac{d E}{d \tau}\;=\; \frac{\partial {\cal H}}{\partial \upsilon},
\hspace{1.0cm} 
\frac{d \upsilon}{d \tau}\;=\; 
-\frac{\partial {\cal H}}{\partial E}\;=\;1  
\label{dfer3}
\end{eqnarray}

Using the condition (\ref{cvar1}) we obtain the differential equations for
the new independent variable $q$
\begin{eqnarray}
{\frac{{d y_i} }{{d q}}} \;=\; 
{\frac{{d y_i} }{{d \tau}}} \cdot 
{\frac{{d \tau}}{{d q}}} \;=\; \bar{J}_{ij}(\vec{y}\,) \:
{\frac{{\partial {\cal H} / \partial y_j}}
{{\partial {\cal H} / \partial p}}}, 
\hspace{0.5cm} 
i \;=\; 1, \ldots, k-2
\label{fer2}
\end{eqnarray}

\begin{eqnarray}
{\frac{{d E}}{{d q}}} \:=\: {\frac{{d E} }{{d \tau}}} \cdot 
{\frac{{d \tau}}{{d q}}} \:=\: 
{\frac{{\partial {\cal H} / \partial \upsilon}}
{{\partial {\cal H} / \partial p}}}, 
\hspace{1.0cm} 
{\frac{{d \upsilon} }{{d q}}} \:=\: 
{\frac{{d \upsilon} }{{d \tau}}} \cdot {\frac{{d \tau} }{{d q}}} \:=\:
-{\frac{{\partial {\cal H} / \partial E} }
{{\partial {\cal H} / \partial p}}}  
\label{fer3}
\end{eqnarray}
We now  wish to show that the equations (\ref{fer2}), (\ref{fer3}) are
the family of nonautonomous Hamiltonian systems defined on the level
surfaces
\begin{eqnarray}
{\cal H}(E,\, \upsilon,\, q,\, p,\, \vec{y}\,) \;=\; c_0
\label{iiii}
\end{eqnarray}
Let
\begin{eqnarray}
p\; =\; K(E,\, \upsilon,\, \vec{y},\, q,\, c_0)  
\nonumber
\end{eqnarray}
be the solution of the equation (\ref{iiii}). According to the implicit
function theorem this solution exists if the condition (\ref{cvar1})
holds. Taking the
derivative with respect to the variable $\:y_i\:$ in the identity
\begin{eqnarray}
{\cal H}(E, \,\upsilon, \,q, \,K(E, \,\upsilon, \,\vec{y}, \,q,\,
 c_0),\,\vec{y}) \;=\; c_0
\label{newH}
\end{eqnarray}
we have
\begin{eqnarray}
0 \;=\; {\frac{{\partial {\cal H}} }{{\partial y_i}}} \:+\: 
{\frac{{\partial {\cal H}} }{{\partial p}}} \cdot 
{\frac{{\partial K} }{{\partial y_i}}}
\label{newH1}
\end{eqnarray}
From (\ref{newH1}) it follows that
\begin{eqnarray}
\frac{\partial K}{\partial y_i}\; =\; -\, 
\frac{\partial {\cal H} / \partial y_j}{\partial {\cal H} / \partial p}  
\label{newH2}
\end{eqnarray}
Similarly we obtain
\begin{eqnarray}
\frac{\partial K}{\partial E} \;=\; - 
\frac{\partial {\cal H} / \partial E}{\partial {\cal H} / \partial p}, 
\hspace{1.0cm} 
\frac{\partial K}{\partial \upsilon} \;=\; - 
\frac{\partial {\cal H} / \partial \upsilon}
{\partial {\cal H} / \partial p}  
\label{newH3}
\end{eqnarray}
Comparing (\ref{fer2}), (\ref{fer3}) and (\ref{newH2}), (\ref{newH3}) and
introducing the Hamiltonian function
\begin{eqnarray}
\hat{H} \;=\; -\,K(E, \,\upsilon,\, \vec{y},\, q,\, c_0)  
\nonumber
\end{eqnarray}
we see that the equations (\ref{fer2}), (\ref{fer3}) become the family of
nonautonomous 
Hamiltonian systems depending on the parameter $\:c_0$
\begin{eqnarray}
\frac{d E}{d q} = \frac{\partial \hat{H}}{\partial \upsilon}, 
\hspace{0.3cm} 
\frac{d \upsilon}{d q} = -\frac{\partial \hat{H}}{\partial E},
\hspace{0.3cm} 
\frac{d y_i}{d q} = 
\bar{J}_{ij}(\vec{y}\,) 
\frac{\partial \hat{H}}{\partial y_j},
\hspace{0.3cm} 
i = 1,\ldots, k-2  
\label{selop}
\end{eqnarray}
After solving the system (\ref{selop}) for a fixed value of $c_0$
we can find the dependence of $\:p\:$ on $\:q\:$ using the identity
\begin{eqnarray}
p(q) \;=\; K(E(q),\, \upsilon(q),\, \vec{y}(q),\, q,\, c_0)  
\nonumber
\end{eqnarray}
and then determine $\:q\:$ as a function of $\:\tau\:$ from the equation
\begin{eqnarray}
\tau \:-\: {\tau}_0 \;=\; 
\int \limits_{q_0}^{q} \frac{d q}{g(q)}
\label{newer}
\end{eqnarray}
where
\begin{eqnarray}
g(q)\;=\;\frac{\partial {\cal H}}{\partial p} 
(E(q),\, \upsilon(q),\, q,\, p(q),\, \vec{y}(q))\; \neq\; 0  
\nonumber
\end{eqnarray}
(equation (\ref{newer}) follows from the first of equations
(\ref{se1})).

Now we have to remember that we did not start from an autonomous system
(\ref{dfer1})-(\ref{dfer3}) but from nonautonomous system (\ref{se1}). 
This means
that we do not need equation (\ref{newer}) because we have the
dependence $\tau(q)$ from the second equation of (\ref{selop}) ($\upsilon$
is just another notation for $\tau$). We also have freedom in the choice
of the initial condition for the variable $E$. This means that we can
choose
a single fixed value of the parameter $c_0$ (usually $c_0 = 0$) and 
replace
the initial nonautonomous Hamiltonian system by the
nonautonomous Hamiltonian system (\ref{selop}) with the same matrix $\hat{J}$
as in (\ref{newj}).

\subsection{Length Along the Design Orbit as 
Independent Variable}

\hspace*{0.5cm}
For a circular accelerator the spin-orbit Hamiltonian is always  a
periodic
function of the variable $z$, but its dependence on time may be more
complicated (for example, in the acceleration mode). This is one of the
reasons for introducing the length along the design orbit as 
the independent
variable. We will assume that
\begin{eqnarray}
\frac{\partial H}{\partial P_{z}} \;\neq\; 0
\label{incon}
\end{eqnarray}
The condition (\ref{incon}) approximately means that during the
motion in the accelerator the particle cannot reverse its direction.

Following the previous subsection we introduce two 
additional canonical
variables ($E$, $\tau$)\footnote{Reserving for a while the new symbol 
$\tau$ for the time $t$ appearing in the spin-orbit Hamiltonian,
and still reserving the symbol $t$ for the independent variable.}
and the new Hamiltonian
\begin{eqnarray}
{\cal H} \; = \; 
H(\tau,\, x,\, P_x,\, y,\, P_y,\, z,\, P_z,\, s_x,\, s_y,\, s_z) 
\:-\: E  
\nonumber
\end{eqnarray}
\begin{eqnarray}
\frac{d E}{d t} \;=\; \frac{\partial {\cal H}}{\partial \tau}, 
\hspace{1.0cm} 
\frac{d \tau}{d t} \;=\; 
-\,\frac{\partial {\cal H}}{\partial E}\; \equiv \;1
\nonumber
\end{eqnarray}
Now we need to solve the equation
\begin{eqnarray}
{\cal H}(E,\, \tau,\, x,\, P_x,\, y,\, P_y,\, z,\, P_z,\, 
s_x,\, s_y,\, s_z)\;=\;0  
\label{oooo}
\end{eqnarray}
with respect to variable $\:P_{z}\:$ to obtain the new Hamiltonian
\begin{eqnarray}
\hat{H} \;=\; -\,P_{z}(E,\, \tau,\, x,\, P_x,\, y,\, 
P_y,\, s_x,\, s_y,\, s_z,\,z)
\label{o1o1o1}
\end{eqnarray}
The dependence of the Hamiltonian ${\cal H}$ on the variable $P_z$ is more
complicated than in the pure orbital case, but nevertheless, we can solve
the equation (\ref{oooo}) with any required precision with respect to spin
variables using the method of successive iterations.

\subsection{The Hamiltonian in New Variables up to 
First Order with Respect to Spin Variables}

\hspace*{0.5cm}
In this subsection we discuss the general 
form of the spin-orbit Hamiltonian
in the new variables $\:(E,\,t,\,x,\,P_x,\,y,\,P_y)$\footnote{Here 
and in the
Hamiltonian (\ref{o1o1o1}) we have reverted
to using the symbol
$t$ instead of $\tau$.}
up to the first order with respect to
spin variables. This Hamiltonian is
\begin{eqnarray}
\hat{H} \;=\; \hat{H}_{orbt}(E,\,t,\,x,\,P_x,\,y,\,P_y,\,z) \:+\: 
\hat{H}_{spin}(E,\,t,\,x,\,P_x,\,y,\,P_y,\,z,\,\vec{s}\,)  
\label{hnew}
\end{eqnarray}
with
\begin{eqnarray}
\hat{H}_{orbt} \;=\; -\,\mbox{\ae}\,(x\,P_y\:-\:y\,P_x) \:-\: 
(1\:+\:h\,x\:+\:\alpha\, y) \cdot 
\nonumber
\end{eqnarray}
\begin{eqnarray}
\cdot 
\left(\frac{e}{c}\,A_{\vec{T}}\:+\: 
\sqrt{\frac{(E \,-\, e\Phi)^2}{c^2}
- m_0^2\,c^2 - 
\left(P_x - \frac{e}{c}\,A_{\vec{N}}\right)^2 -
\left(P_y - \frac{e}{c}\,A_{\vec{B}}\right)^2} \,\right)  
\;=
\nonumber
\end{eqnarray}
\begin{eqnarray}
=\;
-\,
\mbox{\ae} \,x \,\pi_{\vec B}
\:+\:
\mbox{\ae} \,y \,\pi_{\vec N}
\:-\: 
(1 \:+\: h\,x \:+\: \alpha \,y)\, \pi_{\vec T}
\:-\: {e \over c}\, A_z 
\nonumber
\end{eqnarray}
and 
\begin{eqnarray}
\hat{H}_{spin} \;=\; -\, \alpha\, s_x \:+\: h\, s_y \:-\: 
\mbox{\ae}\, s_z \:+
\nonumber
\end{eqnarray}
\begin{eqnarray}
+\: 
\frac{(1\:+\:h\,x\:+\:\alpha \,y) (E\:-\:e\,\Phi)}
{c^2\,
\sqrt{
\frac{(E-e \Phi)^2}{c^2}\,-\,m_0^2 c^2\,-\, 
\left(P_x-\frac{e}{c}A_{\vec{N}}\right)^2 -
\left(P_y-\frac{e}{c}A_{\vec{B}}\right)^2 
}}\:\vec{W} \cdot \vec{s}\;=
\nonumber
\end{eqnarray}
\begin{eqnarray}
=\;
- \,\alpha \, s_x \:+\: h \, s_y \:-\: \mbox{\ae}\, s_z \:+\:
(1\:+\:h\,x\:+\:\alpha\, y) \,
\frac{m_0 \,\gamma}{\pi_{\vec T}} \, \vec{W} \cdot \vec{s}
\nonumber
\end{eqnarray}
Here $\:\vec{W}\:$ has the same form as in (\ref{fom3}) with 
$\:\vec{\pi}$, $\:\vec{\cal B}$, $\:\vec{\cal E}\:$ 
written in terms of projections on the vectors 
$\:\vec{N}$, $\:\vec{B}$, $\:\vec{T}$
\begin{eqnarray}
\pi_{\vec{N}} \;=\; P_x \:-\: \frac{e}{c}\,A_{\vec{N}},
\hspace{1.0cm}
\pi_{\vec{B}} \;=\; P_y \:-\: \frac{e}{c}\,A_{\vec{B}}  
\nonumber
\end{eqnarray}
\begin{eqnarray}
\pi_{\vec{T}} \;=\; 
\sqrt{
\frac{(E - e\Phi)^2}{c^2} - m_0^2 c^2 -
\left(P_x - \frac{e}{c} A_{\vec{N}}\right)^2 - 
\left(P_y - \frac{e}{c} A_{\vec{B}}\right)^2 
}\;=  
\nonumber
\end{eqnarray}
\begin{eqnarray}
=\; 
\left(
{(E - e \Phi)^2 \over {c^2}} \:-\: m_0^2 c^2  \:-\: 
\pi_{\vec N}^2\:-\:\pi_{\vec B}^2
\right)^{1 / 2}
\nonumber
\end{eqnarray}
and we use the notation
\begin{eqnarray}
A_z \;=\; (1\:+\:h\,x\:+\:\alpha\,y) \,A_{\vec{T}} \:+\: 
\mbox{\ae}\, (x\, A_{\vec{B}} \:-\: y\,A_{\vec{N}})
\nonumber
\end{eqnarray}
The value of $\:\gamma\:$ is defined now through the new canonical
variable $\:E$
\begin{eqnarray}
\gamma \;=\; \frac{E \:-\: e \,\Phi}{m_0\, c^2}
\nonumber
\end{eqnarray}
To obtain the Hamiltonian (\ref{hnew}) we have used the condition
(\ref{incon}) in the form
\begin{eqnarray}
\frac{\partial H}{\partial P_z} \;>\; 0
\nonumber
\end{eqnarray}
so that the vector $\;\vec{T}\;$ is chosen with the same
orientation 
as the direction of particle flight in the accelerator.
 
To complete the description we also give the expressions for the
projections
on $\:\vec{N}$, $\:\vec{B}$, $\:\vec{T}$ of the electric and magnetic
fields
in terms of
the vector and scalar potentials.

{\bf The magnetic field:}
\begin{eqnarray}
{\cal B}_{\vec{N}} \;=\; 
\frac{1}{1\:+\:h\,x\:+\:\alpha\,y} \cdot 
\left(
\frac{\partial A_z}{\partial y}\: -\: 
\frac{\partial A_{\vec{B}}}{\partial z}\:-\: 
\mbox{\ae}\,y \, {\cal B}_{\vec{T}} 
\right)  
\nonumber
\end{eqnarray}
\begin{eqnarray}
{\cal B}_{\vec{B}} \;=\; 
\frac{1}{1\:+\:h\,x\:+\:\alpha\,y} \cdot 
\left(
\frac{\partial A_{\vec{N}}}{\partial z} \:-\: 
\frac{\partial A_z}{\partial x}\:+\: 
\mbox{\ae} \,x\, {\cal B}_{\vec{T}}
\right)  
\nonumber
\end{eqnarray}
\begin{eqnarray}
{\cal B}_{\vec{T}} \;=\; 
\frac{\partial A_{\vec{B}}}{\partial x}\: - \:
\frac{\partial A_{\vec{N}}}{\partial y}  
\nonumber
\end{eqnarray}

{\bf The electric field:}
\begin{eqnarray}
{\cal E}_{\vec{N}} \;=\; -\, 
\frac{\partial \Phi}{\partial x}\: -\:
\frac{1}{c}\, 
\frac{\partial A_{\vec{N}}}{\partial t}
\nonumber
\end{eqnarray}
\begin{eqnarray}
{\cal E}_{\vec{B}} \;=\; -\,
\frac{\partial \Phi}{\partial y}\: -\:
\frac{1}{c}\,
\frac{\partial A_{\vec{B}}}{\partial t}  
\nonumber
\end{eqnarray}
\begin{eqnarray}
{\cal E}_{\vec{T}} \;=\; -\,
\frac{1}{1 + hx + \alpha y} 
\left(
\frac{\partial \Phi}{\partial z} \:+\: 
\mbox{\ae} 
\left( 
y\,\frac{\partial \Phi}{\partial x} \:-\: 
x\,\frac{\partial \Phi}{\partial y} 
\right) 
\right) \:-\: 
\frac{1}{c}\, 
\frac{\partial A_{\vec{T}}}{\partial t}
\nonumber
\end{eqnarray}
(Useful formulae which allow us to get the 
equations for vector and
scalar potentials in our curvilinear coordinate 
system may be found in the
Appendix A).

\section{Linear Differential Equations 
of Spin Motion}

\hspace*{0.5cm}
In this section we discuss the situation in which the 
Hamiltonian function is
linear in the spin variables and does not depend on the 
orbit variables.
In particular,  this case includes the
description of the behaviour of the spin vector on the closed (or any
other chosen)
trajectory of orbital motion, using the triangular system.

\subsection{Matrix Representation of the Hamiltonian 
Function and Simple Properties of Solutions}

\hspace*{0.5cm}
The Hamiltonian function, which is linear in spin 
variables and does not
depend on orbit variables, has the form
\begin{eqnarray}
H(\tau, \,\vec{s}\,) \;=\; \vec{w} (\tau) \cdot \vec{s}
\label{mf001}
\end{eqnarray}
We now construct
the skewsymmetric matrix $\:C(\vec{w})\:$ according to the rule
\begin{eqnarray}
C(\vec{w}) \;=\; 
\left(
\begin{array}{ccc}
0 & -w_3 & w_2 \\
w_3 & 0 & -w_1 \\
-w_2 & w_1 & 0
\end{array}
\right)
\nonumber
\end{eqnarray}
where $w_1$, $w_2$, $w_3$ are the components of the vector $\:\vec w$
\begin{eqnarray}
\vec{w} \;=\; (w_1,\, w_2, \,w_3)  
\nonumber
\end{eqnarray}
It is easy to verify that with the help of the 
matrix $\:C(\vec{w})\:$
the Hamiltonian (\ref{mf001}) can be written as follows
\begin{eqnarray}
H(\tau,\, \vec{s}\,) \;=\; 
\frac{1}{2}\, \vec{s} \cdot \mbox{curl}_{\:\vec{s}}
\:\left(C(\vec{w}) \cdot \vec{s}\,\right)
\label{mf002}
\end{eqnarray}
and the equations of motion take the form
\begin{eqnarray}
\frac{d \vec{s}}{d \tau} \;=\;
C\left(\vec{w}(\tau)\right) \cdot \vec{s}
\label{mf003}
\end{eqnarray}

The matrix representation (\ref{mf002}) of the Hamiltonian function
(\ref{mf001}) resembles the representation of the Hamiltonian of linear
orbit motion as a quadratic form and is particularly convenient for linear
nonautonomous canonical transformations of variables. 
Introduce, for example, the new spin variables $\:\vec{u}\:$
by the equation
\begin{eqnarray}
\vec{s} \;=\; A(\tau) \:\vec{u}, 
\hspace{1.0cm} 
A(\tau) \;\in\; \mbox{SO}(3)
\label{mf004}
\end{eqnarray}
Substituting (\ref{mf004}) in the equation of 
motion (\ref{mf003}), we obtain
\begin{eqnarray}
\frac{d \vec{u}}{d \tau}\; =\; 
\left( A^{\top} C A \:-\: A^{\top}
\frac{d A}{d \tau} \right) \cdot \vec{u}  
\nonumber
\end{eqnarray}
where
\begin{eqnarray}
A^{\top} C(\vec{w}) A \:-\: A^{\top} \frac{d A}{d \tau}  
\nonumber
\end{eqnarray}
is a skewsymmetric matrix again. Consequently, the Hamiltonian 
function in the new variables $\:\vec{u}\:$ can be written in the form
\begin{eqnarray}
H(\tau, \,\vec{u}\,) \;=\; 
\frac{1}{2} \,\vec{u} \cdot 
\mbox{curl}_{\:\vec{u}} 
\left(
\left( 
A^{\top} C A \:-\: A^{\top} \frac{d A}{d \tau} 
\right) \cdot 
\vec{u} 
\right)  
\label{mf00*}
\end{eqnarray}

The following simple properties 
of the matrix notation are almost obvious:
\vspace*{0.3cm}

{\bf a)} $\mbox{curl}_{\:\vec{s}}\:
\left(\:A \cdot \vec{s}\:\right) \:+\:
\mbox{curl}_{\:\vec{s}}\:
\left(\:B\cdot \vec{s}\:\right) \;=\;
\mbox{curl}_{\:\vec{s}}\:\left(\:(A + B) \cdot \vec{s}\:\right)$
\vspace*{0.3cm}

{\bf b)} $A^{\top} C(\vec{w})\:A \;=\; 
C\left(A^{\top} \cdot \vec{w}\right)
\hspace{0.4cm}$ for $\hspace{0.4cm} A \;\in\; \mbox{SO}(3) $ 
\vspace*{0.3cm}

Let $\:M(\tau ,\,\tau _0)\:$ be the fundamental matrix solution of
(\ref{mf003}).
It has been mentioned in subsection 3.3 that for any 
$\:\tau \:\geq\: \tau _0$
\begin{eqnarray}
M(\tau ,\,\tau _0)\;\in\; \mbox{SO}(3)
\label{mf005}
\end{eqnarray}
If the vector $\:\vec w\:$ is constant 
(does not depend on $\tau $) and 
$\;\left|\vec w\right| \:\neq\: 0\;$,
then the
matrix $M$ is determined via the matrix $\;C(\vec w)\;$ 
from the formula
\begin{eqnarray}
M(\tau ,\,\tau_0)\;=\;
\exp\left((\tau -\tau_0)\:C(\vec w)\right)\;=  
\nonumber
\end{eqnarray}
\begin{eqnarray}
I + \sin(\mid\vec{w}\mid (\tau -\tau _0)) \cdot 
\frac{C(\vec w)}{\mid \vec{w}\mid} + 
(1-\cos(\mid \vec{w}\mid(\tau -\tau _0))) \cdot 
\left(\frac{C(\vec{w})}{\mid \vec{w}\mid}\right)^2
\label{mf006}
\end{eqnarray}
In the case $\;\left|\vec w\right|\,=\,0\;$ the matrix $M$ is 
equal to the identity matrix.

The  properties of the solutions of the system of 
linear differential
equations (\ref{mf003}) given below follow from the Hamiltonian character 
of the system (\ref{mf003}) (in the  case studied here it is equivalent 
to the fact that the matrix $C(\vec{w})$ in (\ref{mf003}) is real
skewsymmetric):

{\bf a)} The norm of every solution is conserved.

{\bf b)} The angle between any two solutions is conserved.

{\bf c)} If $\;\vec{n}(\tau)\;$ and $\;\vec{m}(\tau)\;$ are solutions of
the system
(\ref{mf003}), then \\
$\;\vec{l}(\tau) \:=\: [\vec{n}(\tau) \times \vec{m}(\tau)]\;$
is a solution too.

{\bf d)} The function $\;\vec{n}(\tau) \cdot \vec{s}\;$ is a constant of
motion of the system (\ref{mf003}) \\
if and only if $\;\vec{n}(\tau)\;$ is a solution.

{\bf e)} If the vector $\;\vec{w}(\tau)\;$ in (\ref{mf001}) has period 
$T$ in $\tau$, then the system (\ref{mf003}) has a 
$T$-periodic solution.

{\bf f)} If the vector $\;\vec{w}(\tau)\;$ in (\ref{mf001}) 
has period $T$ in $\tau$ and the system (\ref{mf003}) has two 
linearly independent $T$-periodic solutions, then any solution 
of the system (\ref{mf003}) is $T$-periodic.

\subsection{Linear Equations of Spin Motion 
with Periodic Coefficients}

\hspace*{0.5cm}
Now we consider a Hamiltonian function (\ref{mf001}) in 
which the vector $\vec{w}$ has period 
$\:2 \pi\:$ in $\:\tau$, that is
\begin{eqnarray}
\vec{w} (\tau + 2 \pi) \;\equiv\; \vec{w} (\tau)  
\nonumber
\end{eqnarray}
We shall attempt to obtain a Hamiltonian function (\ref{mf00*}) of the
simplest form by means of the linear coordinate substitution (\ref{mf004})
with a $2 \pi$-periodic matrix $A(\tau)$.

\subsubsection{Some Properties of SO(3) Matrices}

\hspace*{0.5cm}
Recall that the symbol $\mbox{SO}(3)$ denotes the group of 
$3 \times 3$ real
orthogonal matrices with the determinant equal to 1.

Let $A \in \mbox{SO}(3)$. Since $A^{-1} = A^{\top}$, 
the spectrum of
this
matrix is symmetric about the unit circle (that is, if $\mu$ is an
eigenvalue of a $\mbox{SO}(3)$ matrix, then so is $\mu^{-1}$). From the
fact
that the matrix is real it follows that the spectrum is also symmetric about
the real axes. Taking into account the condition $\mbox{det}\,A \,=\, 1$,
we obtain
that the eigenvalues of the matrix $A$ have the form
\begin{eqnarray}
\mu_1 \;=\; 1, 
\hspace{1.5 cm} 
\mu_{2,\,3} \;=\; \cos \lambda \;\pm\; i\, \sin \lambda
\nonumber
\end{eqnarray}
where
\begin{eqnarray}
\cos \lambda \;=\; \frac{1}{2} \left(\mbox{Tr}(A) \:-\:1\right)
\nonumber
\end{eqnarray}
The next three lemmas give us some additional helpful 
properties of $\mbox{SO}(3)$ matrices.

{\bf Lemma 1}: {\sl A $3 \times 3$ nonsingular real matrix $A$ belongs to
the $\mbox{\rm SO}(3)$ group if and only if every element of the matrix
$A$
is
equal to its own cofactor.}

Note, if every element of a $3 \times 3$ real singular matrix $A$ is equal
to
its own cofactor then $\;A\, =\, 0$.

{\bf Lemma 2}: {\sl All the eigenvalues of a matrix 
$A \in \mbox{\rm SO}(3)$ are
distinct if and only if $A \neq A^{\top}$ (or, equivalently, $A^2 \neq I$).}

{\bf Lemma 3}: {\sl If all the eigenvalues of a matrix 
$A \in \mbox{\rm SO}(3)$
are distinct then the nonzero vector
\begin{eqnarray}
\vec{n} = \frac{\vec{k}}{|\vec{k}|}, \hspace{0.4cm} \vec{k} =
(a_{32}-a_{23}, \hspace{0.3cm} a_{13}-a_{31}, \hspace{0.3cm} a_{21}-a_{12})
\nonumber
\end{eqnarray}
is a unit eigenvector of the matrix $A$ corresponding to the unit
eigenvalue, i.e. $A \vec{n} = \vec{n}$. }

\subsubsection{Real Jordan Canonical Forms of SO(3) Matrices}

\hspace*{0.5cm}
Let $\:\vec{n}\:$ be an eigenvector of $\:A\:$ corresponding to the unit
eigenvalue
\begin{eqnarray}
A \,\vec{n} \;=\; \vec{n}, 
\hspace{1cm} 
\mid \vec{n} \mid \;=\; 1  
\nonumber
\end{eqnarray}
Supplement the vector $\:\vec{n}\:$ 
with two unit
vectors $\:\vec{m}\:$ and $\:\vec{l}\:$ satisfying the condition
\begin{eqnarray}
\vec{m} \cdot \left[\vec{l} \times \vec{n}\,\right] \;=\; 1
\nonumber
\end{eqnarray}
to form an orthogonal 
basis.
Then the matrix
\begin{eqnarray}
B\;=\;\left(\vec{m},\,\vec{l},\,\vec{n}\,\right)\;\in\;
\mbox{SO}(3)
\nonumber
\end{eqnarray}
In the new basis, constructed with vectors 
$\:\vec{m}$, $\:\vec{l}\:$ and $\:\vec{n}\:$,
the matrix $A$ will have the form
\begin{eqnarray}
\bar{A} \;=\; B^{-1} A B \;=\; B^{\top} A B \;=\; 
\left(
\begin{array}{ccc}
A \vec{m} \cdot \vec{m} & A \vec{l} \cdot \vec{m} & 0 \\
A \vec{m} \cdot \vec{l} & A \vec{l} \cdot \vec{l} & 0 \\
0 & 0 & 1
\end{array}
\right) 
\;\in\; \mbox{SO} (3)
\nonumber
\end{eqnarray}
It follows from the condition $\bar{A} \in \mbox{SO}(3)$ 
that for some angle
$\psi$ we can represent this matrix in the form
\begin{eqnarray}
\bar{A} \;=\; \left(
\begin{array}{ccc}
+\cos(\psi) & +\sin(\psi) & 0 \\
-\sin(\psi) & +\cos(\psi) & 0 \\
0 & 0 & 1
\end{array}
\right)
\nonumber
\end{eqnarray}
Using the equality $\;A B \:=\: B \bar{A}\;$, we obtain
\begin{eqnarray}
A \cdot \vec{m} \;=\; \cos(\psi) \cdot \vec{m}\: -\: 
\sin(\psi) \cdot \vec{l} 
\nonumber
\end{eqnarray}
\begin{eqnarray}
A \cdot \vec{l} \;=\; \sin(\psi) \cdot \vec{m}\: +\: 
\cos(\psi) \cdot \vec{l}
\nonumber
\end{eqnarray}
or, in  complex form,
\begin{eqnarray}
A \cdot 
\left(\:\vec{m} \:+\: i\, \vec{l}\:\right) \;=\;
\exp(i\, \psi) \cdot 
\left(\:\vec{m} \:+\: i\, \vec{l}\:\right)
\label{mf007}
\end{eqnarray}
It follows from (\ref{mf007}) that
$\;\exp(\pm\, i \,\psi)\;$ are the eigenvalues
and $\;\vec{m} \:\pm\: i\, \vec{l}\;$ are
the corresponding eigenvectors of the
matrix $A$
(in particular, 
$\;\psi \:=\: \pm \lambda \; (\mbox{mod} \; 2 \pi)$)

The matrix $\bar{A}$ is a
{\bf real Jordan canonical form} of the matrix $A$. We will denote it
by $\:\bar{A}_1$. All possible canonical forms are given by the list
\begin{eqnarray}
\bar{A}_1 = \left(
\begin{array}{ccc}
+\cos(\psi) & +\sin(\psi) & 0 \\
-\sin(\psi) & +\cos(\psi) & 0 \\
0 & 0 & 1
\end{array}
\right), 
\hspace{0.5cm} 
\bar{A}_2 = \left(
\begin{array}{ccc}
+\cos(\psi) & -\sin(\psi) & 0 \\
+\sin(\psi) & +\cos(\psi) & 0 \\
0 & 0 & 1
\end{array}
\right)  
\nonumber
\end{eqnarray}

\begin{eqnarray}
\bar{A}_3 = \left(
\begin{array}{ccc}
+\cos(\psi) & 0 & +\sin(\psi) \\
0 & 1 & 0 \\
-\sin(\psi) & 0 & +\cos(\psi)
\end{array}
\right), 
\hspace{0.5cm} 
\bar{A}_4 = \left(
\begin{array}{ccc}
+\cos(\psi) & 0 & -\sin(\psi) \\
0 & 1 & 0 \\
+\sin(\psi) & 0 & +\cos(\psi)
\end{array}
\right)  
\nonumber
\end{eqnarray}

\begin{eqnarray}
\bar{A}_5 = \left(
\begin{array}{ccc}
1 & 0 & 0 \\
0 & +\cos(\psi) & +\sin(\psi) \\
0 & -\sin(\psi) & +\cos(\psi)
\end{array}
\right), 
\hspace{0.5cm} 
\bar{A}_6 = \left(
\begin{array}{ccc}
1 & 0 & 0 \\
0 & +\cos(\psi) & -\sin(\psi) \\
0 & +\sin(\psi) & +\cos(\psi)
\end{array}
\right)  
\nonumber
\end{eqnarray}
It is clear that all matrices 
$\;\bar{A}_i\;$ are similar one to another. For
example,
\begin{eqnarray}
B^{\top}_i \,\bar{A} \,B_i \;=\; \bar{A}_i, 
\hspace{1.0cm} 
i = 1, \ldots, 6
\nonumber
\end{eqnarray}
where $\;B_i \:\in\: \mbox{SO}(3)\;$ and
\begin{eqnarray}
B_1 = \left(
\begin{array}{rrr}
1 & 0 & 0 \\
0 & 1 & 0 \\
0 & 0 & 1
\end{array}
\right), \hspace{0.3cm} B_2 = \left(
\begin{array}{rrr}
1 & 0 & 0 \\
0 & -1 & 0 \\
0 & 0 & -1
\end{array}
\right), \hspace{0.3cm} B_3 = \left(
\begin{array}{rrr}
1 & 0 & 0 \\
0 & 0 & 1 \\
0 & -1 & 0
\end{array}
\right)  
\nonumber
\end{eqnarray}

\begin{eqnarray}
B_4 = \left(
\begin{array}{rrr}
1 & 0 & 0 \\
0 & 0 & -1 \\
0 & 1 & 0
\end{array}
\right), \hspace{0.3cm} B_5 = \left(
\begin{array}{rrr}
0 & 0 & 1 \\
0 & -1 & 0 \\
1 & 0 & 0
\end{array}
\right), \hspace{0.3cm} B_6 = \left(
\begin{array}{rrr}
0 & 0 & 1 \\
0 & 1 & 0 \\
-1 & 0 & 0
\end{array}
\right)  
\nonumber
\end{eqnarray}

We note that if $\:\psi\: =\: 0\:(\mbox{mod}\:2\pi)$, 
then all matrices $\:\bar A_i\:$
are equal one to another (and equal to the identity matrix). 
If $\:\psi \:=\:\pi\:(\mbox{mod}\:2\pi)$, 
then $\;\bar A_1\:=\:\bar A_2$,
$\;\bar A_3\:=\:\bar A_4\;$ and 
$\;\bar A_5\:=\:\bar A_6$.

\subsubsection{Skew-symmetric Real Logarithm of SO(3) Matrices}

\hspace*{0.5cm}
If the determinant of the matrix $\:A\:$ is nonzero, then
\begin{eqnarray}
\ln(A) \;=\; K  
\nonumber
\end{eqnarray}
is defined as a solution of the equation
\begin{eqnarray}
\exp(K) \;=\; A  
\label{log1}
\end{eqnarray}

We now consider the problem of finding skewsymmetric real solutions of
equation (\ref{log1}) in the case, when $\;A \:\in\: \mbox{SO}(3)$.

{\bf Lemma 4}: {\sl If $\:A\in \mbox{\rm SO}(3)\:$ and 
$\:A\neq A^{\top}\:$, then
all real logarithms of the matrix $A$ are skewsymmetric 
matrices and can be expressed by formula
\begin{eqnarray}
\ln(A)\;=\;\frac{2\,\ln\lambda}{\lambda \:-\:\lambda^{*}}
\left(A\:-\:A^{\top}\right)  
\nonumber
\end{eqnarray}
where $\:\lambda\:$ and $\:\lambda^{*}\:$ are 
complex conjugate eigenvalues of the matrix $\:A\:$
distinct from unity.}

If $A = A^{\top}$ then not all real solutions of equation (\ref{log1}) 
are skewsymmetric matrices. For example
\begin{eqnarray}
\exp \left(
\begin{array}{rrr}
0 & w d & 0 \\
-w/d & 0 & 0 \\
0 & 0 & 0
\end{array}
\right) \;=\; 
\left(
\begin{array}{rrr}
-1 & 0 & 0 \\
0 & -1 & 0 \\
0 & 0 & 1
\end{array}
\right)\; \in\; \mbox{SO}(3)  
\nonumber
\end{eqnarray}
for all real $\;d \:\neq\: 0\;$ 
and all real $\;w \:=\: \pi\:(\mbox{mod} \:2\pi)$.

{\bf Lemma 5}: {\sl Let $\vec{n}$, $\vec{m}$, and $\vec{l}$ be 
the
orthogonal basis 
connected
with matrix $A \in \mbox{\rm SO}(3)$ 
and
defined in the previous
subsubsection. Then all real skewsymmetric 
logarithms of the matrix $A$ are
given by the formula
\begin{eqnarray}
\ln(A) \;=\; \left(\omega \:+\: 2 \pi k \right) \cdot 
\left(
\begin{array}{rrr}
0 & -n_3 & n_2 \\
n_3 & 0 & -n_1 \\
-n_2 & n_1 & 0
\end{array}
\right) \;=\; 
\left(\omega \:+\: 2 \pi k \right) \cdot C(\vec{n})
\nonumber
\end{eqnarray}
where real $\:\omega$ ($0 \:\leq\: \omega \:<\: 2\,\pi$) 
satisfies the equation
\footnote{Here and further on for complex vectors 
$\:\vec u,\,\vec v\:\in\:C^n\;$
we define
$\;\vec u \cdot \vec v\:=\:u_1 \cdot v_1^*+\ldots+u_n\cdot v_n^*$.}
\begin{eqnarray}
\exp(i \omega)\; =\; \frac{1}{2} 
\left(\:\vec{m} \:+\: i\, \vec{l} \:\right) \cdot A
\left(\:\vec{m} \:+\: i\, \vec{l} \:\right)
\label{log2}
\end{eqnarray}
and $\:k\:$ is an arbitrary whole number. }

Note that the right hand side in (\ref{log2}) is just another expression for
one of the eigenvalues 
(which we say is $''$conjugate
with vector $\vec{n}$$''$) of the matrix 
$A$, and hence the equation (\ref{log2}) always has a
real solution.

Below in this paper we will use the notation $\:\ln_s A\:$ for real
skewsymmetric values of the function $\:\ln A$.

{\bf Remark:} In accordance with Lemma 5
\begin{eqnarray}
\ln_s \,I\;=\; 2\,\pi\,k\cdot C(\vec n\,)
\nonumber
\end{eqnarray}
where $\:\vec n\:$ and $\:k\:$ are an arbitrary unit vector
and an arbitrary integer respectively. 

\subsubsection{Normal Forms for Hamiltonians of Linear Equations
of Spin Motion with Periodic Coefficients}

\hspace*{0.5cm}
For simplicity fix $\;\tau_0 \:=\: 0\;$ and let 
$\;M(\tau)\;$ be the
fundamental
matrix of solution of the system (\ref{mf003}). Then
\begin{eqnarray}
M(\tau + 2 \pi) \;=\; M(\tau) \cdot M(2 \pi)  
\nonumber
\end{eqnarray}

{\bf Definition 1:} {\sl A transformation of variables
\begin{eqnarray}
\vec{s}\; =\; A(\tau) \vec{u}, 
\hspace{1cm} 
A(\tau) \;\in\; \mbox{\rm SO}(3)
\nonumber
\end{eqnarray}
is called a {\bf normalizing transformation}, 
if \vspace*{0.3cm} }

{\sl {\bf a)} $A(\tau + 2 \pi) \;\equiv\; A(\tau)$ \vspace*{0.3cm} }

{\sl {\bf b)} the Hamiltonian in the variables $\;\vec{u}\;$ does not
depend on
$\;\tau$ \vspace*{0.3cm} }

{\sl {\bf c)} the matrix 
$\;A^{\top}(2 \pi) \cdot M(2 \pi) \cdot A(2 \pi)\;$ is a
real Jordan canonical form \vspace*{0.3cm} }

{\bf Definition 2:} {\sl The result of the application of the normalizing
transformation to the Hamiltonian (\ref{mf001}) with the periodic vector 
$\vec{w}(\tau)$ is called the {\bf normal form} of the Hamiltonian 
for linear
equations of spin motion with periodic coefficients. }

Let $\bar{M}$ be one of the real Jordan canonical 
forms of the matrix $M(2 \pi)$ and let the matrix 
$\:B \in \mbox{SO}(3)\:$ be such that
\begin{eqnarray}
B^{\top} M(2 \pi)\, B \;=\; \bar{M}  
\nonumber
\end{eqnarray}

{\bf Lemma 6}: {\sl The matrix
\begin{eqnarray}
A(\tau) \;=\; 
M(\tau) \exp \left( -\frac{\tau}{2 \pi} \ln_s M(2 \pi)\right) B
\nonumber
\end{eqnarray}
defines the normalizing transformation, which takes the 
initial Hamiltonian
(\ref{mf001}) into the normal form }
\begin{eqnarray}
\bar{H}(\vec{s}) \;=\; \frac{1}{2} \,\vec{s} \cdot 
\mbox{curl}_{\:\vec{s}}
\left(
\frac{1}{2 \pi} \left(\ln_s \bar{M}\right) \cdot \vec{s}\right)
\label{linnf}
\end{eqnarray}

Examining all possible real Jordan canonical forms of 
the matrix $M(2 \pi)$
and explicitly calculating the logarithm in (\ref{linnf}) we get

{\bf Lemma 7}: {\sl The Hamiltonian (\ref{mf001}) with the 
$2 \pi$-periodic
vector $\vec{w}(\tau)$ can be reduced to the following normal forms
\begin{eqnarray}
\bar{H}(\vec{s}\,) \;=\; (\pm \lambda \:+\: k) \cdot s_m  
\label{linnf1}
\end{eqnarray}
where
\begin{eqnarray}
\lambda \;=\; \frac{1}{2 \pi}\, 
\arccos \left( \frac{1}{2}(\mbox{Tr}(M(2 \pi))-1)
\right)  
\nonumber
\end{eqnarray}
and the sign $^{\prime}+^{\prime}$ or $^{\prime}-^{\prime}$ in front of 
$\lambda$, $m \in \{1,2,3\}$ and the integer $k$ can be chosen
arbitrarily. }


The number $\;\pm \lambda + k\;$ in (\ref{linnf1}) is called 
the {\bf spin tune}.
Using the freedom of the choice of $k$ and of the sign 
$\pm$ in front of $\lambda$ we can normalize the spin tune to 
a value lying in the region from $0$ to $0.5$.

Note, that if $A \neq I$ then the set of Hamiltonians (\ref{linnf1})
contains all possible normal forms. This is easy to see from the fact that
if the Hamiltonian
\begin{eqnarray}
\frac{1}{2}\, \vec{s} \cdot \mbox{curl}_{\:\vec{s}}
\left(C\cdot \vec{s}\,\right)
\nonumber
\end{eqnarray}
is a normal form, then the matrix $\;\exp(2 \pi C)\;$ is a real Jordan
canonical
form of the matrix $M(2 \pi)$.

Another way to bring a $2\pi$-periodic Hamiltonian linear in spin to the
simplest form with help of a $2\pi$-periodic coordinate transformation 
can be extracted from the results of Appendix B.

\subsection{Connection between the SO(3) 
and SU(2) Groups
and the Lax Form of the Equations of Spin Motion}

\hspace*{0.5cm}
The fundamental matrix solution $\:M(\tau,\, \tau_0)\:$ of the linear
differential
equations of spin motion (\ref{mf003}) is a $3 \times 3$ matrix 
and consists
of 9 elements, but since 
$\:M(\tau,\, \tau_0) \:\in\: \mbox{SO}(3)\:$
it can be described completely with the help of a smaller number of
parameters using the connection between the $\mbox{SO}(3)$ and 
$\mbox{SU}(2)$ 
groups. Usually in accelerator physics
this connection is established in the framework of the spinor formalism
\cite{mont,cartan}.
The method we shall use in this subsection (following \cite{tri1}) gives
us
the same results, but does not use the concept of spinors (at least in
explicit
form).

Let us recall the definition of the $\:\mbox{SU}(2)\:$ group: 

a $2\times 2$ matrix $U$ with complex coefficients belongs to the 
$\mbox{SU}(2)$ group if $\:\mbox{det}\,U\,=\,1\:$ and 
$\:U\cdot U^{*}\,=\,I\:$ 
(the asterisk '$*$' indicates complex conjugation of a matrix).

From this definition it follows that any matrix 
$\:U\,\in\, \mbox{SU}(2)\:$ has the
form
\begin{eqnarray}
U\;=\;\left(
\begin{array}{cc}
a & b \\
-b^{*} & a^{*}
\end{array}
\right) ,
\hspace{1.0cm}
a\cdot a^{*}\:+\:b\cdot b^{*}\:=\:1
\label{su201}
\end{eqnarray}
We now define the matrix
\begin{eqnarray}
L \;=\; \left(
\begin{array}{cc}
s_3 & s_1 + i s_2 \\
s_1 - i s_2 & -s_3
\end{array}
\right), 
\hspace{1.0cm} 
L^*\: =\: L  
\label{su202}
\end{eqnarray}
corresponding to the vector $\:\vec{s}\:$ 
and introduce the anti-Hermitian matrix $B$
\begin{eqnarray}
B \;=\; \frac{i}{2} 
\cdot \left(
\begin{array}{cc}
w_3 & w_1 + i w_2 \\
w_1 - i w_2 & -w_3
\end{array}
\right), 
\hspace{1.0cm} 
B^{*} \:=\: -B  
\label{su203}
\end{eqnarray}
By means of the matrices $L$ and $B$ one can write the 
equations of spin
motion (\ref{mf003}) in the form of a {\bf Lax equation}
\begin{eqnarray}
\frac{d L}{d \tau} \;=\; B \cdot L \:-\: L \cdot B
\label{su204}
\end{eqnarray}
Note that the right hand side of equation (\ref{su204}) satisfies the
condition
\begin{eqnarray}
\left( B \cdot L \:-\: L \cdot B \right)^{*} \;=\; 
B \cdot L \:-\: L \cdot B
\nonumber
\end{eqnarray}

{\bf Lemma 8}: {\sl If the matrix 
$\:U(\tau, \,\tau_0)\:$ satisfies the equation
\begin{eqnarray}
\frac{d U}{d \tau} \;=\; B \cdot U, 
\hspace{0.6cm} 
U(\tau_0,\, \tau_0) \:=\: I 
\hspace{0.4cm} 
(\mbox{or} \hspace{0.3cm} U(\tau_0, \,\tau_0) \:=\: -I)
\label{su205}
\end{eqnarray}
then $\:U(\tau, \,\tau_0)\: \in\: \mbox{\rm SU}(2)\:$ and 
the solution of (\ref{su204}) is
given by the formula 
\begin{eqnarray}
L(\tau,\, \tau_0) \;=\; U(\tau,\, \tau_0) \cdot 
L(\tau_0,\, \tau_0) \cdot
U^{*}(\tau,\,\tau_0)  
\label{su206}
\end{eqnarray}
}

{\bf Remark 1}: Because (\ref{su206}) is a similarity transformation it
follows that for any $\:\tau \,\geq \,\tau_0\:$ the matrices 
$\:L(\tau,\,\tau_0)\:$ and $\:L(\tau_0, \,\tau_0)\:$ have the same
eigenvalues. In our case this means that
for any $\:\tau \,\geq\, \tau_0$ 
\begin{eqnarray}
\mid \vec{s}(\tau) \mid \;\equiv\; \mid \vec{s}(\tau_0) \mid
\nonumber
\end{eqnarray}

{\bf Remark 2}: The equation of spin motion (\ref{mf003}) can be written in
the form (\ref{su204}) using another 
choice of the matrices $L$ and $B$, different 
from (\ref{su202}) and (\ref{su203}). See, for example, \cite{tri1}.

The following formulae give us the connection between the matrix 
$\:U(\tau, \,\tau_0)\:$ in (\ref{su206}) and the fundamental matrix 
solution of (\ref{mf003}) $\:M(\tau, \,\tau_0)$: 
\begin{eqnarray}
M \;=\; \left( 
\begin{array}{rrr}
Re(a^2-b^2) & -Im(a^2+b^2) & -2\cdot Re(ab) \\ 
Im(a^2-b^2) & Re(a^2+b^2) & -2\cdot Im(ab) \\ 
2\cdot Re(ab^{*}) & -2\cdot Im(ab^{*}) & aa^{*}-bb^{*}
\end{array}
\right)  
\label{amatr}
\end{eqnarray}
Introduce the quantities 
\begin{eqnarray}
v\;=\;{\frac 12}(m_{11}\:+\:m_{22}\:+\:i\,(m_{21}\:-\:m_{12}))
\;=\;a^2
\nonumber
\end{eqnarray}
\begin{eqnarray}
w\;=\;{\frac 12}(m_{22}\:-\:m_{11}\:-\:i\,(m_{21}\:+\:m_{12}))
\;=\;b^2
\nonumber
\end{eqnarray}
where $\:m_{kl}\:$ are the elements of the matrix $M$. Since for the
orthogonal
matrix  $\:M\:$ we have
\begin{eqnarray}
w\cdot w^{*}\:+\:v\cdot v^{*}
\;=\;{\frac 12} (m_{11}^2\:+\:m_{12}^2\:+\:m_{21}^2\:+\:m_{22}^2)
\;\neq\; 0  
\nonumber
\end{eqnarray}
then using (\ref{su201}) and (\ref{amatr}) we can define 
the elements of the
matrix $U$ by the equations 
\begin{eqnarray}
\left\{
\begin{array}{lll}
a\;=\;\pm \sqrt{v},
&b\;=\;-{\frac 1{2a}}(m_{13}\:+\:i\, m_{23}),
&\mbox{if} \;\;v\:\neq\: 0,\\
\\ 
b\;=\;\pm \sqrt{w},
&a\;=\;-{\frac 1{2b}}(m_{13}\:+\:i\, m_{23}),
&\mbox{otherwise}.
\end{array}
\right.
\nonumber
\end{eqnarray}

{\bf Remark 3}: The freedom of the choice of sign in 
these formulae (and also
in (\ref{su205})) is connected with the fact that the 
$\mbox{SU}(2)$ group
overlaps the $\mbox{SO}(3)$ group twice.

If the vector $\;\vec{w}\;$ does not depend on $\;\tau\,$, then the
solution of
equation (\ref{su205}) has the form 
\begin{eqnarray}
U(\tau,\, \tau_0) \;=\; U(\tau_0,\, \tau_0) 
\left( 
\cos\left(
\frac{|\vec{w}|}{2}(\tau-\tau_0) 
\right) \cdot I 
\:+\: 
\frac{2}{|\vec{w}|} 
\sin\left(
\frac{|\vec{w}|}{2}(\tau-\tau_0)
\right)\cdot B 
\right)  
\nonumber
\end{eqnarray}
where $\:U(\tau_0, \,\tau_0) \:=\: \pm I\:$ and we 
assumed $\:|\vec{w}| \,\neq\, 0\,$. In the
case $\:|\vec{w}|\, =\, 0\:$ the matrix 
$\:U(\tau, \,\tau_0) \:\equiv\: U(\tau_0,\,\tau_0)$.

The usage of the $\mbox{SU}(2)$ representation of $\mbox{SO}(3)$
matrices
not only
allows us to reduce the number of free parameters, but 
is also helpful for
some analytical calculations. Consider, for example, the matrix 

\begin{eqnarray}
\hat L \;=\; i\,(Re(a)\cdot I\:-\:U)\;=\;
\left( 
\begin{array}{cc}
Im(a) & Im(b)-i \cdot Re(b) \\ 
Im(b)+i \cdot Re(b) & -Im(a)
\end{array}
\right)   
\nonumber
\end{eqnarray}
It is easy to verify that this matrix commutes 
with the matrix $U$, i.e. 
\begin{eqnarray}
\hat L\;=\;U\:\hat L\:U^{*}  
\label{suuu}
\end{eqnarray}
Comparing (\ref{suuu}) and (\ref{su206}) we get

{\bf Lemma 9:$\;$} {\sl Let $U$ be a $\mbox{\rm SU}(2)$ matrix, 
corresponding to a
given $\;A \:\in\: \mbox{\rm SO}(3)\:$. If $\:A \,\neq\, I\:$ then 
\begin{eqnarray}
\vec{n} \;=\; 
\frac{1}{\sqrt{1 - (Re(a))^2}} \:
\left(Im(b),\: -Re(b),\: Im(a)\right) 
\nonumber
\end{eqnarray}
is a unit eigenvector of the matrix $A$, corresponding to unit eigenvalue.
}

\section{Normal Forms for the Spin-Orbit Hamiltonian}

\hspace*{0.5cm}
In this section we will consider the Hamiltonian system associated with 
the coupled spin-orbit Poisson bracket in a neighbourhood of a stationary
point (or periodic solution) which can be canonically transformed to be
the origin\footnote{If $\vec{z}_*(\tau)$ is a solution of
a Hamiltonian system associated with 
the coupled spin-orbit Poisson bracket then
the  parallel displacement 
$\vec{z}_{new} = \vec{z}_{old} - \vec{z}_*(\tau)$ 
will be a canonical transformation if and only if
$\vec{s}_*(\tau) \equiv \vec{0}$. The question of what type 
of simple canonical coordinates can be introduced for an arbitrary
solution is studied in Appendix B.}.
In order to
understand the properties of the solutions of such systems it is helpful
firstly to find a coordinate substitution which reduces the original
equations to the simplest possible form. Here we describe an algorithm
which allows us to make coordinate transformations working not with
equations of motion, but directly with the Hamiltonian function.

\subsection{Canonical Transformations of Spin-Orbit Variables 
which Map the Origin into Itself}

\hspace*{0.5cm}
We shall be working with a phase space consisting 
of $\:2n+3\:$ variables 
\begin{eqnarray}
\vec z\;=\;(\vec x,\,\vec s\,)\;=\;
(\vec q,\,\vec p,\,\vec s\,)\;=\;
(q_1,\,\ldots,\,q_n,\,p_1,\,\ldots,\,p_n,\,s_1,\,s_2,\,s_3)
\nonumber
\end{eqnarray}
In this subsection we shall study canonical (i.e. the Poisson bracket (\ref
{fom2}) preserving) nonsingular (i.e. with nondegenerate Jacobian matrix)
maps 
\begin{eqnarray}
\vec z_f\;=\;\vec Z(\vec z_i)  
\label{n1}
\end{eqnarray}
where the components of the real vector function 
$\:\vec Z(\vec z\,)\:$ are power
series in $\:\vec z\:$ without constant terms.

\subsubsection{Linear Canonical Transformations}

\hspace*{0.5cm}
The map which linearizes any symplectic map 
in the neighborhood of a fixed point
is always a linear symplectic map. 
Does this apply to our case, where we are dealing
with maps which are not symplectic but instead are
canonical with respect to the
Poisson bracket (\ref{fom2})? 

{\bf Lemma 1}: {\sl The nonsingular real linear transformation 
\begin{eqnarray}
\vec{z}_f \;=\; A\:\vec{z}_i  
\nonumber
\end{eqnarray}
is canonical if and only if the matrix $\:A\:$ has the form 
}
\begin{eqnarray}
A \;=\; \mbox{diag}\, (A_{orbt},\: A_{spin})
\nonumber
\end{eqnarray}
{\sl
where 
\begin{eqnarray}
A_{orbt} \:\in\: \mbox{\rm Sp}(2n), 
\hspace{1.0cm} 
A_{spin} \:\in\: \mbox{\rm SO}(3)
\nonumber
\end{eqnarray}
}

In more detail the map (\ref{n1}) can be written as 
\begin{eqnarray}
\left\{ 
\begin{array}{l}
\vec{x}_f \;=\; \vec{X}(\vec{x}_i, \:\vec{s}_i) \\ 
\\ 
\vec{s}_f \;=\; \vec{S}(\vec{x}_i, \:\vec{s}_i)
\end{array}
\right.  \nonumber
\end{eqnarray}
Using the condition of spin-orbit Poisson bracket preservation
(\ref{lgtl}) it is possible to show that always
\begin{eqnarray}
\vec{S}\left(\vec{x},\, \vec{0}\,\right)\; =\; \vec{0}  
\nonumber
\end{eqnarray}
Thus we have the following:

{\bf Lemma 2:$\;$} 
{\sl The linearization of the canonical map (\ref{n1}) is
a linear canonical map if and only if 
\begin{eqnarray}
\frac{\partial \vec{X}}{\partial \vec{s}}\:
\left(\vec{0}, \:\vec{0}\,\right) \;=\; 0
\nonumber
\end{eqnarray}
}

So the answer to the above question is negative in the general case. The
linearization of the nonsingular canonical map (in its usual meaning)
is not bound to be a linear canonical transformation.

{\bf Definition 1:$\;$}
{\sl A polynomial $\;P(\vec{z}\,) \:=\: P(\vec{x},\, \vec{s}\,)\;$ is 
{\bf quasi-homogeneous} of degree $\:m\:$ in its arguments 
$\:\vec{x},\:\vec{s}\:$ if
it satisfies the equation
\begin{eqnarray}
P(t\,\vec{x}, \:t^2\,\vec{s}\,) \;=\; 
t^m\cdot P(\vec{x},\, \vec{s}\,)
\nonumber
\end{eqnarray}
for every value of $\:t$.}

The set of all quasi-homogeneous polynomials of degree 
$m$ $(m=0,1,2,\ldots)$
will be denoted by ${\cal H}_s(m)$.

{\bf Definition 2:$\;$}
{\sl The {\bf quasi-linearization} of the map (\ref{n1})
is defined as a linear transformation 
\begin{eqnarray}
\vec{z}_f \;=\; A_s \:\vec{z}_i
\nonumber
\end{eqnarray}
which acts invariantly 
on classes ${\cal H}_s(m)$ 
\footnote{
This means that for an arbitrary $\:m \:=\: 0,\,1,\,2,\,\ldots\:$
and for an arbitrary $\;P(\vec z\,) \:\in \:{\cal H}_s(m)\;$
we have that $\; P(A_s\,\vec z\,) \:\in \:{\cal H}_s(m)$.}
and minimizes the
Euclidean norm of the difference
\begin{eqnarray}
\frac{\partial \vec{Z}}{\partial \vec{z}}\, 
\left(\vec{0}\,\right) \:-\: A_s  
\nonumber
\end{eqnarray}
}

{\bf Lemma 3:$\;$}
{\sl The quasi-linearization of the map (\ref{n1}) exists, is
unique, is a linear canonical map and is defined by the matrix }
\begin{eqnarray}
A_s \;=\; \mbox{diag}\, 
\left( 
\frac{\partial \vec{X}}{\partial \vec{x}}
\left(\vec{0},\: \vec{0}\,\right),\; 
\frac{\partial \vec{S}}{\partial \vec{s}}
\left(\vec{0},\:\vec{0}\,\right) 
\right)  
\label{n2}
\end{eqnarray}

{\bf Remark:} The equality (\ref{n2}) can be used as
definition of quasi-linearization. In this case definition 2 becomes
a statement which must be proven.

\subsubsection{Factorization Theorem}

\hspace*{0.5cm}
The use of the change of coordinates in the form of the power series
(\ref{n1}) is
not too convenient for our purpose because the proof of  canonicity
is reduced to the checking of an infinite number of conditions for
the coefficients of the Taylor expansion of the vector function 
$\vec{Z}(\vec{z})$. Dragt and Finn \cite{DF} have shown that 
any symplectic map in the
neighborhood of a fixed point may be represented in the form of a
composition of a linear symplectic map and a sequence of displacements
along
trajectories of the Hamilton system with Hamiltonians which are homogeneous
polynomials of powers $3,4,5,\ldots$. A similar factorization 
is applicable in
our case too, if instead of homogeneous polynomials we use 
Hamiltonians from the classes ${\cal H}_s(m)$.

{\bf Definition 3:$\;$}
{\sl We will say that the function 
$\;V(\vec{z}\,) \:\in\: {\cal O}_s(m)\:$, if 
\begin{eqnarray}
\exists \;
\lim_{t \rightarrow 0}\: 
\frac{1}{t^m}\: V(t\, \vec{x},\, t^2\, \vec{s}\,)
\;\in\; 
{\cal H}_s(m)
\nonumber
\end{eqnarray}
}

{\bf Factorization Theorem A:$\;$} 
{\sl For every canonical map of the form
(\ref{n1}) and for
every integer $\:m \,\ge\, 3\:$ functions 
$\;F_k(\vec{z}\,)\: \in\: {\cal H}_s(k)\;$ 
$(3 \:\le\: k \:\le\: m)$ can be found, such that 
\begin{eqnarray}
\vec{Z}(\vec{z}\,) \;=_m\; 
:A_s: \exp(:F_3:) \ldots \exp(:F_m:) \,\vec{z}
\nonumber
\end{eqnarray}
Here $A_s$ is the matrix of the quasi-linearization of the map (\ref{n1})
and the symbol '$=_m$' means that the difference of the right and left
parts
is
a function from the class ${\cal O}_s(m)$. }

Here the exponential Lie operators are defined as usual as a power series
(see, for example \cite{Dragt}) involving the spin-orbit Poisson bracket 
(\ref{fom2}), and the action of the operator $\;:A_s:\;$ is defined by
means of the
rules
\begin{eqnarray}
:A_s:\:f(\vec z\,)\;=\;f(:A_s:\vec z\,)  
\nonumber
\end{eqnarray}
for a smooth function $f$, and
\begin{eqnarray}
:A_s:\:\vec z\;=\;A_s\:\vec z  
\nonumber
\end{eqnarray}
for the identity mapping $\vec z$.

The proof of the factorization theorem A is outlined in Appendix C.

Applying theorem A to the canonical map 
$\vec{Z}^{-1}(\vec{z}\,)$ and then
inverting the factorization obtained  we get

{\bf Factorization Theorem B:$\;$}
{\sl For every canonical map of the form 
(\ref{n1}) and for
every integer $\:m \,\ge\, 3\:$ functions 
$\;G_k(\vec{z}\,) \:\in\: {\cal H}_s(k)\;$ 
($3\:\le\: k\: \le\: m$) can be found, such that 
\begin{eqnarray}
\vec{Z}(\vec{z}) \;=_m\; \exp(:G_m:) \ldots \exp(:G_3:) :A_s:\: \vec{z}
\nonumber
\end{eqnarray}
}

{\bf Remark}: The classical Poisson bracket of two homogeneous polynomials
of degree $m$ and $l$ is a homogeneous polynomial of degree $m+l-2$. The
same property holds for functions from classes ${\cal H}_s(m)$ 
and ${\cal H}_s(l)$ with
respect to the spin-orbit Poisson bracket.

\subsection{General Description of the
 Normal Form Algorithm}

\hspace*{0.5cm}
Consider a periodic real
Hamiltonian function $\:H(\tau,\, \vec{z}\,)\:$ 
of period $2 \pi$ with respect to $\tau$
which in a 
neighbourhood of the point
$\:\vec{z}= \vec{0}\:$ can be expanded in quasi-homogeneous 
polynomials in
$\:\vec{z}\:$ beginning with degree $2$ 
\begin{eqnarray}
H(\tau,\,\vec{z}\,) \;=\; H_2(\tau,\,\vec{z}\,) \:+\: 
H_3(\tau,\, \vec{z}\,) \:+
\ldots +\:
H_k(\tau,\, \vec{z}\,) \:+ \ldots 
\label{n21}
\end{eqnarray}
The purpose of this subsection is to find the simplest possible form 
({\bf normal form}) to which the Hamiltonian function (\ref{n21}) 
can be reduced
by means of a nonlinear canonical coordinate transformation of the type 
(\ref{n1}) which depends periodically on $\tau$ as a parameter.

\subsubsection{Extension of Phase Space}

\hspace*{0.5cm}
Introduce two additional variables $E$, $v$ and define the Poisson 
bracket and the new Hamiltonian by the rules: 
\begin{eqnarray}
{\cal H} (v,\,E,\,\vec{z}\,) \;=\; H (v,\,\vec{z}\,) \:-\: E  
\nonumber
\end{eqnarray}
\begin{eqnarray}
\{E,\, v \} \:=\: 1, 
\hspace{0.5cm} 
\{ v,\, z_i \} \:=\: \{E,\, z_i\} \:=\: 0,
\hspace{0.5cm}
i\:=\:1,2, \ldots, 2n+3  
\nonumber
\end{eqnarray}

According to the factorization theorem B, for any given truncation 
order $m$ every canonical transformation of
coordinates of the type (\ref{n1}),  depending periodically on 
$\tau$ as a parameter,  can be
represented with the required precision in the form of an operator
\begin{eqnarray}
\prod\limits_{k=m}^{3} 
\exp(:G_k(\tau,\,\vec{z}\,):):A_s(\tau):
\label{n22}
\end{eqnarray}
In order to extend (\ref{n22}) to the canonical 
transformation in the
extended phase space, we rewrite (\ref{n22}) in the form: 
\begin{eqnarray}
\prod\limits_{k=m}^{3} 
\exp(:G_k( v, \,\vec{z}\,):):A_s(v):  
\label{n23}
\end{eqnarray}
and define the action of the operator $:A_s(v):$ on the identity 
mapping $\left(v,\: \vec{z},\:E\right)$ by the rules 
\begin{eqnarray}
:A_s (v): v \:=\: v, 
\hspace{0.5cm} 
:A_s (v): \vec{z} \:=\: A_s (v) \,\vec{z} 
\nonumber
\end{eqnarray}
\begin{eqnarray}
:A_s(v): E \:=\: E \,-\, 
\frac{1}{2}\, A^{\top}_{orbt} \,J\, A^{^{\prime}}_{orbt}\:
\vec{x} \cdot \vec{x} \:+\: 
\frac{1}{2} \,\vec{s} \cdot \mbox{curl}_{\vec{s}}\:
(A^{\top}_{spin}\, A^{^{\prime}}_{spin}\, \vec{s}\,)  
\nonumber
\end{eqnarray}
Here 
\begin{eqnarray}
A_s(v) \;=\; \mbox{diag}\, (A_{orbt}(v),\: A_{spin}(v))  
\nonumber
\end{eqnarray}
\begin{eqnarray}
A^{^{\prime}}_{orbt} \;=\; \frac{d A_{orbt}}{d v}, 
\hspace{1cm}
A^{^{\prime}}_{spin} \;=\; \frac{d A_{spin}}{d v}  
\nonumber
\end{eqnarray}
and the $\:2n \times 2n\:$ matrix $\:J\:$ is the 
symplectic unit matrix.

The exponential Lie operators are defined again by 
their power series with the
help of the extended Poisson bracket for which we will keep the same
notation $\{*,\,*\}$.

Now, informally speaking, the problem of finding the 
normal form can be expressed as
follows: Find the simplest quasi-homogeneous polynomials
$\:\bar{H}_k \:\in\: {\cal H}_s(k)$ ($2 \leq k \leq m$) such that the
initial Hamiltonian function $\:{\cal H}\:$
can be reduced to the Hamiltonian
\begin{eqnarray}
\bar{\cal H} \;=\; 
\sum \limits _{k=2}^{m} \,\bar{H}_k (v, \,\vec{z}\,) 
\:+\: \tilde{H}_{>m}(v,\, \vec z\,) \:-\: E, 
\hspace{0.5cm} 
\tilde{H}_{>m} \,\in\, {\cal O}_s(m+1)
\nonumber
\end{eqnarray}
by means of a
nonlinear canonical coordinate transformation, which is defined by the
operator (\ref{n23}).
Here informal means that we have not yet defined the exact
sense of the word $''$simplest$''$. The usual way to do this is to
introduce the simplest form for the quasi-quadratic polynomial
$\:\bar{H}_2\:$ axiomatically and then to define $\:\bar{H}_k\:$
as functions which satisfy the condition 
$\:\{\bar{H}_k,\,\bar{H}_2 - E\}\:=\:0$.

\subsubsection{Linear Normalization}

\hspace*{0.5cm}
First we wish to simplify the quasi-quadratic part $H_2$ 
of the Hamiltonian ${\cal H}$ by means of the canonical 
transformation $:A_s:$. Denote
\begin{eqnarray}
H_2 \;=\; \frac{1}{2}\, P(v)\, \vec{x} \cdot \vec{x} 
\:+\: \vec{w}(v) \cdot \vec{s}
\nonumber
\end{eqnarray}
where $P(v)$ is a real symmetric matrix of order 
$2n$ and $P(v+2\pi) \equiv P(v)$, and where the
real vector $\vec{w}(v)$ satisfies
$\vec{w} (v + 2 \pi) \equiv \vec{w}(v)$.

Applying the operator 
$\;:A_s(v):\;$ to the Hamiltonian $\;{\cal H}\;$ we get the
following formula for the quasi-quadratic part $\:\bar{H}_2\:$ 
of the new Hamiltonian 
$\;{\cal H}_{new} \:=\: :A_s: {\cal H}$ 
\begin{eqnarray}
\bar{H}_2 \;=\; \frac{1}{2}\, A^{\top}_{orbt} 
\left(P A_{orbt} \:+\: J A_{orbt}^{\prime} \right) 
\vec{x} \cdot \vec{x} \:+
\nonumber
\end{eqnarray}
\begin{eqnarray}
+\:\frac{1}{2} \,\vec{s} \cdot 
\mbox{curl}_{\,\vec{s}} \,
\left( A^{\top}_{spin} \left( C(\vec{w})\,
A_{spin} \:-\:
A^{\prime}_{spin} \right) \vec{s}\, \right)  
\nonumber
\end{eqnarray}

Below we will study the most important case for applications 
(to which we
will refer later on as the {\bf orbital elliptical case}) when all the
eigenvalues of the one-turn revolution matrix of the system 
\begin{eqnarray}
\frac{d \vec{x}}{d v} \;=\; J P(v)\, \vec{x}  
\nonumber
\end{eqnarray}
are distinct and lie on the unit circle (the general algorithm will be
published elsewhere). Combining the results of the previous section and
the well known
normal form theory of classical linear Hamiltonian systems (see, for example 
\cite{bruno1,yakubo}) we are able to find real 
 matrices $A_{spin} \in \mbox{SO}(3)$
and $A_{orbt} \in \mbox{Sp}(2n)$ $2\pi$-periodic in $v$ such that  
$\bar{H}_2$ finally takes the form 
\begin{eqnarray}
\bar{H}_2 \;=\; \frac{\mbox{\ae}_k}{2}\left(q_k^2 \:+\: p_k^2 \right)
 \:+\: \lambda_s \cdot s_3
\nonumber
\end{eqnarray}
where the values $\:\mbox{\ae}_k\:$ and $\:\lambda_s\:$ are called 
the {\bf characteristic frequencies}
(or the linear orbital and spin tunes).

\subsubsection{Recursive Loop of Nonlinear Normalization}

\hspace*{0.5cm}
Let the functions $\bar{H}_k$ and $G_k$ be already defined for 
$k=3,\ldots,l-1$. Applying 
the operator $\exp(:G_l:)$
to the Hamiltonian 
\begin{eqnarray}
\bar{H}_2 \:+\: \ldots \:+\: \bar{H}_{l-1} \:+\: \tilde{H}_l 
\:+\: \ldots \:-\: E
\;\stackrel{{\rm def}}{=}\;  
\nonumber
\end{eqnarray}
\begin{eqnarray}
\exp(:G_{l-1}:) \ldots \exp(:G_3:) :A_s: 
{\cal H}(v, E, \vec{z}\,)  
\nonumber
\end{eqnarray}
 and collecting the remainders from the 
class ${\cal H}_s(l)$, we get the equation for obtaining 
the functions $\:\bar{H}_l\:$ and $\:G_l\:$ 
\begin{eqnarray}
\frac{\partial G_l}{\partial v} \:+\: \{G_l, \,\bar{H}_2 \} 
\;=\; \bar{H}_l \:-\: 
\tilde{H}_l  
\label{n24}
\end{eqnarray}
This is the so-called {\bf homology equation}. We say that the
quasi-homogeneous polynomial $\:G_l\:$ takes the quasi-homogeneous
polynomial $\:\tilde{H}_l\:$ into the quasi-homogeneous polynomial
$\:\bar{H}_l\:$ if (\ref{n24}) holds.

In order to solve the homology equation 
(\ref{n24}) we introduce complex coordinates 
$\;\vec{w} \:=\: (\vec{\eta}, \,\vec{\xi},\, \vec{u}\,)\;$
related to the old coordinates \\
$\vec{z} \:=\: (\vec{q}, \,\vec{p},\, \vec{s}\,)\:$
by means of a {\bf standard linear transformation} 
$\:\vec{z} \:=\: Q \vec{w}\,$:
\begin{eqnarray}
q_k \;=\; \frac{1 + i}{2}  \, (\eta_k \:+\: \xi_k), 
\hspace{1.0cm} 
p_k \;=\; -\frac{1 - i}{2} \, (\eta_k \:-\: \xi_k) 
\nonumber
\end{eqnarray}
\begin{eqnarray}
k \;=\; 1, \ldots, n  
\nonumber
\end{eqnarray}
\begin{eqnarray}
s_1 \;=\; \frac{1 + i}{2}\, (u_1 \:+\: u_2), 
\hspace{1.0cm} 
s_2 \;=\; -\frac{1 - i}{2}\, (u_1 \:-\: u_2)  
\nonumber
\end{eqnarray}
\begin{eqnarray}
s_3 \;=\; u_3  
\nonumber
\end{eqnarray}
It is easy to verify that the matrix $\;Q\;$ has the following properties:
\begin{eqnarray}
Q Q^* \;=\; I, 
\hspace{1.0cm} 
Q^2 \;=\; - Q^*  
\nonumber
\end{eqnarray}
where the asterisk $''$$*$$''$ indicates complex conjugation of a matrix.

This is a Poisson transformation, but it is not canonical (although it
is symplectic with respect to orbital variables). The nonzero Poisson
brackets of the new basis functions are given by the equalities:
\begin{eqnarray}
\{ \eta_k,\,\xi_k \} \:=\: 1, 
\hspace{0.5cm} 
\{ u_1, \,u_2 \} \:=\: u_3,
\hspace{0.5cm}
\{ u_1, \,u_3 \} \:=\: -i u_1, 
\hspace{0.5cm} 
\{ u_2, \,u_3 \} \:=\: i u_2
\nonumber
\end{eqnarray}
In the new coordinates the equation 
(\ref{n24}) has the form 
\begin{eqnarray}
\frac{\partial g_l}{\partial v} \:+\: 
\{g_l,\, \bar{h}_2 \} \;=\; \bar{h}_l \:-\: \tilde{h}_l  
\label{n2401}
\end{eqnarray}
where we have used the notation 
\begin{eqnarray}
\bar{h}_2 (\vec{w}\,) \:=\: \bar{H}_2 (Q \vec{w}\,)\:=\: 
i \mbox{\ae}_k \eta_k \xi_k \:+\: \lambda_s u_3, 
\hspace{0.5cm} 
\tilde{h}_l (v,\, \vec{w}\,) \:=\: \tilde{H}_l (v,\, Q \vec{w}\,)
\nonumber
\end{eqnarray}
\begin{eqnarray}
g_l (v,\, \vec{w}\,) \:=\: G_l (v,\, Q \vec{w}\,), 
\hspace{0.7cm}
\bar{h}_l (v, \,\vec{w}\,)
\:=\: \bar{H}_l (v,\, Q \vec{w}\,)  
\nonumber
\end{eqnarray}
Expanding the coefficients of the monomials in the polynomials 
$\:g_l\:$, $\:\tilde{h}_l\:$ and $\:\bar{h}_l\:$ 
in Fourier series
with respect to $v$ 
\footnote{We assume that this Fourier series 
converges absolutely.}
\begin{eqnarray}
\left \{ 
\begin{array}{l}
g_l \;=\; g_l^{NIJL} \exp(i N v) \,\eta^I \xi^J u^L \\
\\ 
\bar{h}_l \;=\; \bar{h}_l^{NIJL} \exp(i N v)\, \eta^I \xi^J u^L \\ 
\\
\tilde{h}_l \;=\; \tilde{h}_l^{NIJL} \exp(i N v)\, \eta^I \xi^J u^L
\end{array}
\right.  
\label{n27}
\end{eqnarray}
(here the summation is made over all integers $N$ and over all nonnegative
integer vectors $I,J\in R^n$, $L \in R^3$ satisfying the condition 
$|I| + |J| + 2|L| = l$\footnote{Here the function $|*|$ 
for the integer vector 
$\vec{k} \in R^m$
is defined as $|\vec{k}| = |k_1| + \ldots + |k_m|$.}) 
and substituting them in the equation (\ref{n2401}) we obtain a system of
equations for the coefficients: 
\begin{eqnarray}
i \cdot [N + (I - J) \cdot \vec{\mbox{\ae}} 
+ (L_2 - L_1) \cdot \lambda_s] \cdot
g_l^{NIJL} \;=\; \bar{h}_l^{NIJL} - \tilde{h}_l^{NIJL}  
\label{n28}
\end{eqnarray}

{\bf Definition 4}: {\sl The characteristic frequencies
$\;{\mbox{\ae}}_1, \:\ldots ,\: {\mbox{\ae}}_n, \:\lambda_s\;$ satisfy a
{\bf resonance relation of order} $\:K\:$ if there
exist integers $\:k_l\:$ not all equal to zero such that
\begin{eqnarray}
k_1 {\mbox{\ae}}_1 \:+\: \ldots \:+\: k_n {\mbox{\ae}}_n 
\:+\: k_{n+1} \lambda_s
\;=\; 0 \hspace{0.15cm} (\mbox{mod} \hspace{0.15cm} 1)  
\label{gty}
\end{eqnarray}
\begin{eqnarray}
\mid k_1 \mid \:+\: \ldots \:+\: \mid k_n \mid \:+\: 
\hspace{0.1cm} 2\cdot \mid k_{n+1}\mid \;=\; K
\label{juli}
\end{eqnarray}
}

The number of the linearly independent integer solutions $k_1, \ldots,
k_{n+1}$ of equation (\ref{gty}) is an important characteristic of the
frequencies ${\mbox{\ae}}_1, \ldots , {\mbox{\ae}}_n, \lambda_s$, 
and we will call it the 
{\bf multiplicity of the resonance}.

Note that our definition of the order of a resonance 
(multiplier 2 in front of $k_{n+1}$ in (\ref{juli})) 
is different from the usual one, and corresponds to the
definition of quasi-homogeneity.

{\bf Definition 5}: {\sl A quasi-homogeneous polynomial
\begin{eqnarray}
\bar{h}_l (v, \,\vec{w}\,) \;=\; \bar{h}_l^{NIJL} \exp(i N v)\, 
\eta^I \xi^J u^L
\nonumber
\end{eqnarray}
is said to be a {\bf complex normal form} if 
\begin{eqnarray}
\mid N + (I - J) \cdot \vec{\mbox{\ae}} + 
(L_2 - L_1) \cdot \lambda_s \mid \cdot
\mid \bar{h}_l^{NIJL} \mid \;=\; 0  
\nonumber
\end{eqnarray}
that is, the normal form contains only {\bf resonant terms}.}

In each quasi-homogeneous polynomial 
\begin{eqnarray}
h_l \;=\; h_l^{NIJL} \exp(i N v)\, \eta^I \xi^J u^L  
\label{dfr}
\end{eqnarray}
we isolate the {\bf resonant part} $h_l^{\prime}$, 
which contains all those
and only those terms of (\ref{dfr}) whose indices satisfy 
\begin{eqnarray}
N \:+\: (I - J) \cdot \vec{\mbox{\ae}} \:+\: (L_2 - L_1) \cdot 
\lambda_s \;=\;0 
\nonumber
\end{eqnarray}
A quasi-homogeneous polynomial $h_l^{\prime \prime} = h_l - h_l^{\prime}$ 
is called the {\bf nonresonant part} of the quasi-homogeneous polynomial
$h_l$. Obviously, $h_l = h_l^{\prime} + h_l^{\prime \prime}$. We also
introduce the
resonant $\;H_l^{\prime}\;$ and nonresonant 
$\;H_l^{\prime \prime}\;$ parts of
a quasi-homogeneous polynomial $\:H_l(v,\, \vec{z}\,)\:$ as
\begin{eqnarray}
H_l^{\prime} (v, \,\vec{z}\,) \;=\; h_l^{\prime}(v,\, Q^* \vec{z}\,),
\hspace{1.0cm}
H_l^{\prime \prime} (v, \,\vec{z}\,) \;=\; 
h_l^{\prime \prime}(v,\, Q^*\vec{z}\,) 
\nonumber
\end{eqnarray}
where $\:h_l^{\prime}(v, \,\vec{w}\,)\:$ and 
$\:h_l^{\prime \prime}(v,\,\vec{w}\,)\:$ are
the resonant and nonresonant parts of the 
quasi-homogeneous polynomial 
$\:h_l(v, \,\vec{w}\,) \:=\: H_l(v,\, Q \vec{w}\,)$.

According to the definition 5, if the quasi-homogeneous polynomial 
$\:\bar{h}_l\:$ is a complex normal form, 
then $\:\bar{h}_l^{\prime} \:=\:\bar{h}_l\,$.

{\bf Definition 6}: {\sl A quasi-homogeneous polynomial 
\begin{eqnarray}
\bar{h}_l (v,\, \vec{w}\,) \;=\; 
\bar{h}_l^{NIJL} \exp(i N v) \,\eta^I \xi^J u^L 
\nonumber
\end{eqnarray}
is said to be a {\bf nonresonant complex normal form} if 
\begin{eqnarray}
\mid \mid N \mid + \mid I - J \mid + \mid L_2 - 
L_1 \mid \mid \cdot \mid 
\bar{h}_l^{NIJL} \mid \;=\; 0  
\nonumber
\end{eqnarray}
that is, the normal form contains only 
{\bf trivial resonant terms}. }

{\bf Definition 7}: {\sl A quasi-homogeneous polynomial 
$\bar{H}_l (v, \vec{z})$ is said to be a 
{\bf (nonresonant) normal form} if after the standard
transformation $\vec{z} = Q \vec{w}$ we obtain the quasi-homogeneous
polynomial $\bar{h}_l (v, \vec{w}) = \bar{H}_l (v, Q \vec{w})$ 
which is a (nonresonant) complex normal form}.

It is easy to see (comparing definitions 5 and 7, and equation (\ref{n28}))
that for every given quasi-homogeneous polynomial $\tilde{H}_l$ there is the
quasi-homogeneous polynomial $G_l$ which takes $\tilde{H}_l$ into the normal
form $\bar{H}_l$. Here the nonresonant part $G_l^{\prime \prime}$ and the
normal form $\bar{H}_l$ are uniquely determined, and the resonant 
part $G_l^{\prime}$ can be specified arbitrarily.

Among the quasi-homogeneous polynomials $G_l$ which take $\tilde H_l$ into
the normal form $\bar H_l$ we select the one for which 
$\;G_l^{\prime}\:=\:0\;$ and we call it {\bf basic}.

Remembering that the initial Hamiltonian function ${\cal {H}}$ is real (the
values of the function are real for real values of arguments) and 
that after 
a real
canonical transformation $:A_s (v):$ we again have the real Hamiltonian
function $:A_s (v): {\cal {H}}$, we wish to find conditions on the functions
$G_l$ that will guarantee the reality of the transformation from $:A_s (v): 
{\cal {H}}$ to normal form $\bar{{\cal {H}}}$. Denote by $\hat{h}(v, \vec{w})
$ the quasi-homogeneous polynomial in which all the coefficients of the
powers of $\vec{w}$ are complex conjugates (as functions of $v$) of the
corresponding coefficients in the quasi-homogeneous polynomial $h(v, \vec{w})
$.

{\bf Lemma 4}: {\sl Suppose that in 
the homology equation (\ref{n24}) $\bar{H}_2$
and $\tilde{H}_l$ are real functions. Then 
the normal form $\bar{H}_l$ will
be a real function
and the necessary and sufficient conditions that
the quasi-homogeneous polynomial $G_l$ (which
takes $\tilde{H}_l$ into the normal form $\bar{H}_l$) will be a real
function are that its
resonant part $\:G_l^{\prime}(v,\, \vec{z}\,)\:$ satisfies the condition
\begin{eqnarray}
g^{\prime}_l(v,\, \vec{w}\,) \;=\; 
\hat{g}^{\prime}_l(v,\, - Q^{*} \vec{w}\,)
\hspace{0.5cm} \mbox{{\sl where}} \hspace{0.5cm} 
g^{\prime}_l(v,\, \vec{w}\,)\;=\;
G^{\prime}_l(v,\, Q \vec{w}\,)  
\label{kond}
\end{eqnarray}
In particular, the basic solution of the homology equation $G_l$ 
($G_l^{\prime} = 0$) which takes $\tilde{H}_l$ into the normal 
form $\bar{H}_l$
will be a real function.}

\subsubsection{Normalization Theorem}

\hspace*{0.5cm}
Now we are ready to summarize the results of this section as follows

{\bf Normalization Theorem}:  
{\sl In the orbital elliptical case,
for a real initial Hamiltonian
of the form
\begin{eqnarray}
{\cal H}(v,\,E,\,\vec{z}\,) \;=\; \sum \limits_{k=2}^{\infty} 
H_k (v,\,\vec{z}\,) \:-\: E, 
\hspace{1.0cm} H_k \:\in\: {\cal H}_s(k)  
\nonumber
\end{eqnarray}
and for every integer $m \geq 3$
there exists a 
real canonical transformation
depending periodically on $v$ and
defined by an operator
\begin{eqnarray}
\prod\limits_{k=m}^{3} \exp(:G_k( v,\, \vec{z}\,):)\,:A_s(v):  
\nonumber
\end{eqnarray}
under which the initial Hamiltonian
becomes a real Hamiltonian
\begin{eqnarray}
\bar{\cal H}(v,\,E,\,\vec{z}\,) \;=\; 
\frac{\mbox{\ae}_k}{2}\left(q_k^2 + p_k^2 \right)
 \:+\: \lambda_s \cdot s_3 \:+\:
\sum \limits_{k=3}^{m} 
\bar{H}_k (v,\, \vec{z}\,) \:+\: \tilde{H}_{>m} (v, \,\vec{z}\,) 
\:-\: E 
\nonumber
\end{eqnarray}
such that 
$\;\tilde{H}_{>m} \:\in\: {\cal O}_s(m+1)\;$ 
and all quasi-homogeneous polynomials 
$\;\bar{H}_k \:\in\: {\cal H}_s(k)\;$ are normal forms.
Here the resonant parts of the quasi-homogeneous polynomials
$G_k$ may be chosen arbitrarily satisfying condition
(\ref{kond}); then the nonresonant parts of $G_k$ and
$\bar{H}_k$ are uniquely defined.
}

\section{General Properties of Systems whose
Hamiltonians are Normal Forms}

\hspace*{0.5cm}
A Hamiltonian system whose Hamiltonian function is a normal form has, as a
rule, an abundance of symmetries, and this enables the order to be
lowered. This is the advantage of a normal form over an arbitrary
Hamiltonian function. We will not discuss the technical details of the
procedure of lowering the order here, but concentrate on the question of
finding invariant functions (conservations laws). 
For convenience we will
use the complex coordinates $\vec{w}$, but it is helpful to remember
that in the original coordinates $\vec{z}$ all functions that we use
will be functions with real values.

\subsection{Normal Forms and Invariant Functions}

\hspace*{0.5cm}
Writing a normal form $\;\bar{\cal H}\;$ in complex coordinates 
$\:\vec w\:$ we have 
\begin{eqnarray}
\bar h= \bar{h}(v, E, \vec{w}) = \bar{{\cal H}}(v, E, Q \vec{w}) 
= 
i\mbox{\ae}_k\eta _k\xi _k
+\lambda _su_3+\sum\limits_{k=3}^\infty \bar h_k-E
\label{circu1}
\end{eqnarray}
where 
\begin{eqnarray}
\bar h_k\;=\;\bar h_k^{NIJL}\exp (iNv)\,\eta ^I\xi ^Ju^L  
\nonumber
\end{eqnarray}
and the summation is made over all integers $N$ and over all nonnegative
integer vectors $I$, $J$, $L$ satisfying the conditions
\begin{eqnarray}
|I|+|J|+2|L|\;=\;k
\nonumber
\end{eqnarray}
\begin{eqnarray}
N\:+\:(I-J)\cdot \vec{\mbox{\ae}} \:+\:(L_2-L_1)\cdot \lambda _s
\;=\;0
\nonumber
\end{eqnarray}

We denote by ${\cal L}$ the linear subspace of $R^{n+1}$ that is the linear
hull of all integer solutions of equation (\ref{gty}). 
The dimension of ${\cal L}$ is equal to the multiplicity of the resonance. 
Let a vector $\vec m\in R^{n+1}$ be orthogonal to ${\cal L}$, 
i.e. $\vec m\perp {\cal L}$. Then the function 
\begin{eqnarray}
F_{\vec m}\;=\;i\,m_k \,\eta_k\, \xi_k\:+\:m_{n+1}\,u_3  
\label{tree1}
\end{eqnarray}
satisfies the condition $\:\{ F_{\vec m},\, \bar{h}\}\:=\:0\:$ 
and is hence an
invariant. So we have $\;n\,+\,1\,-\,\mbox{dim}\,{\cal L}\;$ functionally
independent
invariants of the type (\ref{tree1}).

Besides that an arbitrary normal form commutes with the function 
\begin{eqnarray}
F \;=\; \bar{h} \:-\: \left( i\, \mbox{\ae}_k \,\eta_k \,\xi_k 
\:+\: \lambda_s\, u_3 \:-\:E \right)
\label{tree2}
\end{eqnarray}
and with the Casimir function
\begin{eqnarray}
s_1^2 \:+\: s_2^2 \:+\: s_3^2 \;=\; 2\, i\, u_1\, u_2 
\:+\: u_3^2  
\label{tree3}
\end{eqnarray}

{\bf Remark:} And, of course, the Hamiltonian (\ref{circu1}) is always
a constant of motion (in extended phase space).

\subsection{Action-Angle Type Coordinates}

\hspace*{0.5cm}
Instead of the standard cartesian coordinates 
\begin{eqnarray}
\eta_1, \ldots, \eta_n, 
\hspace{0.3cm}
\xi_1, \ldots, \xi_n, 
\hspace{0.3cm}
u_1, u_2, u_3 
\nonumber
\end{eqnarray}
for normal form analysis and construction
we can use action-angle type canonical polar coordinates 
\begin{eqnarray}
\rho_1, \ldots, \rho_{n+2}, 
\hspace{0.3cm}
\varphi_1, \ldots,
\varphi_{n+1}
\label{cpcor}
\end{eqnarray}
introduced with the help of the formulae
\begin{eqnarray}
\eta_l \;=\; -\frac{1+i}{2}\sqrt{2 \rho_l} \exp(i \varphi_l),
\hspace{0.5cm}
\xi_l \;=\; \frac{1+i}{2}\sqrt{2 \rho_l} \exp(-i \varphi_l)
\nonumber
\end{eqnarray}
\begin{eqnarray}
l \;=\; 1, \ldots, n
\nonumber
\end{eqnarray}
\begin{eqnarray}
u_1 \;=\; \frac{1-i}{2}\sqrt{2 \rho_{n+1}} \exp(-i \varphi_{n+1}),
\hspace{0.5cm}
u_2 \;=\; \frac{1-i}{2}\sqrt{2 \rho_{n+1}} \exp(i \varphi_{n+1})
\nonumber
\end{eqnarray}
\begin{eqnarray}
u_3 \;=\; \rho_{n+2}
\nonumber
\end{eqnarray}

Note that this change of variables is not defined on the set
\begin{eqnarray}
\eta_1 \cdot \ldots \cdot \eta_n \cdot
\xi_1 \cdot \ldots \cdot \xi_n
\cdot u_1 \cdot u_2 \;=\; 0
\nonumber
\end{eqnarray} 
and, as in the case of the standard linear transformation $\:Q$, 
it is a Poisson transformation, but it is not canonical (and is not even
symplectic with respect to the orbital 
variables\footnote{The symplecticity with respect to the orbital
variables can be easily achieved, if necessary, by interchanging
the places of the orbital actions and angles in (\ref{cpcor}).}). 

The nonzero Poisson brackets of the new basis functions are given by the
equalities
\begin{eqnarray}
\{\varphi_l, \,\rho_l \} \;=\; 1, 
\hspace{1.0cm} 
l \;=\; 1, \ldots, n
\nonumber
\end{eqnarray}
\begin{eqnarray}
\{\rho_{n+1},\,\varphi_{n+1} \} \;=\; \rho_{n+2}, 
\hspace{1.0cm} 
\{\varphi_{n+1},\, \rho_{n+2} \} \;=\; 1
\nonumber
\end{eqnarray}
and the Hamilton equations with the Hamiltonian 
function $H(\vec{\rho},\, \vec{\varphi})$ take the form
\begin{eqnarray}
\frac{d \rho_l}{dt} \;=\; -
\frac{\partial H}{\partial \varphi_l},
\hspace{1cm}
\frac{d \varphi_l}{dt} \;=\;
\frac{\partial H}{\partial \rho_l}
\nonumber
\end{eqnarray}
\begin{eqnarray}
l \;=\; 1, \ldots, n
\nonumber
\end{eqnarray}
\begin{eqnarray}
\frac{d \rho_{n+1}}{dt} \;=\;
\rho_{n+2} \cdot
\frac{\partial H}{\partial \varphi_{n+1}},
\hspace{1cm}
\frac{d \varphi_{n+1}}{dt} \;=\;
-\rho_{n+2} \cdot
\frac{\partial H}{\partial \rho_{n+1}} \:+\:
\frac{\partial H}{\partial \rho_{n+2}}
\nonumber
\end{eqnarray}
\begin{eqnarray}
\frac{d \rho_{n+2}}{dt} \;=\;
-\frac{\partial H}{\partial \varphi_{n+1}}
\nonumber
\end{eqnarray}

The direct transition from the original variables $\vec{z}$
to the polar coordinates introduced above has the form
\begin{eqnarray}
q_l \;=\; \sqrt{2 \rho_l} \sin(\varphi_l),
\hspace{1.0cm}
p_l \;=\; \sqrt{2 \rho_l} \cos(\varphi_l)
\nonumber
\end{eqnarray}
\begin{eqnarray}
l \;=\; 1, \ldots, n
\nonumber
\end{eqnarray}
\begin{eqnarray}
s_1 \;=\; \sqrt{2 \rho_{n+1}} \cos(\varphi_{n+1}),
\hspace{1.0cm}
s_2 \;=\; \sqrt{2 \rho_{n+1}} \sin(\varphi_{n+1})
\nonumber
\end{eqnarray}
\begin{eqnarray}
s_3 \;=\; \rho_{n+2}
\nonumber
\end{eqnarray}

\subsection{Normal Form in Nonresonant Case}

\hspace*{0.5cm}
In the {\bf nonresonant case} ($\mbox{dim}\,{\cal L}=0$) the normal form
(\ref{circu1})
contains only those monomials for which
\begin{eqnarray}
I\;=\;J,
\hspace{0.5cm}
L_1\;=\;L_2,
\hspace{0.5cm}
N\;=\;0  
\nonumber
\end{eqnarray}
and hence we can rewrite it as
\begin{eqnarray}
\bar h\;=\;i\,\mbox{\ae}_k\,\eta_k\,\xi_k\:+\:\lambda_s\,u_3
\:+\:\bar{h}_{IL_1L_3}\,
(\eta \xi)^I\,(u_1u_2)^{L_1}\,u_3^{L_3}\:-\:E
\label{circu2}
\end{eqnarray}
where the summation is made over all nonnegative integer vectors $I$ and
over all nonnegative integers $L_1$, $L_3$ satisfying the condition 
\begin{eqnarray}
2|I| \:+\: 4 L_1 \:+\: 2 L_3 \;\geq\; 3  
\nonumber
\end{eqnarray}

According to (\ref{tree1}) the functions
\begin{eqnarray}
i\eta _1\xi _1,
\hspace{0.5cm}
\ldots ,
\hspace{0.5cm}
i\eta _n\xi _n,
\hspace{0.5cm}
u_3  
\label{circu3}
\end{eqnarray}
are constants of motion. Taking into account that $\bar h$ is independent
of $v$, and subtracting from the Casimir function (\ref{tree3}) the
invariant $u_3^2$ we get two additional invariants 
\begin{eqnarray}
E,
\hspace{0.5cm}
iu_1u_2  
\label{circu4}
\end{eqnarray}
Altogether (\ref{circu3}) and (\ref{circu4}) give us $n+3$ independent
integrals of motion. This allows us to consider the system with
the Hamiltonian (\ref{circu2}) as integrable. 
Moreover it is easy to find the general
solution by quadrature. Let us do this in the original real variables.

In the variables $\;\vec z\;$ the Hamiltonian 
(\ref{circu2}) can be written as 
\begin{eqnarray}
\bar{\cal H}\;=\; 
\bar{H}\left(J_1,\,\ldots,\,J_n,\,I_1,\,I_2\right)\:-\:E
\end{eqnarray}
where 
\begin{eqnarray}
J_k\;=\;\frac 12\left( q_k^2+p_k^2\right),
\hspace{0.5cm}
I_1\;=\;\frac 12\left(s_1^2+s_2^2\right),
\hspace{0.5cm}
I_2\;=\;s_3  
\nonumber
\end{eqnarray}
are the functions (\ref{circu3}), (\ref{circu4}) expressed in 
variables $\:\vec z\:$, and the equations of motion take the form 
\begin{eqnarray}
\frac{dq_k}{d\tau }\;=\;\omega_k\,p_k,
\hspace{1.0cm}
\frac{dp_k}{d\tau}\;=\;-\omega_k\,q_k  
\label{star1}
\end{eqnarray}
\begin{eqnarray}
\frac{ds_1}{d\tau}\;=\;-\Omega\, s_2,
\hspace{0.5cm}
\frac{ds_2}{d\tau}\;=\;\Omega \,s_1,
\hspace{0.5cm}
\frac{ds_3}{d\tau}\;=\;0  
\label{star2}
\end{eqnarray}
Here we have used the notations
\begin{eqnarray}
\omega_k\left(J_1,\,\ldots,\,J_n,\,I_1,\,I_2\right) \;=\; 
\frac{\partial \bar{H}}{\partial J_k},
\label{star4_d}
\end{eqnarray}
\begin{eqnarray}
\Omega \left(J_1,\,\ldots,\,J_n,\,I_1,\,I_2\right) \;=\;
\frac{\partial \bar{H}}{\partial I_2}\:-\: 
\frac{\partial \bar{H}}{\partial I_1} \cdot I_2  
\label{star4}
\end{eqnarray}
and we have omitted the trivial equations for the variables $v$ and $E$.

Fixing for simplicity $\tau_0=0$ and taking into account that the functions
(\ref{star4_d}) and (\ref{star4})
are constants of motion, we get the solution of the system 
(\ref{star1}), (\ref{star2}) in the form
\begin{eqnarray}
q_k(\tau)\;=\;\cos(\omega_k^0\tau)\,q_k(0)\:+\:
\sin(\omega_k^0\tau)\,p_k(0)
\nonumber
\end{eqnarray}
\begin{eqnarray}
p_k(\tau)\;=\;-\sin(\omega_k^0\tau)\,q_k(0)\:+\:
\cos(\omega_k^0\tau)\,p_k(0) 
\nonumber
\end{eqnarray}
\begin{eqnarray}
s_1(\tau)\;=\;\cos(\Omega^0\tau)\,s_1(0)\:-\:
\sin(\Omega^0\tau)\,s_2(0)
\nonumber
\end{eqnarray}
\begin{eqnarray}
s_2(\tau)\;=\;\sin(\Omega^0\tau)\,s_1(0)\:+\:
\cos(\Omega^0\tau)\,s_2(0)
\nonumber
\end{eqnarray}
\begin{eqnarray}
s_3(\tau)\;=\;s_3(0)
\nonumber
\end{eqnarray}
where $\;\omega_k^0\;$ and $\;\Omega^0\;$ are the values 
of the functions (\ref{star4_d}) and (\ref{star4}) 
calculated at the initial time $\;\tau \:=\:0\,$.

Obviously, these formulae give us the solution of 
the triangular system too,
if we redefine $\:\omega_k\:$ and $\:\Omega\:$ in 
(\ref{star4_d}), (\ref{star4}) as
\begin{eqnarray}
\omega_k \left(J_1,\ldots,J_n\right) \;=\; 
\left. \frac{\partial \bar{H}}{\partial J_k}\right|_{I_1=I_2=0}, 
\hspace{0.5cm} 
\Omega\left(J_1,\ldots,J_n\right) \;=\; 
\left. \frac{\partial \bar{H}}{\partial I_2}\right|_{I_1=I_2=0}  
\nonumber
\end{eqnarray}
The values $\omega_k$ and $\Omega$ are called the ({\bf nonlinear})
{\bf orbital} and {\bf spin tunes} respectively.

\subsection{Normal Form in the Case of
a Single Isolated Resonance}

\hspace*{0.5cm}
We shall say that the Hamiltonian function $\;\bar{{\cal H}}\;$ 
is a {\bf generalized single 
resonance normal form}\footnote{In order to keep the size
of this paper within reasonable bounds we shall not discuss the
formal aspects of constructing of the real canonical
transformation which will bring the initial Hamiltonian to
a single resonance normal form because this procedure
is almost the same as that described in subsection 6.2.}
if there exist integers  
\begin{eqnarray}
k_1^0, \hspace{0.5cm} \ldots, \hspace{0.5cm} k_{n+1}^0, 
\hspace{0.5cm} N^0 
\nonumber
\end{eqnarray}
where not all $\hspace{0.15cm}k_l^0\hspace{0.15cm}$ 
are zero such that when written
in complex coordinates $\vec w$ this function has the
form
\begin{eqnarray}
\bar h\;=\;i\,\mbox{\ae}_k\,\eta_k\,\xi_k \:+\: \lambda_s\, u_3 \:+\:
\bar h^{NIJL} \exp(i N v)\, \eta^I \xi^J u^L \:-\: E
\label{circu22}
\end{eqnarray}
and so that in the expansion (\ref{circu22}) there are only those
terms whose indices for some integer 
$\hspace{0.15cm}m\:=\:m(N,I,J,L)\hspace{0.15cm}$ satisfy
\begin{eqnarray}
I_l \:-\: J_l \;=\; m \cdot k_l^0, 
\hspace{1cm} 
l \;=\; 1, \ldots, n
\nonumber
\end{eqnarray}
\begin{eqnarray}
L_2 \:-\: L_1 \;=\; m \cdot k_{n+1}^0, 
\hspace{0.5cm} 
N \;=\; m \cdot N^0,
\hspace{0.5cm} 
|I|+|J|+2|L| \;\geq\; 3  
\nonumber
\end{eqnarray}
Thus, the normal form (\ref{circu22}) contains only
those monomials for which
\begin{eqnarray}
N \:+\: (I-J) \cdot \vec{\mbox{\ae}} \:+\: (L_2-L_1) \cdot \lambda_s 
\;= 
\nonumber
\end{eqnarray}
\begin{eqnarray}
=\;m \cdot \left(
N^0 \:+\: k_1^0 \cdot \mbox{\ae}_1 \:+\: \ldots \:+\:
k_n^0 \cdot \mbox{\ae}_n
\:+\: k_{n+1}^0 \cdot \lambda_s \right) 
\nonumber
\end{eqnarray}
and we do not require the condition
\begin{eqnarray}
N^0 \:+\: k_1^0 \cdot \mbox{\ae}_1 \:+\: \ldots \:+\:
k_n^0 \cdot \mbox{\ae}_n
\:+\: k_{n+1}^0 \cdot \lambda_s \;=\; 0 
\nonumber
\end{eqnarray}
to be satisfied.

Let a vector $\:\vec m\in R^{n+1}\:$ be orthogonal to the
vector $\:\vec{k}^0\:$. Then the function 
\begin{eqnarray}
F_{\vec m}\;=\;i\,m_k\, \eta_k\,\xi_k\:+\:m_{n+1}\,u_3  
\label{esr1}
\end{eqnarray}
is an invariant. So, together with the Casimir function
(\ref{tree3}) and the Hamiltonian (\ref{circu22}),
we have $n+2$ functionally independent
integrals of motion, and in the remainder of this subsection
we shall find an additional invariant that will allow us
to consider the single resonance problem as integrable.

\subsubsection{Single Resonance Between Orbital Tunes}

\hspace*{0.5cm}
If we have a single resonance between
orbital tunes (i.e. $\;k^0_{n+1} \:=\: 0$) 
then the required additional invariant is given by
an expression
\begin{eqnarray}
F \;=\; i\, \varepsilon \,k_k^0\, \eta_k \,\xi_k \:+\:
\left(\bar h \:+\: E \:-\: i\,\mbox{\ae}_k\, \eta_k\,\xi_k \:-\: 
\lambda_s\,u_3
\right) 
\label{circu221}
\end{eqnarray}
where the quantity
\begin{eqnarray}
\varepsilon \;=\; 
\frac{\mbox{\ae}_k k_k^0\: +\: N^0}{\parallel \vec{k}^0\parallel^2}
\nonumber
\end{eqnarray}
is called the {\bf distance from orbital resonance}.
If the integral (\ref{circu221}) is independent of the integrals
of the form (\ref{esr1}) and of the Casimir function 
(\ref{tree3}), the single resonance normal form (\ref{circu22})
is completely integrable. If not, then $\bar h + E$ is a 
series in 
$i\eta _1\xi _1,\ldots ,i\eta _n\xi _n,iu_1u_2,u_3$  
only, and the additional integral will be
$F =i k^0_k \eta_k \xi_k$. Hence, we are always in the
situation of complete integrability. 

The triangular truncated equations of spin motion corresponding to 
the Hamiltonian (\ref{circu22})
\begin{eqnarray}
\frac{du_1}{d\tau }=-i\Omega u_1,
\hspace{0.4cm}
\frac{du_2}{d\tau }= i\Omega u_2,
\hspace{0.4cm}
\frac{du_3}{d\tau }= 0,  
\hspace{0.4cm}
\Omega  = 
\left. \frac{\partial \bar{h}}{\partial u_3}
\right|_{u_1=u_2=u_3=0}  
\label{sssr}
\end{eqnarray}
can be easily integrated by quadrature if 
we know the solution of the orbital part
of equations of motion. In the original variables
the system (\ref{sssr}) will take the form 
\begin{eqnarray}
\frac{ds_1}{d\tau }=-\Omega s_2,
\hspace{0.4cm}
\frac{ds_2}{d\tau }= \Omega s_1,
\hspace{0.4cm}
\frac{ds_3}{d\tau }= 0,  
\hspace{0.4cm}
\Omega  = 
\left. \frac{\partial \bar{{\cal H}}}{\partial s_3}
\right|_{s_1=s_2=s_3=0}  
\nonumber
\end{eqnarray}
and this allows us to draw a number of conclusions.
For example, we can state that in principle,
it is possible to organize slow extraction of
a polarized proton beam using a third-integer orbital
resonance without significant loss of polarization.

{\bf Remark:} Note that the case considered here
includes, in our opinion, the correct treatment of 
the situation when the synchrotron frequency is very 
small compared to the other frequencies in the system 
(as, for example, 
in the case of HERA-p ring \cite{wash2_1}).

\subsubsection{Single Spin-Orbit Resonance}

\hspace*{0.5cm}
If $\;k^0_{n+1} \:\neq\: 0\;$ then as an additional 
invariant we can take the function
\begin{eqnarray}
F \;=\; \varepsilon \,u_3 \:+\:
\left(\bar h \:+\: E \:-\: i\,\mbox{\ae}_k\, \eta_k\,\xi_k 
\:-\: \lambda_s\, u_3
\right) 
\label{circu2211}
\end{eqnarray}
where the quantity
\begin{eqnarray}
\varepsilon \;=\; 
\frac{\mbox{\ae}_k\, k_k^0 \:+\: \lambda_s\, k_{n+1}^0 \:
+\: N^0}{k_{n+1}^0}
\nonumber
\end{eqnarray}
is called the {\bf distance from spin-orbit resonance}.
If the integral (\ref{circu2211}) is independent of the integrals
of the form (\ref{esr1}) and of the Casimir function 
(\ref{tree3}), the single resonance normal form (\ref{circu22})
is completely integrable. If not, then $\bar h + E$ is a 
series in 
$i\eta _1\xi _1,\ldots ,i\eta _n\xi _n,iu_1u_2,u_3$  
only, and the additional integral will be
$F = u_3$. Hence, again we are in the
situation of complete integrability. 

In more detail the invariant (\ref{circu2211}) can be written as
\begin{eqnarray}
F(v,\, \vec{\eta}, \,\vec{\xi},\, \vec{u}\,) \;=\; \varepsilon \,u_3 
\:+\:
\bar h^{NIJL} \exp(i N v)\, \eta^I \xi^J u^L 
\nonumber
\end{eqnarray}
and we see that in general it does not satisfy the 
condition $F(v, \vec{\eta}, \vec{\xi}, \vec{0}) \equiv 0$,
which is very important 
for application to the triangular system. This situation can be
easily changed if we choose $n$ invariants of the type
(\ref{esr1}) in the form
\begin{eqnarray}
V_1 \;=\; i \eta_1 \xi_1 - \frac{k_1^0}{k^0_{n+1}}\, u_3 ,
\hspace{0.7cm}
\ldots,
\hspace{0.7cm}
V_n \;=\; i \eta_n \xi_n - \frac{k_n^0}{k^0_{n+1}}\, u_3 
\label{stergef}
\end{eqnarray}
and instead of $F$ use the invariant
\begin{eqnarray}
\bar F \;=\; F \:-\: \sum \limits_{|I| > 1} \bar h^{0II0} \:
i^{-|I|} \:V^I 
\nonumber
\end{eqnarray}
So the triangular system will have the spin dependent integral of
motion  
\begin{eqnarray}
\left.\frac{\partial \bar F}{\partial u_1} \right|_{\vec u = \vec 0}
\cdot u_1 \;+\;
\left.\frac{\partial \bar F}{\partial u_2} \right|_{\vec u = \vec 0}
\cdot u_2 \;+\;
\left.\frac{\partial \bar F}{\partial u_3} \right|_{\vec u = \vec 0}
\cdot u_3
\nonumber
\end{eqnarray}
and hence the solution for spin can be obtained by
quadrature following, for example, the procedure described in
Appendix D (a detailed analysis will be published in a separate
paper).

{\bf Remark:} The set of invariants (\ref{stergef}) show us
that a proton beam cannot be split into two
polarized parts with different amplitudes
using the  classical Stern-Gerlach effect 
at a single spin-orbit resonance as has been suggested
in \cite{russm}. For related comments see \cite{derb,barber2}.

\section{Methods of Numerical Integration}

\hspace*{0.5cm}
Today {\bf symplectic tracking}
methods for orbital motion (methods which conserve the classical Poisson
bracket\footnote{Numerical integration schemes for differential equations
which preserve other {\bf qualitative properties} (like 
autonomous Hamiltonian or,
more generally, first integral conserving algorithms, and so forth)
are discussed in \cite{marsden}, and more recent references may be
found in \cite{quispel}}) are common tools in accelerator physics
and we believe that such methods, which conserve the Poisson
bracket (\ref{fom2}) exactly or with high accuracy, are the most suitable
ones for numerical simulation of the equations (\ref{f31})-(\ref{f33}).
For simplicity we consider the case of
autonomous Hamiltonian
functions. In the nonautonomous case we can repeat all the steps,
introducing two additional canonical variables to obtain an autonomous
Hamiltonian system in a higher dimensional phase space.

Those methods, of course, are mainly of theoretical interest,
but in application to the triangular system (practical interest) the
Hamiltonian approach allows us to reduce the initial problem of
numerical integration of the system (\ref{vk1})-(\ref{vk2}) to that of
symplectic integration
of the equations of orbital motion (\ref{vk1}) only.

\subsection{Canonical Numerical Methods 
for Hamiltonian Equations}

\hspace*{0.5cm}
Let $K$ be a compact subset of the manifold $M$, and the autonomous
Hamiltonian function $H(\vec{z}\,)$ be zero onto the set $M \setminus
N$ (here $N$ is some compact subset of $M$ containing $K$). Then the
solutions of the Hamiltonian system (\ref{fom1}) generate the one-parameter
group of the canonical transformations of phase space
\begin{eqnarray}
T(\tau) : M \;\rightarrow\; M, 
\hspace{1.0cm} 
T(0) \;=\; I  
\nonumber
\end{eqnarray}

Consider a one-parameter family of canonical transformations
\begin{eqnarray}
T_m (\tau) : M \;\rightarrow\; M  
\nonumber
\end{eqnarray}
defined for $\:\mid \tau \mid \;<\; \tau_0\:$ ($0 \;<\; \tau_0$).  

We will say that $\:T_m(\tau)\:$ {\bf approximates} 
$\:T(\tau)\:$ in a
neighbourhood
of the identity mapping with the order 
$m$ (with respect to $K$), if for all
$\;\vec{z} \:\in\: K\;$
\begin{eqnarray}
\parallel T(\tau)\,\vec{z} \:-\: T_m(\tau)\,\vec{z} \parallel 
\;=\;
O(\tau^{m+1})
\nonumber
\end{eqnarray}
Here $\:\parallel \cdot \parallel\:$ is the norm in Euclidean space,
containing
the manifold $M$.

A one-parameter family $\:T_m(\tau)\:$ is said to be 
{\bf locally computable} if
for each $\:\vec{z}_0 \,\in\, K\:$ there is a chart $U$ 
(from the atlas of the manifold $M$) containing all points
\begin{eqnarray}
\vec{z}_{\tau} \;=\; T_m(\tau)\, \vec{z}_0  
\nonumber
\end{eqnarray}
for $\:\mid \tau \mid \,<\, \tau_0\:$ and a vector function 
$\:\vec{F}_{U}\:$ such that $\:\vec{z}_{\tau}\:$ is defined 
uniquely by the equation
\begin{eqnarray}
\vec{F}_{U} (\tau, \,\vec{z}_0,\, \vec{z}_{\tau}) \;=\; \vec{0}
\nonumber
\end{eqnarray}

{\bf Definition: } {\sl A locally computable one-parameter family of
canonical transformations $\:T_{m}(\tau)\:$ is said to be a {\bf canonical
integrator} of order $\:m\:$ of the group $\:T(\tau)$, if $\:T_m(\tau)\:$ 
approximates
$\:T(\tau)\:$ with the order $\:m$.}

\subsubsection{Recursively Generated High Order 
Canonical Integrators}

\hspace*{0.5cm}
Let us assume that by some method we find 
a canonical integrator of order $\,m\,$
of the group $\,T\,$, which has in each of the charts $\,U\,$ 
the formal representation\footnote{
The representation (\ref{d001}) holds if, for example, 
we assume the constancy of the rank of the Poisson bracket,
the possibility of introducing Darboux coordinates globally in each 
of the charts $U$ and some smoothness properties of 
the integrator considered.}
\begin{eqnarray}
T_m(\tau) \;=\; \exp(:-\tau\, H\:+\: \tau^{m+1}R_{m}(\tau,\, U):)
\:+\: O(\tau^{r+1})
\label{d001}
\end{eqnarray}
where $\:m \:<\: r$. 

The purpose of the present section is to demonstrate 
a general approach for constructing higher order integrators 
starting from integrator (\ref{d001}) of order $\,m$.  

Following \cite{suz1,suz2} we consider the mapping 
\begin{eqnarray}
P(\tau) \;=\; \prod\limits_{j = 1}^{n} \:
T_{m}^{q_j} (p_j \,\tau)
\label{d002}
\end{eqnarray}
where the exponents $\:q_j\:$ are nonzero integers and 
the coefficients 
$\:p_j\:$ are real numbers. Substituting
the representation (\ref{d001}) in (\ref{d002}) and combining 
the product of exponential operators into a single Lie exponent 
using the Campbell-Baker-Hausdorff formula we have
\begin{eqnarray}
P(\tau) \;=\;O(\tau^{r+1}) \;+
\nonumber
\end{eqnarray}
\begin{eqnarray}
+\:
\exp(:-\tau (q_j p_j) H \:+\: \tau^{m+1} (q_j p_j^{m+1}) R_{m}(0, U) 
\:+\:
O(\tau^{m+2}):) 
\label{d003}
\end{eqnarray}

If $\:q_j\,$, $\:p_j\:$ satisfy the conditions 
\begin{eqnarray}
q_j\cdot p_j \;=\; 1, 
\hspace{1cm} 
q_j\cdot p_j^{m+1} \;=\; 0  
\label{gggg1}
\end{eqnarray}
and $\:m+1 \,\geq\, r\:$
then from (\ref{d003}) it follows that the mapping (\ref{d002}) 
is a canonical integrator of order $\,m+1\,$ of the group $\:T(\tau)\,$.

Besides that, if $\,m\,$ is an even number, $\:m+2 \,\geq\, r\,$,
$\:T_m(\tau)\:$ is {\bf time reversible}, so that
\begin{eqnarray}
T_m(\tau)\: T_m(-\tau) \;=\; I  
\nonumber
\end{eqnarray}
and
\begin{eqnarray}
q_{n + 1 - j} \;=\; q_j, 
\hspace{1cm} 
p_{n + 1 - j} \;=\; p_j  
\nonumber
\end{eqnarray}
then (\ref{d002}) gives us a
time reversible canonical integrator of order
$m+2$.

Obviously, the scheme (\ref{d002}) may be recursively applied  
any number of times to obtain a canonical integrator of
required order $\:l\:$ 
(of course, with the assumption that this order $\:l\:$
is not bigger than $\:r$).

{\bf Example :} Let 
\begin{eqnarray}
a_m \;=\; 
\left(
r_m \:+\: l_m
\left(-\frac{r_m}{l_m}\right)^{\frac{1}{m+1}}
\right)^{-1}, 
\hspace{1cm} 
b_m \;=\; 
\left(
-\frac{r_m}{l_m}
\right)^{\frac{1}{m+1}} a_m  
\nonumber
\end{eqnarray}
where nonzero integers 
$\;r_m, \:l_m\; (r_m \:\neq \:\pm l_m)\;$ for odd $\,m\,$ 
satisfy the condition
\begin{eqnarray}
r_m\, l_m \;<\; 0  
\nonumber
\end{eqnarray}
Then
\begin{eqnarray}
\left\{
\begin{array}{lllll}
r_m \cdot a_m  &+ & l_m \cdot b_m &= &1\\
\\
r_m \cdot a_m^{m+1} &+& l_m \cdot b_m^{m+1} &=&0
\end{array}
\right.
\nonumber
\end{eqnarray}
so that (\ref{gggg1}) is satisfied for $\:n\,=\,2$ and hence
the integrator of order $\:m+1\:$ can be chosen in the form
\begin{eqnarray}
T_{m+1}(\tau) \;=\; T_m^{r_m}(a_m \,\tau)\: T_m^{l_m}(b_m \,\tau)
\nonumber
\end{eqnarray}

If additionally $\:m\:$ and $\:|r_m|\:$ are even numbers, 
and $\:T_m(\tau)\:$ is time reversible,
then the integrator of order $\:m+2\:$ can be taken as
\begin{eqnarray}
T_{m+2}(\tau) \;=\; T_m^{\frac{r_m}{2}} (a_m \,\tau) \;
T_m^{l_m}(b_m\,\tau) \;
 T_m^{\frac{r_m}{2}}(a_m \,\tau)  
\nonumber
\end{eqnarray}

{\bf Remark :} Note that the recursive scheme considered is slightly more
general than that used in \cite{suz1,suz2}. Here it is not necessary
that the integers $\:q_j\:$ in (\ref{d002}) are positive and the initial
integrator is
not assumed to be time reversible.

\subsubsection{Low Order Canonical Integrators for Equations 
of Classical Spin-Orbit Motion}

\hspace*{0.5cm}
In order to be able to start the recursive procedure described above,
we now turn to the problem of constructing low order canonical
integrators for a Hamiltonian system. We will give a few examples, and
other
helpful recipes (and references) may be found in 
\cite{marsden,sanz-serna,f3,f2}.

{\bf Example 1:} If $\:\hat{J}\:$ the structure matrix of Poisson bracket  
is constant, then for any constant matrix $\:A\,$, satisfying the
condition 
$\;\hat{J}\:=\: A \hat{J} \:+\: \hat{J} A^{\top}\;$, the mapping
\begin{eqnarray}
\vec{z}_{\tau} \;=\; \vec{z}_{0} \:+\: \tau\, \vec{\varphi}
(A\,\vec{z}_{\tau} \:+\: (I-A)\,\vec{z}_0)  
\nonumber
\end{eqnarray}
where
\begin{eqnarray}
\vec{\varphi}(\vec{z}\,) \;=\; 
\hat{J} \hspace{0.15cm} \mbox{grad}_{\:\vec{z}} 
\,H(\vec{z}\,)  
\nonumber
\end{eqnarray}
is a canonical integrator of first order. In the special case 
when $\:A\,=\,\frac{1}{2} I\,$, this integrator is a second order 
approximation, is
time reversible, and is known as the {\bf mid-point rule} \cite{marsden}.

{\bf Example 2:} The property that the integrator is canonical is
preserved
under Poisson transformations. This simple remark combined with example 1
allows us to construct first or second order canonical integrators for
the equations of classical spin-orbit motion, because we know the 
explicit form
of the transition formulae to Darboux coordinates. Note that although this
approach is applicable for arbitrary dependence of the Hamiltonian function
on orbit and spin variables, its computer realization requires the use of
more than one (at least two) local coordinate systems.

{\bf Example 3:} Consider a Hamiltonian function which depends on spin
variables only in the linear combination $\;\vec i\cdot \vec s$
\begin{eqnarray}
H\;=\;H\left(\vec x,\,\vec i\cdot \vec s\,\right)  
\nonumber
\end{eqnarray}
Without loss of generality we can assume that 
$\:\left|\vec i\right|\, =\,1$.
Supplement the vector $\,\vec i\,$ with two unit
vectors $\,\vec j\,$ and $\,\vec k\,$ satisfying the condition
\begin{eqnarray}
\vec i\cdot \left[\:\vec j\times \vec k\:\right]\;=\;1  
\nonumber
\end{eqnarray}
so that the triplet $\vec i$, $\vec j$, $\vec k$ forms an 
orthogonal coordinate system.

Then the mapping given by the system of equations
\begin{eqnarray}
\vec s_\tau \;=\;\left(
\begin{array}{c}
\vec i \\
\vec j \\
\vec k
\end{array}
\right) ^{\top }\left(
\begin{array}{rrr}
1 & 0 & 0 \\
0 & \cos (\alpha \tau ) & -\sin (\alpha \tau ) \\
0 & \sin (\alpha \tau ) & \cos (\alpha \tau )
\end{array}
\right) \left(
\begin{array}{c}
\vec i \\
\vec j \\
\vec k
\end{array}
\right) \hspace{0.1cm}\vec s_0  
\nonumber
\end{eqnarray}
\begin{eqnarray}
\vec x_\tau \;=\;\vec x_0\:+\:\tau\, \vec \varphi 
\left( {\frac{{\vec x_\tau +\vec x_0}}2},
\hspace{0.2cm}\vec i\cdot \vec s_0\right)  
\nonumber
\end{eqnarray}
is a time reversible canonical integrator of second order defined in the
initial variables (global ones in 9-dimensional phase space). Here $J$ is
the symplectic unit and
\begin{eqnarray}
\vec \varphi \left(\vec x,\:\vec i\cdot \vec s\:\right)
\;=\;J
\hspace{0.15cm}\mbox{grad}_{\:\vec x}\:
H\left(\vec x,\:\vec i\cdot \vec s\:\right)
\nonumber
\end{eqnarray}
\begin{eqnarray}
\omega \left(\vec x,\:\vec i\cdot \vec s\:\right)\;=\;
\vec i\cdot
\mbox{grad}_{\:\vec s}\:H\left(\vec x,\:\vec i\cdot \vec s\:
\right)  
\nonumber
\end{eqnarray}
\begin{eqnarray}
\alpha \;=\;\omega 
\left( {\frac{{\vec x_\tau +\vec x_0}}2},\;
\vec i\cdot \vec s_0\right)  
\nonumber
\end{eqnarray}

{\bf Example 4:} If the Hamiltonian function 
$H(\vec{x}, \,\vec{s}\,)$ can be
represented in the form of the sum
\begin{eqnarray}
H(\vec{x},\, \vec{s}\,) \;=\; 
H_1\left(\vec{x}, \:\vec{i}_1 \cdot \vec{s}\:\right) \:+\:
\ldots \:+\: 
H_m\left(\vec{x},\: \vec{i}_m \cdot \vec{s}\:\right)  
\nonumber
\end{eqnarray}
then the mapping 
\begin{eqnarray}
A^{\vec{i}_m}_{H_m}\left(\frac{\tau}{2}\right) 
\ldots A^{\vec{i}_2}_{H_2}\left(\frac{\tau}{2}\right) 
A^{\vec{i}_1}_{H_1}\left(\tau\right) 
A^{\vec{i}_2}_{H_2}\left(\frac{\tau}{2}\right) \ldots 
A^{\vec{i}_m}_{H_m}\left(\frac{\tau}{2}\right)  
\nonumber
\end{eqnarray}
where $\:A^{\vec{i}}_{H}\:$ is the integrator of example 3, gives us a
time
reversible canonical integrator of second order. It solves (at least in
theory) the problem of canonical integration for 
Hamiltonians linearly dependent on spin.

\subsection{Approximately Canonical Numerical 
Methods for Short-Term Tracking 
(Numerical Methods for the Triangular System)}

\hspace*{0.5cm}
The length of the vector $\:\vec s\:$ is commensurate with Planck's
constant 
$\:\hbar $. If we perform the renormalization 
\begin{eqnarray}
\vec s_{old}\;\rightarrow\; {\frac \hbar 2}\cdot \vec s_{new},
\hspace{1.0cm}
\mid \vec s_{new}\mid \;=\;1  
\nonumber
\end{eqnarray}
then  the equations of spin-orbit motion corresponding to the
Hamiltonian function $\;H(\vec x,\,\vec s\,)\;$  become
\begin{eqnarray}
{\frac{{d\vec x}}{{d\tau }}}\;=\;J\hspace{0.15cm}
\mbox{grad}_{\:\vec x}\:H_{orbt}(\vec x)\:+\:O(\hbar)  
\nonumber
\end{eqnarray}
\begin{eqnarray}
{\frac{{d\vec s_{new}}}{{d\tau }}}\;=\;
\left[\:\vec W(\vec x\,)\times \vec s_{new}\:\right]\:+\:O(\hbar)
\nonumber
\end{eqnarray}
where 
\begin{eqnarray}
H_{orbt}(\vec x)\;=\;
H\left(\vec x,\,\vec 0\,\right)
\hspace{0.5cm}
{\mbox{and}}
\hspace{0.5cm}
\vec W(\vec x)\;=\;\left. \left(\mbox{grad}_{\:\vec s}
\:H(\vec x,\,\vec s\,)
\right)\right|_{\vec s=\vec 0}  
\nonumber
\end{eqnarray}
Consequently both the effect of spin on the orbit motion 
and the induced nonlinear
influence of spin on itself, are very small and 
we will neglect them, if the
integration interval $T$ satisfies the condition 
\begin{eqnarray}
T\;<<\;\hbar ^{-1}  
\nonumber
\end{eqnarray}
This means that with high accuracy we can reduce the initial problem of
numerical integration of the equations of spin-orbit motion to that of
symplectic integration of orbital motion equations only. 
We will now demonstrate this in
a few steps.

First, for any vector $\vec U(\vec x)$ we can approximate
the action of the mapping 
\begin{eqnarray}
\exp \left(-:\vec U(\vec x)\cdot \vec s \:+\: 
O\left(|\vec s|^2\right):\right)
\nonumber
\end{eqnarray}
on phase space variables with required precision by means of 
the explicit
formulae 
\begin{eqnarray}
\vec x_f\;=\;\vec x_i  
\label{sodx}
\end{eqnarray}
\begin{eqnarray}
\vec s_f\;=\;\left(I\,+\,
\frac{\sin \left(\mid \vec U\mid \right)}{\mid \vec U\mid }
\, A(\vec x_i)\;+\;
\frac{1\,-\,\cos\left(\mid\vec U\mid\right)}
{\mid \vec U\mid^2}
\, A^2(\vec x_i)
\right)\: \vec s_i  
\label{sods}
\end{eqnarray}
where 
\begin{eqnarray}
A\;=\;\left( 
\begin{array}{ccc}
0 & -U_3 & U_2 \\ 
U_3 & 0 & -U_1 \\
-U_2 & U_1 & 0
\end{array}
\right)   
\nonumber
\end{eqnarray}

Second, for any given positive integer $k$ and using the
Campbell-Baker-Hausdorff formula it is possible 
to find vectors 
\begin{eqnarray}
\vec W^1(\vec x,\,\tau),
\hspace{1.0cm}
\vec W^2(\vec x,\,\tau),
\hspace{1.0cm}
\vec W^3(\vec x,\,\tau),
\hspace{1.0cm}
\vec W^4(\vec x,\,\tau)  
\nonumber
\end{eqnarray}
so that the following decomposition formulae obtain 
\begin{eqnarray}
\exp \left( -:\tau H:\right) \;=\;
\exp \left( -:\tau \left( H_{orbt}+\vec{W}\cdot 
\vec{s}+O\left(|\vec s|^2\right)\right) :\right) \;=^k\;  
\nonumber
\end{eqnarray}
\begin{eqnarray}
\exp \left( -:{\frac \tau 2} H_{orbt}:\right) 
\exp \left( -:\tau \left(\vec W^1\cdot \vec s + 
O\left(|\vec s|^2\right)\right):\right) 
\exp \left( -:{\frac \tau 2} H_{orbt}:\right) =^k 
\nonumber
\end{eqnarray}
\begin{eqnarray}
=^k\;
\exp \left( -:{\frac \tau 2}\left( \vec W^2\cdot \vec s 
+ O\left(|\vec s|^2\right)\right):\right) 
\exp \left(-:\tau H_{orbt}:\right) \cdot
\nonumber
\end{eqnarray}
\begin{eqnarray}
\cdot \exp \left( -:{\frac \tau 2} \left(
\vec W^2\cdot \vec s + O\left(|\vec s|^2\right)\right):\right) \;=^k
\nonumber
\end{eqnarray}
\begin{eqnarray}
=^k\;
\exp \left( -:\tau H_{orbt}:\right) 
\exp \left( -:\tau \left( \vec W^3\cdot \vec s 
+ O\left(|\vec s|^2\right)\right):\right)\; 
=^k  
\nonumber
\end{eqnarray}
\begin{eqnarray}
=^k\;
\exp \left( -:\tau \left( \vec W^4\cdot \vec s 
+ O\left(|\vec s|^2\right)\right):\right) 
\exp \left( -:\tau H_{orbt}:\right)  
\label{ppp}
\end{eqnarray}
Here $\tau$ is the size of the integration step and $=^k$ indicates that
the differences between the right and left side have at least the order 
$O(\mid \tau \mid ^{k+1})$.  

Finally, let us select a decomposition formula from (\ref{ppp}) and hence
one of the vectors $\vec{W}^{i}(\vec{x}), \hspace{0.1cm} i = 1, 2,
3, 4$. If we use the combination of some symplectic integration method of
order $k$ for orbital motion and the formulas (\ref{sodx}), (\ref{sods}) for
the mapping 
\begin{eqnarray}
\exp\left(:\lambda \left(\vec{W}^{i}(\vec{x}, \tau) \cdot \vec{s} +
O\left(|\vec s|^2\right)\right):\right)  \nonumber
\end{eqnarray}
($\lambda = \tau / 2$ for $i=2$, or $\lambda = \tau$ for $i = 1,3,4$)
we will obtain a numerical method which has the order $k$, is symplectic
for the orbital motion and automatically maintains the length of a spin
vector equal to its
initial value.

{\bf Example:} 
\begin{eqnarray}
\vec{W}^1 \;=\; \vec{W} \:+\: {\frac{\tau^2 }{24}} \cdot
\left\{H_{orbt},\:\left\{H_{orbt},\:\vec{W}\right\}\right\} \:-\:
{\frac{\tau^2}{12}} \cdot \left[\left\{H_{orbt},\:
\vec{W}\right\} \times
\vec{W}\right]
\nonumber
\end{eqnarray}
\begin{eqnarray}
\vec{W}^2 \;=\; \vec{W} \:+\: {\frac{\tau^2 }{24}} \cdot
\left[\left\{H_{orbt},\:\vec{W}\right\}
\times \vec{W}\right] \:-\: {\frac{\tau^2 }{12}} \cdot
\left\{H_{orbt},\:\left\{H_{orbt},\:
\vec{W}\right\}\right\}  
\nonumber
\end{eqnarray}
for $\;k = 4\;$ and 
\begin{eqnarray}
\vec{W}^3 \;=\; \vec{W} \:+\: {\frac{\tau}{2}} \cdot
\left\{H_{orbt},\:\vec{W}\right\}, 
\hspace{1cm} 
\vec{W}^4 \;=\; \vec{W} \:-\: {\frac{\tau}{2}} \cdot 
\left\{H_{orbt},\:\vec{W}\right\}  
\nonumber
\end{eqnarray}
for $\;k = 2\,$. 

Here the notation 
$\;\left\{H_{orbt},\:\vec{U}\right\}$ means the vector
with
components:
\begin{eqnarray}
\left(\left\{H_{orbt},\:U_1\right\},
\hspace{0.3cm}
\left\{H_{orbt},\:U_2\right\},
\hspace{0.3cm}
\left\{H_{orbt},\:U_3\right\}\right)  
\nonumber
\end{eqnarray}

\section{Additional Transformations 
of the Spin-Orbit Hamiltonian}

\hspace*{0.5cm}
One of the most important problems in the design of accelerators 
is the problem of providing electric and magnetic fields 
which can hold a charged particle beam in a sufficiently small 
neighbourhood of some geometrical line  which we already introduced in 
subsection 4.3 under the name of closed design 
orbit\footnote{In fact  in this paper we never use 
the condition for the closed design orbit to be a {\bf closed} curve,
and so the equations derived can be used in applications to
linear accelerators, cyclotrons and etc. Note also that in order
to be more consistent in a coupled spin-orbit formalism 
the closed design orbit should be considered as a curve in the 
six-dimensional space of three orbital and three spin coordinates,
but we do it in fact by assuming that for the spin part this curve
satisfies the equation $\vec{s} = \vec{0}$ for all the times.}.
One of the commonly used approaches to find a solution is to 
create electric and magnetic fields for which the given
geometrical line will be the projection on $R^3$ of the six-dimensional
trajectory of orbital motion in these fields,
such that this trajectory has to be stable with respect to small 
perturbations of
the initial conditions and of the values of electric and magnetic
fields. We will now call this trajectory the {\bf reference particle},
and all transformations of the spin-orbit Hamiltonian in this section
will be more or less  connected with this concept.

\subsection{Closed Design Orbit and Reference Particle}

\hspace*{0.5cm}
Since in the final form of the equations we would like to keep  the
possibility to treat effects like misalignments of different 
electromagnetic elements of the accelerator, fluctuations
in the values of electric and magnetic fields, influence of space 
charge, etc., we will assume that the reference particle
is not a trajectory resulting 
from the Hamiltonian function (\ref{hnew}), 
but is a solution of the system with Hamiltonian function
\begin{eqnarray}
\tilde H \;=\; \tilde{H}_{orbt} + \vec{0} \cdot \vec{s} \;=\;
-\mbox{\ae} x \tilde{\pi}_{\vec B}
+\mbox{\ae} y \tilde{\pi}_{\vec N}
- (1 + hx + \alpha y) \tilde{\pi}_{\vec T}
- {e \over c} \tilde{A}_z 
\label{e0}
\end{eqnarray}
where
\begin{eqnarray}
\tilde{\pi}_{\vec N} \;=\; 
 P_x \:-\: {e \over c} \,\tilde{A}_{\vec{N}},
\hspace{1.0cm}
\tilde{\pi}_{\vec B} \;=\; 
 P_y \:-\: {e \over c}\, \tilde{A}_{\vec{B}} 
\nonumber
\end{eqnarray}
\begin{eqnarray}
\tilde{\pi}_{\vec T} \;=\; \left(
{(E - e \tilde{\Phi})^2 \over {c^2}}\: -\: 
m_0^2 c^2  \:-\: 
\tilde{\pi}_{\vec N}^2 \:-\: 
\tilde{\pi}_{\vec B}^2 
\right)^{1 / 2}
\nonumber
\end{eqnarray}
and the components of the vector potential
$\:\tilde{A}_{\vec{N}}\,$, $\:\tilde{A}_{\vec{B}}\,$, 
$\:\tilde{A}_z\,$, and 
the scalar potential  $\:\tilde{\Phi}\,$, generally speaking, are
not equal to the corresponding values in the 
Hamiltonian function (\ref{hnew})\footnote{Sometimes it is
helpful to assume that even
the charge and the rest mass of the reference
particle are different from the corresponding values in the 
Hamiltonian function (\ref{hnew}) too.}.

Now we must discuss 
the conditions under which 
the functions
\begin{eqnarray}
x_0(z), \hspace{0.5cm} P_x^0(z), \hspace{0.5cm} 
y_0(z), \hspace{0.5cm} P_y^0(z), \hspace{0.5cm} 
E_0(z), \hspace{0.5cm} t_0(z)
\label{e1}
\end{eqnarray}
will be the solution of the equations of motion corresponding to
the Hamiltonian (\ref{e0}), and the conditions under which
the projection of this solution
on $R^3$ will coincide with the closed design orbit.

Since $x$ and $y$ were introduced in subsection 4.3 as 
the transverse deviations from the closed design orbit 
it is necessary that 
\begin{eqnarray}
x_0 (z) \;\equiv\; 0
\hspace{1.0cm}
\mbox{and}
\hspace{1.0cm}
y_0 (z) \;\equiv\; 0
\label{e2}
\end{eqnarray}
Substituting (\ref{e1}) and (\ref{e2}) in the equations
\begin{eqnarray}
{d x \over d z} & = & {{\partial \tilde{H}_{orbt}} 
\over {\partial P_x}} = \hspace{0.3 cm} 
\mbox{\ae} y + {(1+hx+\alpha y) \over \tilde{\pi}_{\vec T}}
\,\tilde{\pi}_{\vec N}
\nonumber \\
{d y \over d z} & = & {{\partial \tilde{H}_{orbt}} 
\over {\partial P_y}} =
- \mbox{\ae} x + {(1+hx+\alpha y) \over \tilde{\pi}_{\vec T}}
\,\tilde{\pi}_{\vec B}
\nonumber
\end{eqnarray}
we obtain that the momenta $\:P_x^0\:$ and $\:P_y^0\:$ have to be
\begin{eqnarray}
P_x^0(z) \;=\; {e \over c}\, \tilde{A}_{\vec{N}}^0(z)
\hspace{1.0cm}
\mbox{and}
\hspace{1.0cm}
P_y^0(z) \;=\; {e \over c}\, \tilde{A}_{\vec{B}}^0(z)
\label{e3}
\end{eqnarray}
where
\begin{eqnarray}
\tilde{A}_{\vec{N}}^0(z) \;=\;
\tilde{A}_{\vec{N}}(x_0(z),\, y_0(z),\, z,\, t_0(z))
\nonumber
\end{eqnarray}
\begin{eqnarray}
\tilde{A}_{\vec{B}}^0(z) \;=\;
\tilde{A}_{\vec{B}}(x_0(z),\, y_0(z),\, z,\, t_0(z))
\nonumber
\end{eqnarray}
Then putting (\ref{e1})-(\ref{e3}) in the equations
\begin{eqnarray}
{d E \over d z}  \;=\;   
{\partial \tilde{H}_{orbt} \over \partial t},
\hspace{1.0cm}
{d t \over d z} \; =\;  
-{\partial \tilde{H}_{orbt} \over \partial E}
\nonumber
\end{eqnarray}
we obtain that $\:E_0(z)\:$ and $\:t_0(z)\:$ have to satisfy
the equalities 
\begin{eqnarray}
{d E_0 \over d z} \;=\; 
-{e \over c} 
\left( {{\partial \tilde{A}_z} \over {\partial t}} \right)^0 \:+\:
{e \over \tilde{\pi}_{\vec T}^0}
\cdot {{E_0 - e \tilde{\Phi}_0} \over {c^2}} \cdot 
\left({{\partial \tilde{\Phi}} \over {\partial t}} \right)^0
\label{e3_1}
\end{eqnarray}
\begin{eqnarray}
{d t_0 \over d z}  \;=\;  {1 \over \tilde{\pi}_{\vec T}^0}\, 
{{E_0 - e \tilde{\Phi}_0} 
\over {c^2}}
\label{e4}
\end{eqnarray}
where
\begin{eqnarray}
\tilde{\Phi}_0(z) \:=\: \tilde{\Phi}(x_0(z),\, y_0(z),\, z,\, t_0(z)),
\hspace{1.0cm}
\tilde{\pi}_{\vec T}^0 \:=\: \sqrt{ {{(E_0 - e \tilde{\Phi}_0)^2} \over
c^2} 
- m_0^2c^2}
\nonumber
\end{eqnarray}
\begin{eqnarray}
\left({{\partial \tilde{\Phi}} \over {\partial t}} \right)^0
\;=\;\left.{{\partial \tilde{\Phi}} \over {\partial t}}
\right|_{\stackrel{x = x_0, y = y_0,}{z = z, t = t_0}},
\hspace{1.0cm}
\left({{\partial \tilde{A}_z} \over {\partial t}} \right)^0
\;=\;\left.{{\partial \tilde{A}_z} \over {\partial t}}
\right|_{\stackrel{x = x_0, y = y_0,}{z = z, t = t_0}}
\nonumber
\end{eqnarray}
The two remaining equations
\begin{eqnarray}
{d P_x \over d z} \;=\;
-{{\partial \tilde{H}_{orbt}} \over \partial {x}},
\hspace{1.0cm}
{d P_y \over d z} \;=\;
-{{\partial \tilde{H}_{orbt}} \over \partial {y}}
\nonumber
\end{eqnarray}
give us two additional conditions 
which have to be satisfied too
\begin{eqnarray}
{e \over c}\, {d \tilde{A}_{\vec{N}}^0 \over d z} \;=\;
h \tilde{\pi}_{\vec T}^0 \:+\: {e \over c} \left(
{\partial \tilde{A}_z \over \partial x} \right)^0 \:-\:
{e \over \tilde{\pi}_{\vec T}^0} \cdot {{E_0 - e \tilde{\Phi}_0}
\over c^2}
\left( {\partial \tilde{\Phi} \over \partial x} \right)^0
\label{edco1}
\end{eqnarray}
\begin{eqnarray}
{e \over c} {d \tilde{A}_{\vec{B}}^0 \over d z} \;=\;
\alpha \tilde{\pi}_{\vec T}^0 \:+\: {e \over c} \left(
{\partial \tilde{A}_z \over \partial y} \right)^0 \:-\:
{e \over \tilde{\pi}_{\vec T}^0} 
{{E_0 - e \tilde{\Phi}_0} \over c^2}
\left( {\partial \tilde{\Phi} \over \partial y} \right)^0
\label{edco2}
\end{eqnarray}
Here
\begin{eqnarray}
\left({{\partial \tilde{\Phi}} \over {\partial x}} \right)^0
=\left.{{\partial \tilde{\Phi}} \over {\partial x}}
\right|_{\stackrel{x = x_0, y = y_0,}{z = z, t = t_0}},
\hspace{1.3cm}
\left({{\partial \tilde{\Phi}} \over {\partial y}} \right)^0
=\left.{{\partial \tilde{\Phi}} \over {\partial y}}
\right|_{\stackrel{x = x_0, y = y_0,}{z = z, t = t_0}}
\nonumber
\end{eqnarray}
\begin{eqnarray}
\left({{\partial \tilde{A}_z} \over {\partial x}} \right)^0
=\left.{{\partial \tilde{A}_z} \over {\partial x}}
\right|_{\stackrel{x = x_0, y = y_0,}{z = z, t = t_0}},
\hspace{1.0cm}
\left({{\partial \tilde{A}_z} \over {\partial y}} \right)^0
=\left.{{\partial \tilde{A}_z} \over {\partial y}}
\right|_{\stackrel{x = x_0, y = y_0,}{z = z, t = t_0}}
\nonumber
\end{eqnarray}
So we find that the necessary and sufficient conditions for the functions
(\ref{e1}) to be a reference particle with respect to a given closed
design
orbit and electromagnetic field are the conditions 
(\ref{e2})-(\ref{edco2}). These conditions 
(especially (\ref{edco1}) and (\ref{edco2})) looks
rather complicated, but in practice they can often  be
satisfied with the help of a very simple model of the closed design
orbit (a sequence of line segments and arcs) and a corresponding 
piecewise constant model of the electromagnetic fields.
We will not discuss in this paper the conditions of
stability of this trajectory (and different definitions
of stability) because this is one of the major problems 
of accelerator physics and it deserves special detailed 
consideration.

The reference particle is a trajectory, but for this trajectory
just
as for some real particle, we define quantities such as
Lorentz factor $\gamma_0$, relative velocity $\beta_0$, and 
modulus of the velocity and kinetic momentum vectors, which we
will denote by  $v_0$ and $\pi_0$ respectively: 

\begin{eqnarray}
\gamma_0 \;=\; {{E_0 - e \tilde{\Phi}_0} \over {m_0 c^2}},
\hspace{1.0cm}
\beta_0 \;=\; \sqrt{1 - \frac{1}{\gamma_0^2}},
\hspace{1.0cm}
v_0 \;=\; \beta_0 c
\nonumber
\end{eqnarray}
\begin{eqnarray}
\pi_0 = m_0 \gamma_0 v_0 =
m_0 \gamma_0 \beta_0 c =
m_0 \gamma_0 c \sqrt{1 - {1 \over {\gamma_0^2}} } =
\sqrt{{{(E_0 - e \tilde{\Phi}_0})^2 \over {c^2}}  
- m_0^2 c^2} = \tilde{\pi}_{\vec T}^0
\nonumber
\end{eqnarray}
In these new notations the equations (\ref{e3_1}) and
(\ref{e4}) can be rewritten as
\begin{eqnarray}
{d E_0 \over d z} \;=\; 
-{e \over c} 
\left( {{\partial \tilde{A}_z} \over {\partial t}} \right)^0 \:+\:
{e \over \beta_0 c}
\left({{\partial \tilde{\Phi}} \over {\partial t}} \right)^0,
\hspace{1.2cm}
{d t_0 \over d z}  \;=\;  {1 \over \beta_0 c} 
\;=\;  {1 \over v_0 }
\label{e4_1}
\end{eqnarray}

Note that sometimes it is more convenient to introduce
the reference particle axiomatically just as an arbitrary
set of six functions
(\ref{e1}) satisfying only the conditions
\begin{eqnarray}
x_0 (z) \;\equiv\; 0,
\hspace{1.0cm}
y_0 (z) \;\equiv\; 0,
\hspace{1.0cm}
{d t_0 \over d z} \; \neq\; 0 
\nonumber
\end{eqnarray}
In this case
we define the value of $\:v_0\:$ using the second 
equation in (\ref{e4_1}), and after this $\:\beta_0\,$, 
$\:\gamma_0\:$ and $\:\pi_0\:$ will be defined as follows
\begin{eqnarray}
\beta_0 \;=\; \frac{v_0}{c},
\hspace{1.0cm}
\gamma_0 \;=\; {1 \over \sqrt{1 - \beta_0^2}},
\hspace{1.0cm}
\pi_0 \;=\; m_0 \,\gamma_0\, \beta_0\, c
\nonumber
\end{eqnarray}
In order, in the following, to have a uniform treatment for
both ways of introducing the reference particle,
for the axiomatic case we will define also 
$\:\tilde{A}_{\vec{N}}^0\,$, $\:\tilde{A}_{\vec{B}}^0\:$ and
$\:\tilde{\Phi}_0\:$ as
\begin{eqnarray}
\tilde{A}_{\vec{N}}^0 \;=\; \frac{c}{e} P_x^0, 
\hspace{1.0cm}
\tilde{A}_{\vec{B}}^0 \;=\; \frac{c}{e} P_y^0, 
\hspace{1.0cm}
\tilde{\Phi}_0 \;=\; \frac{E_0 - m_0 \gamma_0 c^2}{e}
\nonumber
\end{eqnarray}

\subsection{Deviations of Orbital Coordinates from 
the Solution for the Reference Particle}

\hspace*{0.5cm}
Since the reference particle is a reflection 
of our understanding (or of our desire) of how 
the accelerator should behave 
it is natural to introduce as new coordinates
the deviations of orbit variables
from the trajectory of the reference particle. 
Let us define new coordinates ($x, y, \sigma$)
and new momenta ($P_x, P_y, \varepsilon$) as
\begin{eqnarray}
\begin{array}{llllll}
x_{new} &=& x_{old}, 
&
\hspace*{1.0cm}
P_x^{new} &=& P_x^{old} - P_x^0 \;=\; P_x^{old} - 
{e \over c} \tilde{A}_{\vec{N}}^0\\
\\
y_{new} &=& y_{old}, 
&
\hspace*{1.0cm}
P_y^{new} &=& P_y^{old} - P_y^0 \;=\; P_y^{old} -
{e \over c} \tilde{A}_{\vec{B}}^0\\
\\
\sigma &=& - \left( t - t_0 \right),
&
\hspace*{1.0cm}
\varepsilon &=& E \:-\: E_0
\end{array}
\nonumber
\end{eqnarray}
The orbit Hamiltonian in new variables 
has the form\footnote{
Note, that before this transformation $E$ was a generalized 
coordinate and $t$ was a generalized momentum, but now 
we choose
$\sigma = -(t - t_0)$ as the new generalized coordinate and 
$\varepsilon = E - E_0$ becomes a generalized momentum.}
\begin{eqnarray}
\hat{H}_{orbt}  \;=\;  
- \mbox{\ae}\, x \,\pi_{\vec B}
\:+\: \mbox{\ae}\, y \,\pi_{\vec N}
\:-\: (1 + h x + \alpha y)\, \pi_{\vec T} \: -\: 
{e \over c} A_z \:+
\nonumber
\end{eqnarray}
\begin{eqnarray}
+\: \varepsilon \,{{d t_0} \over {d z}} 
\:+\:\sigma\, {d E_0 \over d z} 
\:+\: \frac{e}{c}\left( x\, {d \tilde{A}_{\vec{N}}^0 \over d z} 
\:+\: y\, {d \tilde{A}_{\vec{B}}^0 \over d z}\right)
\nonumber
\end{eqnarray}
where now
\begin{eqnarray}
\vec{\pi} \;=\; \left(
\pi_{\vec N} , \;
\pi_{\vec B} , \;
\pi_{\vec T} 
\right)
\nonumber
\end{eqnarray}
\begin{eqnarray}
\pi_{\vec N} \;=\;
P_x \:-\: {e \over c}\, \Delta A_{\vec{N}}, 
\hspace{1.0cm}
\pi_{\vec B} \;=\;
P_y \:-\: {e \over c}\, \Delta A_{\vec{B}} 
\nonumber
\end{eqnarray}
\begin{eqnarray}
\pi_{\vec T}  \;=\;  \left(
m_0^2 c^2 \left(\gamma^2 -1\right)  
\:-\: \pi_{\vec N}^2
\:-\: \pi_{\vec B}^2
\right)^{1 / 2}
\nonumber
\end{eqnarray}
\begin{eqnarray}
\Delta A_{\vec{N}} \;=\; A_{\vec{N}} \:-\: \tilde{A}_{\vec{N}}^0,
\hspace{0.6cm}
\Delta A_{\vec{B}} \;=\; A_{\vec{B}} \:-\:  \tilde{A}_{\vec{B}}^0,
\hspace{0.6cm}
\Delta \Phi \;=\; \Phi \:-\: \tilde{\Phi}_0
\nonumber
\end{eqnarray}
\begin{eqnarray}
\gamma \;=\; \gamma_0 \:+\: \frac{\varepsilon - e \Delta \Phi} 
{m_0 c^2}
\nonumber
\end{eqnarray}
and the spin part keeps the same form as in subsection 4.6 with
the component of the vector $\;\vec{\pi}\;$ and $\;\gamma\;$
defined above.

The variable $\sigma$ describes the difference in arrival times 
at the position $z$ between a given particle and the reference particle, 
and the quantity $\varepsilon$ is their energy deviation.
It seems to be more suitable to use another pair of canonical variables 
\begin{eqnarray}
\sigma_{new} \;=\; \beta_0\, c \,\sigma_{old} 
\hspace{1.0cm}
\mbox{and}
\hspace{1.0cm}
\varepsilon_{new} \; =\; {\varepsilon_{old} \over {\beta_0 c}} 
\nonumber
\end{eqnarray}
where the new $\sigma$ describes the longitudinal 
separation of the particle from the centre of the bunch.
This canonical transformation does not concern any another 
variables and the new orbital part of the Hamiltonian becomes
\begin{eqnarray}
\hat{H}_{orbt}  \;=\;  
- \mbox{\ae}\, x\, \pi_{\vec{B}} 
\:+\: \mbox{\ae} \,y \,\pi_{\vec{N}} 
\:-\: (1 + h x + \alpha y)\, \pi_{\vec{T}}
\:-\: {e \over c} A_z  \:+  
\nonumber
\end{eqnarray}
\begin{eqnarray}
+\: \varepsilon 
\:+\: {\sigma \over {\beta_0 c}}\, {d E_0 \over d z} 
\:+\: {{\sigma \varepsilon} \over {\pi_0 \gamma_0^2}}
{{d \pi_0} \over {d z}} 
\:+\: {e \over c} \left( 
  x {d \tilde{A}_{\vec{N}}^0 \over d z} 
\:+\: y {d \tilde{A}_{\vec{B}}^0 \over d z} \right)
\nonumber
\end{eqnarray}
whereas the spin part $\:\hat{H}_{spin}\:$ remains the same
if we take into account the new expression for $\gamma$
\begin{eqnarray}
\gamma \;=\; \gamma_0 \:+\:
\frac{\beta_0}{m_0 c}
\left( \varepsilon - {e \over {\beta_0 c}} \Delta \Phi \right)
\nonumber
\end{eqnarray}

\subsection{Scaling of the Orbital Variables}

\hspace*{0.5cm}
The canonical scaling of the orbital variables
which we use in this paper is
\begin{eqnarray}
x_{new} \;=\; \sqrt{\pi_0} x_{old},
\hspace{1.0cm}
P_x^{new} \;=\; {{P_x^{old}} \over {\sqrt{\pi_0}}}
\nonumber
\end{eqnarray}
\begin{eqnarray}
y_{new} \;=\; \sqrt{\pi_0} y_{old},
\hspace{1.0cm}
P_y^{new} \;=\; {{P_y^{old}} \over {\sqrt{\pi_0}}}
\nonumber
\end{eqnarray}
\begin{eqnarray}
\sigma_{new} \;=\; \sqrt{\pi_0} \sigma_{old},
\hspace{1.3cm}
\varepsilon_{new}\; =\; {{\varepsilon_{old}} \over {\sqrt{\pi_0}}}
\nonumber
\end{eqnarray}

This scaling is different from that usually used  in accelerator
physics (see Appendix E) and is applicable for both
storage and acceleration regimes.

\subsection{The General Form of the Spin-Orbit Hamiltonian 
in New Coordinates
up to First Order with Respect to Spin Variables}

\hspace*{0.5cm}
Before writing out the final form of the spin-orbit
Hamiltonian we would like to note that various
authors use various coordinate systems for the treatment 
of fully coupled transverse and longitudinal motion 
(synchro-betatron motion). Our variables are closest
in their physical meaning to the coordinates used  
in \cite{mais, barber}.  

In the variables introduced above  the spin-orbit Hamiltonian
takes the following final (at least for this paper) form
\begin{eqnarray}
\hat{H} \;=\;  
\hat{H}_{orbt} \;+\;  
\hat{H}_{spin}   
\nonumber 
\end{eqnarray} 
\begin{eqnarray}
\hat{H}_{orbt} \;=\;  
- \mbox{\ae} x \left(\frac{\pi_{\vec B}}{\sqrt{\pi}_0}\right)
\:+\: \mbox{\ae} y \left(\frac{\pi_{\vec N}}{\sqrt{\pi}_0}\right)
\:-\: (\sqrt{\pi_0} + h x + \alpha y) 
\left(\frac{\pi_{\vec T}}{\sqrt{\pi_0}}\right) 
\:-\: {e \over c} A_z \:+  
\nonumber 
\end{eqnarray}
\begin{eqnarray}
+\: \sqrt{\pi_0} \,\varepsilon  
\:+\: {\sigma \over {\sqrt{\pi_0}\, \beta_0\, c}}\, {d E_0 \over d z}
\:+\: {e \over {\sqrt{\pi_0}\, c}}  \left(
  x \,\frac {d \tilde{A}_{\vec{N}}^0} {d z} 
\:+\: y\, \frac {d \tilde{A}_{\vec{B}}^0} {d z} 
\right) \:+
\nonumber 
\end{eqnarray}
\begin{eqnarray}
+\: \frac{1}{\pi_0}
\left(
\frac{\sigma \,\varepsilon}{\gamma_0^2}
\:+\: \frac{x\, P_x \:+\: y\, P_y \:+\: \sigma\, \varepsilon}{2}
\right)
\frac{d \pi_0}{d z}
\nonumber
\end{eqnarray}
\begin{eqnarray}
\hat{H}_{spin} \;=\; -\alpha\, s_x \:+\: h\, s_y \:-\: 
\mbox{\ae}\, s_z \:+\:
(\sqrt{\pi_0} \:+\: h\,x \:+\:\alpha\, y) \, 
\frac{m_0\, \gamma}{\sqrt{\pi_0}\, \pi_{\vec T}}
\, \vec{W} \cdot \vec{s}
\nonumber
\end{eqnarray}
where
\begin{eqnarray}
\vec{W} = -\frac{e}{m_0 \gamma c} \left(
\left( 1 + \gamma G \right) \vec{{\cal B}}
- \frac{G \left(\,\vec{\pi} \cdot \vec{\cal B}\,\right) \vec{\pi}}  
{m_0^2 c^2 (1 + \gamma)} 
- {1 \over {m_0 c}} \left( G + {1 \over {1 + \gamma}} \right)
\left[\vec{\pi} \times \vec{\cal E}\,\right]
\right)
\nonumber
\end{eqnarray}
\begin{eqnarray}
\vec{\pi} \;=\; \left(
\pi_{\vec N} , \;
\pi_{\vec B} , \;
\pi_{\vec T} 
\right)
\nonumber
\end{eqnarray}
\begin{eqnarray}
\pi_{\vec{N}}  \;=\; \sqrt{\pi_0} \left(
P_x\: -\: {e \over {\sqrt{\pi_0}}\,c}\, \Delta A_{\vec{N}} \right),
\hspace{1.0cm}
\pi_{\vec{B}} \;=\; \sqrt{\pi_0} \left( 
P_y \:-\: {e \over {\sqrt{\pi_0}}\,c}\, \Delta A_{\vec{B}} \right)
\nonumber
\end{eqnarray}
\begin{eqnarray}
\pi_{\vec{T}}  \;=\; \sqrt{ m_0^2 c^2 (\gamma^2 - 1)
\:-\: \pi_{\vec{N}}^2 \:-\: \pi_{\vec B}^2}
\nonumber
\end{eqnarray}
\begin{eqnarray}
\gamma \;=\; 
\gamma_0 \:+\: \frac{\sqrt{\pi_0}\, \beta_0}{m_0\, c}
\left( \varepsilon \:-\: {e \over {\sqrt{\pi_0}\, \beta_0\, c}}\,
\Delta \Phi \right)
\nonumber
\end{eqnarray}
\begin{eqnarray}
\gamma_0 \;=\; \frac{E_0 \:-\: e\, \tilde{\Phi}_0} {m_0\, c^2},
\hspace{1.0cm}
\pi_0 \;=\; m_0 \,\gamma_0\, \beta_0\, c
\nonumber
\end{eqnarray}
Remembering that the electric and magnetic fields, 
and the vector and scalar potentials are supposed
to be defined in the curvilinear coordinate system 
connected with the closed design orbit 
(see subsection 4.6 and appendix A for more details)
and using the conversion formulae
\begin{eqnarray}
\begin{array}{llllll}
x' &=& x\, / \sqrt{\pi_0}, 
&\hspace{1.0cm}
P'_x &=& \sqrt{\pi_0}\,P_x \:+\: (e / c)\, \tilde{A}_{\vec N}^0\\
\\
y' &=& y \,/ \sqrt{\pi_0}, 
&\hspace{1.0cm}
P'_y &=& \sqrt{\pi_0}\,P_y \:+\: (e / c)\, \tilde{A}_{\vec B}^0\\
\\
t' &=& t_0 \:-\: \sigma\, / \left(\sqrt{\pi_0}\, \beta_0\, c\right),
&\hspace{1.0cm}
E' &=& E_0 \:+\: \left(\sqrt{\pi_0}\, \beta_0\, c\right)\, \varepsilon\\
\\
z' &=& z
\end{array}
\nonumber 
\end{eqnarray}
from these variables, denoted here as
$\;z',\, x',\, P'_x,\, y',\, P'_y,\, E',\, t'\,$,
to our final canonical coordinates
$\;z,\, x,\, P_x,\, y,\, P_y,\, \sigma,\, \varepsilon\;$
we obtain the rule for the substitution of the arguments  
of the mentioned above functions
\begin{eqnarray}
F(t',\, x',\, y',\, z' )\; \rightarrow \;
F \left( t_0-\frac{\sigma}{\sqrt{\pi_0} \beta_0 c},
\hspace{0.2cm}
\frac{x}{\sqrt{\pi_0}}, 
\hspace{0.2cm}
\frac{y}{\sqrt{\pi_0}}, 
\hspace{0.2cm} z
\right)  
\nonumber 
\end{eqnarray}

In our final variables the relations between 
fields and potentials become:

{\bf The magnetic field:} 
\begin{eqnarray}
{\cal B}_{\vec{N}}\; =\; 
\frac{\sqrt{\pi_0}}{\sqrt{\pi_0}\: +\: h\, x \:+\: \alpha\,y}
\cdot \left(
\sqrt{\pi_0}\, \frac{\partial A_z}{\partial y} 
\:-\: \frac{\partial A_{\vec{B}}}{\partial z} 
\:-\: \frac{\mbox{\ae}\, y}{\sqrt{\pi_0}}\, {\cal B}_{\vec{T}}
\right) 
\nonumber
\end{eqnarray}
\begin{eqnarray}
{\cal B}_{\vec{B}} \;=\; 
\frac{\sqrt{\pi_0}}{\sqrt{\pi_0}\: +\: h\, x \:+\: \alpha \,y}\,
\cdot \left(
  \frac{\partial A_{\vec{N}}}{\partial z} 
\:-\: \sqrt{\pi_0}\, \frac{\partial A_z}{\partial x} 
\:+\: \frac{\mbox{\ae}\, x}{\sqrt{\pi_0}}\, {\cal B}_{\vec{T}}
\right) 
\nonumber
\end{eqnarray}
\begin{eqnarray}
{\cal B}_{\vec{T}} \;=\; \sqrt{\pi_0}\,
\left(
  \frac{\partial A_{\vec{B}}}{\partial x} 
\:-\: \frac{\partial A_{\vec{N}}}{\partial y} 
\right) 
\nonumber
\end{eqnarray}
where
\begin{eqnarray}
A_z \;=\; \frac{1}{\sqrt{\pi_0}}\, \left(
\left(\sqrt{\pi_0}\: +\: h\, x \:+\: \alpha\, y\right)\: A_{\vec{T}}
\:+\: \mbox{\ae}\: (x\, A_{\vec{B}}\: -\: y\, A_{\vec{N}}) 
\right)
\nonumber
\end{eqnarray}

{\bf The electric field:} 
\begin{eqnarray}
{\cal E}_{\vec{N}}\; =\; \sqrt{\pi_0}\, 
\left(\beta_0\, \frac{\partial A_{\vec{N}}}{\partial \sigma}
\:-\:
\frac{\partial \Phi}{\partial x} \right)
\nonumber
\end{eqnarray}
\begin{eqnarray}
{\cal E}_{\vec{B}} \;=\; \sqrt{\pi_0}\, 
\left(\beta_0 \,\frac{\partial A_{\vec{B}}}{\partial \sigma}
\:-\:
\frac{\partial \Phi}{\partial y} \right)
\nonumber
\end{eqnarray}
\begin{eqnarray}
{\cal E}_{\vec{T}}\; =\; \sqrt{\pi_0}\, \left(
\beta_0\, \frac{\partial A_{\vec{T}}}{\partial \sigma} \:-\:
\frac{1}{\sqrt{\pi_0}\: +\: h\, x \:+\: \alpha\, y}
\left(
\frac{\partial \Phi}{\partial z}
\:+\: \mbox{\ae} \left(
  y\, \frac{\partial \Phi}{\partial x}
\:-\: x\, \frac{\partial \Phi}{\partial y}
\right) \right)\right)
\nonumber
\end{eqnarray}

To have the spin part of the Hamiltonian in more
detailed form, let us introduce a vector $\;\vec{\Omega}\;$
by means of the equality
$\;\hat{H}_{spin} \:=\: \vec{\Omega} \cdot \vec{s}\;$
and write out its components $\:\Omega_x\,$, $\:\Omega_y\:$
and $\:\Omega_z\:$:

\begin{eqnarray}
\Omega_x  \;=\;  -\alpha \:+\: 
(\sqrt{\pi_0} \:+\: h\,x \:+\: \alpha\, y)\,
\frac{e}{\sqrt{\pi_0}\, \pi_{\vec T}\, c}
\, \cdot
\nonumber 
\end{eqnarray}
\begin{eqnarray}
\cdot \, \left[
- (1 \:+\: \gamma \,G)\: {\cal B}_{\vec{N}}
\:+\: \frac{G 
\left(
  \pi_{\vec N}\, {\cal B}_{\vec{N}}
\:+\: \pi_{\vec B}\, {\cal B}_{\vec{B}}
\:+\: \pi_{\vec T}\, {\cal B}_{\vec{T}} 
\right) 
\pi_{\vec N}}{m_0^2\, c^2\, (1\: +\: \gamma)} \;+ 
\right.
\nonumber 
\end{eqnarray}
\begin{eqnarray}
\left.
+\; \frac{1}{m_0\, c}\, 
\left( G \:+\: \frac{1}{1\: +\: \gamma} \right)
\left(
  \pi_{\vec B}\, {\cal E}_{\vec{T}} 
\:-\: \pi_{\vec T}\, {\cal E}_{\vec{B}}
\right)
\right]
\nonumber
\end{eqnarray}

\hspace{0.3cm}

\begin{eqnarray}
\Omega_y \; = \; h \:+\: 
(\sqrt{\pi_0} \:+\: h\,x \:+\: \alpha\, y)\,
\frac{e}{\sqrt{\pi_0}\, \pi_{\vec T}\, c}
\, \cdot
\nonumber 
\end{eqnarray}
\begin{eqnarray}
\cdot \, \left[
- (1\: +\: \gamma\, G)\, {\cal B}_{\vec{B}}
\:+\: \frac{G 
\left(
  \pi_{\vec N}\, {\cal B}_{\vec{N}}
\:+\: \pi_{\vec B}\, {\cal B}_{\vec{B}}
\:+\: \pi_{\vec T}\, {\cal B}_{\vec{T}} 
\right) 
\pi_{\vec B}}{m_0^2 c^2 (1 + \gamma)} \;+ 
\right.
\nonumber 
\end{eqnarray}
\begin{eqnarray}
\left.
+\; \frac{1}{m_0\, c}\, 
\left( G\: +\: \frac{1}{1\: +\: \gamma} \right)
\left(
  \pi_{\vec T}\, {\cal E}_{\vec{N}} 
\:-\: \pi_{\vec N}\, {\cal E}_{\vec{T}}
\right)
\right]
\nonumber
\end{eqnarray}

\hspace{0.3cm}

\begin{eqnarray}
\Omega_z  \;=\;  -\mbox{\ae} \:+\: 
(\sqrt{\pi_0} \:+\: h\,x \:+\: \alpha\, y)\,
\frac{e}{\sqrt{\pi_0}\, \pi_{\vec T}\, c}
\, \cdot
\nonumber 
\end{eqnarray}
\begin{eqnarray}
\cdot \, \left[
- (1 \:+\: \gamma\, G)\, {\cal B}_{\vec{T}}
\:+\: \frac{G 
\left(
  \pi_{\vec N}\, {\cal B}_{\vec{N}}
\:+\: \pi_{\vec B}\, {\cal B}_{\vec{B}}
\:+\: \pi_{\vec T}\, {\cal B}_{\vec{T}} 
\right)\, 
\pi_{\vec T}}{m_0^2\, c^2\, (1\: +\: \gamma)} \;+ 
\right.
\nonumber 
\end{eqnarray}
\begin{eqnarray}
\left.
+\; \frac{1}{m_0\, c} 
\left( G \:+\: \frac{1}{1 \:+\: \gamma} \right)
\left(
  \pi_{\vec N}\, {\cal E}_{\vec{B}} 
\:-\: \pi_{\vec B}\, {\cal E}_{\vec{N}}
\right)
\right]
\nonumber
\end{eqnarray}

\section{Acknowledgments}

\hspace*{0.5cm}
The authors are very grateful to the DESY MPY group for hospitality
and support. 
We wish to thank Desmond P. Barber for stimulating discussions and for
continued encouragement. 
We thank A.M. Kondratenko, Ya.S. Derbenev and H. Mais
for many interesting and useful discussions. 
The careful reading of the manuscript by D.P. Barber is gratefully
acknowledged.

\appendix

\section{How to Transform the Equations 
of the Electromagnetic Field to Curvilinear 
Coordinates Associated with 
the Closed Design Orbit}

\hspace*{0.5cm}
The purpose of this appendix is to point out the form which differential
operators entering in the Maxwell system of equations for the
electromagnetic field 
will have in the  curvilinear coordinates considered
(they could be equations for the vector and scalar
potentials, or direct equations for the electric $\vec{\cal E}$ and
magnetic $\vec{\cal B}$ fields)

In the coordinate system used every vector $\vec{A}$ can be uniquely
represented in the form
\begin{eqnarray}
\vec{A} \;=\; A_{\vec{N}} \cdot \vec{N} \:+\: 
A_{\vec{B}} \cdot \vec{B} \:+\:
A_{\vec{T}} \cdot \vec{T}  
\nonumber
\end{eqnarray}
(at least in a sufficiently small neighbourhood of closed design orbit).

As {\bf physical Components} of $\;\vec{A}\;$ we understand the values
\begin{eqnarray}
A_{\vec{N}} \:=\: A_{\vec{N}}(t,\, x,\, y,\, z), 
\hspace{0.5cm}
A_{\vec{B}}\: =\:A_{\vec{B}}(t,\, x,\, y,\, z), 
\hspace{0.5cm} 
A_{\vec{T}} \:=\: A_{\vec{T}}(t,\, x,\, y, \,z)
\nonumber
\end{eqnarray}
Introduce the vectors
\begin{eqnarray}
\vec{T}_z \;=\; \left(1 + h x + \alpha y \right) \vec{T} \:+\: 
\mbox{\ae}
 \left(x \vec{B} \:-\:
y \vec{N} \right)  
\nonumber
\end{eqnarray}
\begin{eqnarray}
\vec{N}_z \;=\; \left(1 + h x + \alpha y \right) \vec{N} \:+\: 
\mbox{\ae} y \vec{T}
\nonumber
\end{eqnarray}
\begin{eqnarray}
\vec{B}_z \;=\; \left(1 + h x + \alpha y \right) \vec{B} \:-\: 
\mbox{\ae} x \vec{T}
\nonumber
\end{eqnarray}
and define $\:A_z\:$, $\:A_x\:$, and $\:A_y\:$ as projections of 
the vector
$\:\vec{A}\:$ on
the vectors $\:\vec{T}_z\:$, $\:\vec{N}_z\:$ and $\:\vec{B}_z\:$
respectively
\begin{eqnarray}
A_{z} \;=\; \vec{A} \cdot \vec{T}_z\,, 
\hspace{1.0cm} 
A_{x} \;=\; \vec{A}\cdot \vec{N}_z\,, 
\hspace{1.0cm} 
A_{y} \;=\; \vec{A} \cdot \vec{B}_z  
\nonumber
\end{eqnarray}
Using the quantities introduced above we get the 
following formulae:

{\bf The Gradient of a Scalar Function:}

\begin{eqnarray}
\mbox{grad}\: \phi \;=\; \nabla \phi \;=\; 
\frac{1}{1 + h x + \alpha y}
\left[
\frac{\partial \phi}{\partial z} \cdot \vec{T} \:+\: 
\frac{\partial \phi}{\partial x} \cdot \vec{N}_z \:+\:
\frac{\partial \phi}{\partial y} \cdot \vec{B}_z \right]
\nonumber
\end{eqnarray}

{\bf The Laplacian of a Scalar Function:}

\begin{eqnarray}
\Delta \phi = \nabla^2 \phi = \frac{1}{1 + hx + \alpha y} 
\left[\frac{\partial}{\partial z} 
\left(\nabla \phi \cdot \vec{T} \right) +
\frac{\partial }{\partial x} 
\left(\nabla \phi \cdot \vec{N}_z \right) +
\frac{\partial}{ \partial y} 
\left(\nabla \phi \cdot \vec{B}_z \right) \right]
\nonumber
\end{eqnarray}

{\bf The Divergence of a Vector Field:}

\begin{eqnarray}
\mbox{div} \vec{A} \;=\; \nabla \cdot \vec{A} \;=\; 
\frac{1}{1 + h x + \alpha y}
\left[ \frac{\partial A_{\vec{T}}}{\partial z} \:+\: 
\frac{\partial A_x}{\partial x} \:+\:
\frac{\partial A_y}{\partial y}\right] 
\nonumber
\end{eqnarray}

{\bf The Curl of a Vector Field:}

\begin{eqnarray}
\mbox{curl} \vec{A} \;=\; \nabla \times \vec{A} \;=  
\nonumber
\end{eqnarray}

\begin{eqnarray}
=\;
\frac{1}{1 + h x + \alpha y} 
\left[ 
\left(\frac{\partial A_{\vec{B}}}{\partial x} \:-\:
\frac{\partial A_{\vec{N}}}{\partial y} \right) \vec{T}_z \:+\:
\left( \frac{\partial A_z}{\partial y} \:-\:
\frac{\partial A_{\vec{B}}}{\partial z} \right) \vec{N} \:+
\right.
\nonumber
\end{eqnarray}

\begin{eqnarray}
+\:
\left. \left( \frac{\partial A_{\vec{N}}}{\partial z} \:-\:
\frac{\partial A_z}{\partial x} \right) \vec{B} \right]  
\nonumber
\end{eqnarray}

{\bf The Laplacian of a Vector Field:}

The Laplacian of a vector field, if it is needed, can be expressed
using the above formulae using the equality

\begin{eqnarray}
\Delta \vec{A} \;=\; \nabla (\nabla \cdot \vec{A}) \;-\; 
\nabla \times (\nabla \times \vec{A})  
\nonumber
\end{eqnarray}

\section{Simple Canonical Coordinates 
for the Periodic Solution}

\hspace*{0.5cm}
Let the differential equations of spin-orbit motion
\begin{eqnarray}
{\frac{{d \vec{z}} }{{d \tau}}} \;=\; 
\{\vec{z},\:H(\tau, \,\vec{z}\,)\}
\nonumber
\end{eqnarray}
$T$-periodic in $\tau$ and with possible nonlinear dependence 
of the Hamiltonian function $H(\tau, \,\vec{z}\,)$ 
on the variables $\vec{s}$ 
have the $T$-periodic solution 
$\;\vec{z}_{*}(\tau)$
\begin{eqnarray}
\vec{x}(\tau) \;=\; \vec{x}_*(\tau), 
\hspace{1.0cm} 
\vec{s}(\tau) \;=\;\vec{s}_*(\tau)  
\label{grid1}
\end{eqnarray}
satisfying the condition
\begin{eqnarray}
|\vec{s}_*(0)| \;\neq\; 0  
\label{qustar}
\end{eqnarray}
Since the parallel displacement 
$\;\vec{z}_{new} \:=\: \vec{z}_{old} \:-\: \vec{z}_*(\tau)\;$ 
which converts the
solution (\ref{grid1}) into the origin $\vec{0}$
is not a canonical transformation with respect to the coupled spin-orbit
Poisson bracket if (\ref{qustar}) holds, in this Appendix we introduce
other simple canonical 
coordinates for the periodic solution (\ref{grid1}).

Introduce the real skewsymmetric matrix
\begin{eqnarray}
C(\tau,\, \vec{z}\,) \;=\; \left(
\begin{array}{rrr}
0 & -\frac{\partial H}{\partial s_3} & \frac{\partial H}{\partial s_2} \\
& & \\
\frac{\partial H}{\partial s_3} & 0 & -\frac{\partial H}{\partial s_1} \\
& & \\
-\frac{\partial H}{\partial s_2} & \frac{\partial H}{\partial s_1} & 0
\end{array}
\right)  
\nonumber
\end{eqnarray}
and write $\;\hat{C}(\tau) \:=\: C(\tau, \,\vec{z}_*(\tau))\,$.

It is easy to check that the unit vector
\begin{eqnarray}
\vec{n}(\tau) \;=\; \frac{1}{|\vec{s}_*(0)|} \:\vec{s}_*(\tau)  
\nonumber
\end{eqnarray}
periodically dependent on $\tau$
satisfies the equation
\begin{eqnarray}
\frac{d \vec{n}}{d \tau} \;=\; \hat{C}(\tau)\, \vec{n}  
\label{grid2}
\end{eqnarray}

Let us assume that we have found two  
unit vectors $\:\vec m(\tau)\:$ and $\:\vec l(\tau)\:$ 
$T$-periodic in $\tau $
which supplement the vector $\:\vec n(\tau)\:$ 
to form an orthogonal coordinate system satisfying the condition
\begin{eqnarray}
\left[\, \vec m(\tau )\times \vec l(\tau )\,\right]
\cdot \vec n(\tau)\;\equiv\; 1
\label{grid3}
\end{eqnarray}
Introduce new coordinates $\:\vec y$, $\:\vec u\:$ with the help of
the canonical transformation
\begin{eqnarray}
\vec x\;=\;\vec x_{*}(\tau )\:+\:\vec y\,,
\hspace{1.0cm}
\vec s\;=\;A(\tau)\,\vec u
\nonumber
\end{eqnarray}
\begin{eqnarray}
A(\tau)\;=\;\left(\vec m(\tau),\:\vec l(\tau),\:\vec n(\tau)
\right)\;\in\; \mbox{SO}(3)
\nonumber
\end{eqnarray}
In the variables $\:\vec y$, $\:\vec u\:$ the Hamiltonian function 
takes the form
\begin{eqnarray}
H_{new}=H_{old}\left(\tau,\:\vec x_{*}(\tau)+\vec y,\:
A(\tau )\vec u\right) +
\vec y\cdot J\frac{d\vec x_{*}}{d\tau }-
\frac 12\vec u\cdot 
\mbox{curl}_{\,\vec u}
\left( A^{\top }\frac{dA}{d\tau}\right)   
\nonumber
\end{eqnarray}
and the periodic solution 
$\;\vec z_{*}(\tau)\;$ is now expressed as follows
\begin{eqnarray}
\vec y_{*}\;=\;\vec 0,
\hspace{1.0cm}
\vec u_{*}\;=\;(0,\:0,\:|\vec s_{*}(0)|)
\nonumber
\end{eqnarray}

Now we wish to discuss the problem of existence and possible freedom of
choice of the  vectors $\:\vec{m}(\tau)\:$ and $\:\vec{l}(\tau)\:$
introduced above.

Taking the derivative with respect to $\tau$ in the identities
\begin{eqnarray}
\vec{m}(\tau)\cdot\vec{n}(\tau) \;=\; 0, 
\hspace{1.0cm} 
\vec{m}(\tau)\cdot\vec{m}(\tau) \;=\; 1  
\nonumber
\end{eqnarray}
we have
\begin{eqnarray}
\left(\frac{d \vec{m}}{d \tau} \:-\: 
\hat{C}(\tau) \vec{m}\right) \cdot \vec{n}
\;=\; 0 
\hspace{1.0cm} 
\mbox{and} 
\hspace{1.0cm} 
\frac{d \vec{m}}{d \tau} \cdot \vec{m} \;=\; 0  
\label{grid4}
\end{eqnarray}
Subtracting 
the identity $\;\hat{C}\vec{m}\cdot\vec{m}\:=\:0\;$ 
from the second of the equalities (\ref{grid4}) 
we find that the vector
\begin{eqnarray}
\frac{d \vec{m}}{d \tau} \;-\; \hat{C}(\tau)\: \vec{m}  
\nonumber
\end{eqnarray}
is orthogonal to the vectors $\:\vec{n}\:$ and $\:\vec{m}\:$ 
for all values of $\tau$,
and hence can be represented in the form
\begin{eqnarray}
\frac{d \vec{m}}{d \tau} \:-\: \hat{C}(\tau)\, \vec{m} 
\;=\; \psi_1 (\tau)\,
\vec{l}
\nonumber
\end{eqnarray}
Similarly we have
\begin{eqnarray}
\frac{d \vec{l}}{d \tau} \:-\: \hat{C}(\tau)\, \vec{l} 
\;=\; \psi_2 (\tau)\,
\vec{m}
\nonumber
\end{eqnarray}
From the condition
\begin{eqnarray}
\frac{d}{d \tau} \left(\,\vec{m} \cdot \vec{l}\,\right) \;=\; 0  
\nonumber
\end{eqnarray}
it follows that 
$\;\psi_2(\tau) \:=\: -\psi_1(\tau) \:\stackrel{{\rm def}}{=}\:
\psi(\tau)\;$
, i.e. the vectors $\:\vec{m}$, $\:\vec{l}\:$ satisfy the system of
differential
equations
\begin{eqnarray}
\frac{d \vec{m}}{d \tau} \:=\: \hat{C}(\tau) \vec{m} - \psi (\tau)
\vec{l},
\hspace{1.0cm}
\frac{d \vec{l}}{d \tau}\: =\: \hat{C}(\tau) \vec{l} + \psi (\tau)
\vec{m}
\label{grid5}
\end{eqnarray}
The solution of (\ref{grid5}), written in complex notation, is
\begin{eqnarray}
\vec{m}(\tau) + i \vec{l}(\tau) \;=\; 
\exp\left(i \int \limits_{0}^{\tau}
{\psi(\eta) d \eta} \right) D(\tau)(\vec{m}(0) + i \vec{l}(0))  
\label{grid6}
\end{eqnarray}
where $\:D(\tau)\:$ is the fundamental matrix solution of (\ref{grid2})
so that 
\begin{eqnarray}
\frac{d D}{d \tau} \;=\; \hat{C}(\tau) \,D, 
\hspace{1.0cm} D(0) \;=\; I
\nonumber
\end{eqnarray}
Remembering that the vectors $\:\vec{m}\:$ and $\:\vec{l}\:$ must be
$T$-periodic in 
$\tau$ we have \\
from (\ref{grid6})
\begin{eqnarray}
D(T) (\vec{m}(0) + i \vec{l}(0)) \;=\; \exp\left(-i \int \limits_{0}^{T}
{\psi(\eta) d \eta} \right) (\vec{m}(0) + i \vec{l}(0))
\label{grid7}
\end{eqnarray}
so that
\begin{eqnarray}
\exp\left(-i \int \limits_{0}^{T} {\psi(\eta) d \eta}\right)  
\nonumber
\end{eqnarray}
is an eigenvalue and $\;\vec{m}(0) + i \vec{l}(0)\;$ is the corresponding
eigenvector
of the matrix $D(T)$. Multiplying (\ref{grid7}) 
onto the vector $\;\vec{m}(0) + i\vec{l}(0)\;$ we get
\begin{eqnarray}
\exp\left(-i \int \limits_{0}^{T} {\psi(\eta) d \eta} \right) 
\;=\; \frac{1}{2}
D(T) (\vec{m}(0) + i \vec{l}(0)) \cdot (\vec{m}(0) + i \vec{l}(0))
\label{grid8}
\end{eqnarray}

This arguments can be reversed 
to show that the
differentiable  unit vectors 
$\:\vec{m}(\tau)\:$ and $\:\vec{l}(\tau)\:$ 
$T$-periodic in $\tau$
will supplement the vector $\:\vec{n}(\tau)\:$ to form an orthogonal basis
satisfying the condition (\ref{grid3}) if and only if they do so for
$\:\tau \,=\, 0\:$ and then satisfy the differential equations
(\ref{grid5})
for some continuous function $\psi(\tau)$ for which  the equality 
(\ref{grid8}) holds.

Applying the transformation $\;\vec s\:=\:A(\tau )\,\vec u\;$ 
described above
to 
the Hamiltonian
\begin{eqnarray}
H\;=\;\vec w(\tau )\cdot \vec s,
\hspace{1.0cm}
\vec w(\tau +2\pi )\;\equiv\; \vec w(\tau)  
\nonumber
\end{eqnarray}
discussed
in subsubsection $5.2.4$ 
we find that in the variables $\:\vec u\:$ this Hamiltonian becomes
\begin{eqnarray}
H\;=\;\psi (\tau )\cdot u_3  
\nonumber
\end{eqnarray}
and is hence a normal form for all possible 
functions $\psi (\tau)$
satisfying (\ref{grid8}), which are independent of $\tau$.

\section{Sketch of the Proof 
of the Factorization Theorem}

\hspace*{0.5cm}
The purpose of this appendix is to prove the factorization theorem A.

The matrix of the quasi-linearization $A_s$ is uniquely defined by the
formula (\ref{n2}). Introduce maps
\begin{eqnarray}
{\cal M}_2 \;=\; :A_s:\:, 
\hspace{0.5cm} 
{\cal M}_k \;=\; {\cal M}_{k-1}\,\exp(:F_k:)\,, 
\hspace{0.5cm} 
k\;=\;3, \ldots, m  
\nonumber
\end{eqnarray}
Using induction in $k$, we will show that the 
functions $\;F_k \:\in\: {\cal H}_s(k)\;$ can be chosen 
in such a way that
the conditions 
\begin{eqnarray}
\left\{
\begin{array}{l}
{\cal M}^{-1}_k \:\vec{X}(\vec{x},\,\vec{s}\,) \;=_{k+1}\; 
\vec{x} \:+\:\vec{X}_k(\vec{x},\, \vec{s}\,) \\
\\
{\cal M}^{-1}_k \:\vec{S}(\vec{x},\,\vec{s}\,) \;=_{k+2}\; 
\vec{s} \:+\:\vec{S}_{k+1}(\vec{x},\,\vec{s}\,)
\end{array}
\right.  
\label{a1}
\end{eqnarray}
are fulfilled
where 
$\;\vec{X}_k \:\in\: {\cal H}_s(k)\;$ and 
$\;\vec{S}_{k+1} \:\in\: {\cal H}_s(k+1)$.

It is easy to see that the equality (\ref{a1}) for $k=m$ gives the proof
of
the theorem.

Obviously, (\ref{a1}) is correct at $k=2$. Applying the operator
\begin{eqnarray}
\exp(-:F_l(\vec{z}):)
\nonumber
\end{eqnarray}
to both parts of (\ref{a1}) for $k=l-1$ 
we obtain that (\ref{a1}) will be correct for $k=l$, 
if the function
$F_l$
satisfies the equations
\begin{eqnarray}
\{F_l,\: \vec{x}\} \;=\; \vec{X}_{l-1}, 
\hspace{1.0cm} 
\{F_l,\: \vec{s}\} \;=\; \vec{S}_{l}  
\label{a2}
\end{eqnarray}
The structural matrix of the spin-orbit Poisson bracket has the form
\begin{eqnarray}
\hat{J}(\vec{z}\,) \;=\; \mbox{diag}\, (J, \,J_s(\vec{s}\,))
\nonumber
\end{eqnarray}
where the $2n \times 2n$ matrix $J$ is the symplectic unit and
\begin{eqnarray}
J_s(\vec{s}\,) \;=\; \left(
\begin{array}{rrr}
0 & s_3 & -s_2 \\
-s_3 & 0 & s_1 \\
s_2 & -s_1 & 0
\end{array}
\right)  
\nonumber
\end{eqnarray}
With the help of the matrix $\hat{J}$ the equations (\ref{a2}) can be
written in the form of the system
\begin{eqnarray}
J \cdot \mbox{grad}_{\,\vec{x}}\, F_l \;=\; - \vec{X}_{l-1}
\label{a3}
\end{eqnarray}
\begin{eqnarray}
J_s(\vec{s}\,) \cdot \mbox{grad}_{\,\vec{s}} \,F_l \;=\; -\vec{S}_l
\label{a4}
\end{eqnarray}
Represent the vector function $\;\vec{X}_{l-1}\;$ in the form of a sum
\begin{eqnarray}
\vec{X}_{l-1} \;=\; \sum \limits_{k=0}^{\left[{\frac{{l-1} }{2}}\right]}
\vec{X}
_{l-2k-1}^{H}
\nonumber
\end{eqnarray}
where $\;\vec{X}_k^H(\vec{x},\,\vec{s}\,)\;$ are homogeneous polynomials
of degree 
$k$ in the variables $\vec{x}$ and the symbol $\left[ m \right]$ denotes
the biggest
integer which is smaller or equal to $m$.

{\bf Lemma 1}: {\sl Any solution $F_l(\vec{z}) \in {\cal H}_s(l)$ 
of the
system (\ref{a3})-(\ref{a4}) is given by the formula
\begin{eqnarray}
F_l \;=\; \vec{x} \cdot 
\sum \limits_{k=0}^{\left[{\frac{{l-1}}{2}}\right]} 
{\frac{1}{{l-2k}}} \,J\, \vec{X}_{l-2k-1}^H \:-\: V(\vec{s}\,)
\nonumber
\end{eqnarray}
where $\;V(\vec{s}\,) \:\in\: {\cal H}_s(l)\;$ satisfies the equation
\begin{eqnarray}
J_s(\vec{s}) \cdot \mbox{grad}_{\,\vec{s}} \:V \;=\;
\vec{S}_l\left(\vec{0},\,\vec{s}\,\right)
\label{a5}
\end{eqnarray}
}

The proof of this and following Lemmas is 
left as an exercise for the interested reader.

Starting from this point, we can forget about the existence of 
the orbital
variables $\vec{x}$, because the proof of Theorem 1 has been reduced to 
finding the solution of equation (\ref{a5}), depending only on
variables $\vec{s}$.

Let ${\bf H}_s(k)$ be the class of homogeneous
polynomials of degree $k$ in the variables $\vec{s}$.
Our current task is to find the solution $V(\vec{s}) \in {\bf H}_s(k)$ of
the equation
\begin{eqnarray}
J_s(\vec{s}) \cdot \mbox{grad} \:V \;=\; \vec{f}(\vec{s}\,)  
\label{a6}
\end{eqnarray}
where the vector function 
$\;\vec{f} \:\in\: {\bf H}_s(k),\;\; k \:\ge\: 1\;$ satisfies the
condition
\begin{eqnarray}
\mbox{grad} \left(\vec{s} \cdot \vec{f}\,\right) \;=\; 
\mbox{div}\left(\vec{f}\,\right) \cdot \vec{s}
\label{a7}
\end{eqnarray}
The condition (\ref{a7}) follows from 
requiring that
the map (\ref{n1}) is canonical.

{\bf Lemma 2}: {\sl If the solution $V(\vec{s}) \in {\bf H}_s(k)$ of the
system (\ref{a6}) exists, then it satisfies the equation
\begin{eqnarray}
k(k+1)\,V \:-\: |\vec{s}\,|^2 \,\Delta V \;=\; R(\vec{s}\,)  
\label{a8}
\end{eqnarray}
where
\begin{eqnarray}
R(\vec{s}\,) \;=\; \vec{s} \cdot \mbox{curl}\, \vec{f}, 
\hspace{1.0cm} 
\Delta \;=\; 
{\frac{{\partial^2}}{{\partial s_1^2}}} \:+\: 
{\frac{{\partial^2}}{{\partial s_2^2}}} \:+\: 
{\frac{{\partial^2}}{{\partial s_3^2}}}
\nonumber
\end{eqnarray}
} 

{\bf Lemma 3}: {\sl If the solution of the equation (\ref{a8}) exists,
then for odd $k$ it is unique and for even $k$ the difference 
between any two
solutions is given by the formula
\begin{eqnarray}
V_1(\vec{s}\,) \:-\: V_2(\vec{s}\,) \;=\; c \cdot |\vec{s}\,|^k
\nonumber
\end{eqnarray}
where $c$ is an arbitrary constant. }

{\bf Lemma 4}: {\sl The equation (\ref{a8}) has a solution if and only if
\begin{eqnarray}
\Delta ^{\left[{\frac{k+1 }{2}} \right]} R(\vec{s}\,) \;=\; 0  
\label{bgfty}
\end{eqnarray}
If (\ref{bgfty}) holds then the function
\begin{eqnarray}
V \;=\; \sum \limits_{j=0}^{\left[{\frac{{k-1} }{2}}\right]} 
\,a_j\,
|\vec{s}\,|^{2j}\:
\Delta^j \:R(\vec{s}\,)  
\label{a9}
\end{eqnarray}
where
\begin{eqnarray}
a_j \;=\; \prod \limits_{i=0}^j {\frac{1 }{{(k-2i+1)(k-2i)}}}  
\nonumber
\end{eqnarray}
satisfies the equation (\ref{a8}).
}

{\bf Lemma 5}: {\sl
\begin{eqnarray}
\Delta^j \left(\vec{s} \cdot \mbox{curl}\, \vec{f}\,\right) 
\;=\; \vec{s}
\cdot \mbox{curl}\,
\left(\Delta^j \vec{f}\,\right)  
\nonumber
\end{eqnarray}
and therefore,
\begin{eqnarray}
\Delta^{\left[{\frac{k+1 }{2}} \right]}
\left(\vec{s} \cdot \mbox{curl} \,\vec{f}\,\right) \;=\;
0
\nonumber
\end{eqnarray}
}

So in Lemmas 2 - 5 we have established that if the solution of the system
(\ref{a6}) exists, then it is unique (up to an additive Casimir
function of the spin-orbit Poisson bracket (\ref{fom2}) $c \cdot |\vec{s}|^k$
at even $k$) and is given by the formula (\ref{a9}). Substituting (\ref{a9})
in (\ref{a6}), we obtain:

{\bf Lemma 6}: {\sl The necessary and sufficient conditions for the
solvability of the equation (\ref{a6})  are that
\begin{eqnarray}
\vec{s} \cdot \vec{f} \;=\; 0 
\hspace{1.0cm} \mbox{and} 
\hspace{1.0cm}
\mbox{div}\,
\vec{f} \;=\; 0  
\label{a10}
\end{eqnarray}
}

The complete solution of the problem is

{\bf Lemma 7}: {\sl If $\vec{f} \in {\bf H}_s(k)$, then conditions (\ref{a10}
) and (\ref{a7}) are equivalent. }

{\bf Remark}: From (\ref{a1}) it follows that the actual precision of the
representation of spin variables 
(functions $\vec{S}(\vec{x},\,\vec{s}\,)$) in
theorems 1 and 2 is higher by one order 
(in the sense of classes 
${\cal O}_s(m)$) than the precision of the 
representation of orbital variables.

\section{How to Integrate Linear 
Equations of Spin Motion in Quadratures if We Know their 
Partial Solution}

\hspace*{0.5cm}
For general linear homogeneous systems of ordinary 
differential equations of order $n$ knowledge 
of one nontrivial solution allows us to lower the 
order of the system by one unit. So, in order to be 
able to find the general solution in quadratures we 
need to know $n-1$ linearly independent solutions.
In this appendix we will show that due to a special
symmetry, third order linear equations of spin motion 
can be integrated completely if we know only one 
partial solution.

Let the vector
\begin{eqnarray}
\vec{n}(\tau) \;=\;
(n_1(\tau),\hspace{0.1cm}n_2(\tau),\hspace{0.1cm}n_3(\tau)),
\hspace{1.0cm} 
|\vec{n}(\tau)| \;=\; 1 
\nonumber
\end{eqnarray}
satisfy the system 
\begin{eqnarray}
\frac{d \vec{s}}{d \tau} \;=\; \vec{W}(\tau) \times \vec{s} \;=\; 
C\left( \vec{W}(\tau)\right) \vec{s}
\label{apqua1}
\end{eqnarray}
Without loss of generality we will consider the 
$\tau$-interval on which one of components
of the vector $\vec n (\tau)$ is not equal to zero.
Let it be $n_2(\tau)$. If on the considered interval 
$n_1(\tau) \neq 0$ 
(or $n_3(\tau) \neq 0$) 
we can begin by making a coordinate transformation 
$\;\vec{s}_{new} \:=\: B \vec{s}_{old}\;$ where the matrix 
\begin{eqnarray}
B \;=\;
\left(
\begin{array}{rrr}
0 & -1& 0 \\
1 &  0& 0 \\
0 &  0& 1
\end{array}
\right) 
\hspace*{1.0cm}
\left(
\hspace{0.05cm}
\mbox{or}
\hspace{0.15cm}
B \;=\;
\left(
\begin{array}{rrr}
1 &  0& 0 \\
0 &  0& 1 \\
0 & -1& 0
\end{array}
\right)
\right) 
\nonumber
\end{eqnarray}
 Introduce the vectors
\begin{eqnarray}
\vec{m}(\tau) \;=\; \frac{1}{\sqrt{1-n_3^2(\tau)}}
\:(n_2(\tau), \hspace{0.1cm}-n_1(\tau), \hspace{0.1cm}0)
\nonumber
\end{eqnarray}
\begin{eqnarray}
\vec{l}(\tau) = \vec{n}(\tau) \times \vec{m}(\tau) =
\frac{1}{\sqrt{1 - n_3^2(\tau)}}
(n_1(\tau)n_3(\tau), \hspace{0.1cm}n_2(\tau)n_3(\tau), 
\hspace{0.1cm}n_3^2(\tau)-1) 
\nonumber
\end{eqnarray}
\begin{eqnarray}
|\vec{m}(\tau)| \;=\; |\vec{l}(\tau)| \;=\; 1 
\nonumber
\end{eqnarray}
and the new variables 
\begin{eqnarray}
\vec u \;=\;
\left(
\begin{array}{r}
\vec{m}(\tau) \\
\vec{l}(\tau) \\
\vec{n}(\tau)
\end{array}
\right) \cdot \vec s
\;\stackrel{{\rm def}}{=}\;
A(\tau) \cdot \vec s
\nonumber
\end{eqnarray}
In the new variables the system (\ref{apqua1})
becomes
\begin{eqnarray}
\frac{d \vec{u}}{d \tau} \;=\; 
\left(\frac{d A}{d \tau} \cdot A^{\top}
\:+\: A \cdot C\left( \vec{W}\right) \cdot A^{\top} \right)
\cdot \vec{u} 
\;\stackrel{{\rm def}}{=}\; C \left(\vec b (\tau) \right)
\cdot \vec{u} 
\label{apqua2}
\end{eqnarray}
where the components of the vector $\vec b (\tau)$ are
\begin{eqnarray}
b_1 \;=\;
\frac{d \vec n}{d \tau}\cdot \vec l 
\:+\: \vec n \cdot C\left( \vec{W}\right) \vec l,
\hspace{1.0cm}
b_2 \;=\; -\frac{d \vec n}{d \tau}\cdot \vec m 
\:-\: \vec n \cdot C\left( \vec{W}\right) \vec m
\nonumber
\end{eqnarray}
\begin{eqnarray}
b_3 \;=\; \frac{d \vec l}{d \tau}\cdot \vec m 
\:+\: \vec l \cdot C\left( \vec{W}\right) \vec m
\nonumber
\end{eqnarray}
Since the vector $\vec n (\tau)$ satisfies the system (\ref{apqua1}),
then $\;b_1 (\tau) \equiv 0\;$ and 
$\;b_2 (\tau) \equiv 0\;$. After some simple manipulations we get
for $\;\omega = b_3\;$
\begin{eqnarray}
\omega(\tau) \;=\; \frac{1}{1 - n_3^2(\tau)}
\left(W_1(\tau)\,n_1(\tau) \:+\: W_2(\tau)\,n_2(\tau)\right)
\label{vecont}
\end{eqnarray}
and hence the solution of the system (\ref{apqua2}) is
\begin{eqnarray}
\vec u (\tau)=
\left(
\begin{array}{rrr}
\cos\Psi(\tau, \tau_0) & -\sin\Psi(\tau, \tau_0) & 0 \\
\sin\Psi(\tau, \tau_0) &  \cos\Psi(\tau, \tau_0) & 0 \\
0 & 0& 1
\end{array}
\right) \cdot \vec u (\tau_0)
\stackrel{{\rm def}}{=}
K(\tau, \tau_0) \cdot \vec u (\tau_0)
\nonumber
\end{eqnarray}
where
\begin{eqnarray}
\Psi (\tau, \tau_0) \;=\; 
\int \limits_{\tau_0}^{\tau} \omega(\varepsilon) \:d \varepsilon
\nonumber
\end{eqnarray}
So the solution of the initial system (\ref{apqua1})
can be expressed now as follows
\begin{eqnarray}
\vec s (\tau) \;=\;
A^{\top}(\tau)\, K(\tau, \tau_0)\,A(\tau_0) \,\vec s (\tau_0)
\nonumber
\end{eqnarray}
where the  matrix $A(\tau)$ has the form
\begin{eqnarray}
A (\tau)\;=\;
\left(
\begin{array}{ccc}
\frac{n_2(\tau)}{\sqrt{1 - n_3^2(\tau)}} & 
-\frac{n_1(\tau)}{\sqrt{1 - n_3^2(\tau)}} & 0 \\
\\
\frac{n_1(\tau)n_3(\tau)}{\sqrt{1 - n_3^2(\tau)}} & 
\frac{n_2(\tau)n_3(\tau)}{\sqrt{1 - n_3^2(\tau)}} & 
-\sqrt{1 - n_3^2(\tau)} \\
\\
n_1(\tau) & n_2(\tau)& n_3(\tau)
\end{array}
\right)
\nonumber
\end{eqnarray}

{\bf Remark:}
It is easy to see that in fact the procedure described above is 
applicable for condition weaker than $\;n_2(\tau) \neq 0\;$, 
$\;n_1^2(\tau) + n_2^2(\tau) \neq 0\;$ 
(which is equivalent to $\left|n_3(\tau)\right| < 1$).

{\bf Example:}
The equations of spin motion generated by the
model Hamiltonian of the stationary single resonance problem
\begin{eqnarray}
H(\tau, \,\vec s\,) \;=\;
\varepsilon \cos(\Omega \tau + \phi) \cdot s_1
\:+\: \varepsilon \sin(\Omega \tau + \phi) \cdot s_2
\:+\: \lambda \cdot s_3
\label{nuinu}
\end{eqnarray}
for $\:\varepsilon \neq 0\:$ have the nontrivial 
$\:(2 \pi / \Omega)$-periodic solution
\begin{eqnarray}
\vec{n}(\tau) \;=\; 
\frac{1}{\sqrt{\varepsilon^2 + (\lambda - \Omega)^2}}\:(
\varepsilon  \cos(\Omega \tau + \phi), 
\hspace{0.2cm}
\varepsilon  \sin(\Omega \tau + \phi),
\hspace{0.2cm}
\lambda - \Omega 
)
\nonumber
\end{eqnarray}
which satisfies the condition $\:\left|n_3(\tau)\right| < 1\:$
for all values of $\tau$. 

The $\:(2 \pi / \Omega)$-periodic matrix $\,A(\tau)\,$ 
defined by this solution 
has the form 
\begin{eqnarray}
A (\tau)\;=\;
\left(
\begin{array}{rrr}
 \frac{\varepsilon}{|\varepsilon|} \sin(\Omega \tau + \phi) &
-\frac{\varepsilon}{|\varepsilon|} \cos(\Omega \tau + \phi) & 
 0 \\
\\
\frac{\varepsilon}{|\varepsilon|} 
\frac{\lambda - \Omega}{\omega} 
\cos(\Omega \tau + \phi) & 
\frac{\varepsilon}{|\varepsilon|} 
\frac{\lambda - \Omega}{\omega} 
\sin(\Omega \tau + \phi) & 
-\frac{|\varepsilon|}{\omega}
\\
\\
\frac{\varepsilon}{\omega} \cos(\Omega \tau + \phi) & 
\frac{\varepsilon}{\omega} \sin(\Omega \tau + \phi) &
\frac{\lambda - \Omega}{\omega} 
\end{array}
\right)
\nonumber
\end{eqnarray}
where $\omega$, calculated in accordance with (\ref{vecont}), is 
\begin{eqnarray}
\omega \;=\; 
\sqrt{\varepsilon^2 + (\lambda - \Omega)^2} \;=\; const
\nonumber
\end{eqnarray}
So the fundamental matrix solution of the 
stationary single resonance problem
\begin{eqnarray}
M(\tau,\, \tau_0) \;=\; A^{\top}(\tau) \:K(\tau,\, \tau_0)\: A(\tau_0) 
\nonumber
\end{eqnarray}
expressed in terms of matrix elements $\:m_{ij}\:$ takes 
the following form
\begin{eqnarray}
\left\{
\begin{array}{l}
m_{11} = \alpha \sin(\Omega \tau_0 + \phi) + 
\left(
\frac{\lambda - \Omega}{\omega} \beta  + 
\left(\frac{\varepsilon}{\omega}\right)^2
\cos(\Omega \tau + \phi) \right) \cos(\Omega \tau_0 +\phi) \\
\\
m_{12} = \left(
\frac{\lambda - \Omega}{\omega} \beta +
\left(\frac{\varepsilon}{\omega}\right)^2
\cos(\Omega \tau + \phi) \right)\sin(\Omega \tau_0 +\phi)
-\alpha \cos(\Omega \tau_0 +\phi) \\
\\
m_{13} = 
\frac{\varepsilon}{\omega} 
\left(
\frac{\lambda - \Omega}{\omega}
\cos(\Omega \tau + \phi)-
\beta \right) \\
\\
m_{21} = \zeta \sin(\Omega \tau_0 +\phi) + 
\left(
\frac{\lambda - \Omega}{\omega} \gamma + 
\left(\frac{\varepsilon}{\omega}\right)^2
\sin(\Omega \tau + \phi) \right) \cos(\Omega \tau_0 +\phi) \\
\\
m_{22} = \left(
\frac{\lambda - \Omega}{\omega} \gamma +
\left(\frac{\varepsilon}{\omega}\right)^2
\sin(\Omega \tau + \phi) \right)\sin(\Omega \tau_0 +\phi)
-\zeta \cos(\Omega \tau_0 +\phi)\\
\\
m_{23} = 
\frac{\varepsilon}{\omega} 
\left(
\frac{\lambda - \Omega}{\omega}
\sin(\Omega \tau + \phi)-
\gamma \right) \\
\\
m_{31} =  \frac{\varepsilon}{\omega} 
\left(
\vartheta
\cos(\Omega \tau_0 +\phi)- 
\sin(\omega (\tau-\tau_0)) \sin(\Omega \tau_0 +\phi)\right) \\
\\
m_{32} =  \frac{\varepsilon}{\omega} 
\left(
\sin(\omega (\tau-\tau_0)) \cos(\Omega \tau_0 +\phi) 
+\vartheta
\sin(\Omega \tau_0 +\phi)\right)\\
\\
m_{33} =  
\left(
\frac{\varepsilon}{\omega} 
\right)^2 
\cos(\omega (\tau-\tau_0)) +
\left(
\frac{\lambda - \Omega}{\omega} 
\right)^2
\end{array}
\right.
\nonumber
\end{eqnarray}
where we have used the notations
\begin{eqnarray}
\left\{
\begin{array}{l}
\alpha =
\cos(\omega (\tau-\tau_0)) \sin(\Omega \tau + \phi) 
+ \frac{\lambda - \Omega}{\omega}
\sin(\omega (\tau-\tau_0)) \cos(\Omega \tau + \phi)\\
\\ 
\beta =
\frac{\lambda - \Omega}{\omega}
 \cos(\omega (\tau-\tau_0)) \cos(\Omega \tau + \phi)
-\sin(\omega (\tau-\tau_0)) \sin(\Omega \tau + \phi) \\
\\
\gamma =
\sin(\omega (\tau-\tau_0)) \cos(\Omega \tau + \phi) 
+ \frac{\lambda - \Omega}{\omega}
\cos(\omega (\tau-\tau_0)) \sin(\Omega \tau + \phi) \\
\\
\zeta =
\frac{\lambda - \Omega}{\omega}
 \sin(\omega (\tau-\tau_0)) \sin(\Omega \tau + \phi)
-\cos(\omega (\tau-\tau_0)) \cos(\Omega \tau + \phi) \\
\\
\vartheta =
\frac{\lambda - \Omega}{\omega} 
\left(1 - \cos(\omega (\tau-\tau_0))\right) 
\end{array}
\right.
\nonumber
\end{eqnarray}

\section{Noncanonical Scaling of the Orbital 
Variables in the Case of the Simplest 
Storage Regime}

\hspace*{0.5cm}
In the case of the simplest storage regime, that is 
\begin{eqnarray}
\pi_0(z) \equiv const > 0,
\nonumber
\end{eqnarray}
it is typical in accelerators physics to keep the orbital 
position variables unchanged and to normalize the corresponding 
momenta to the value of the design kinetic momentum $\pi_0$. 
So, following tradition, we would like to have the possibility
to describe the orbital motion using the new  variables
\begin{eqnarray}
\left\{
\begin{array}{lllllllll}
x_{new} &=& x_{old}, \hspace{1.0cm} &
y_{new} &=& y_{old}, \hspace{1.0cm} &
\sigma_{new} &=& \sigma_{old} \\
\\
P_x^{new} &=& P_x^{old} / \pi_0,  &
P_y^{new} &=& P_y^{old} / \pi_0,  &
\varepsilon_{new} &=& \varepsilon_{old} / \pi_0
\end{array}
\right.
\label{itis2}
\end{eqnarray}
where the superscript $'old'$ indicates variables of subsection 9.2.

Unfortunately the coordinate transformation (\ref{itis2}) 
is not symplectic and even if for the study of the orbital motion alone
we can treat this transformation as canonical using a rescaled Hamiltonian
(see remark 2 in subsection 4.1), we cannot proceed in the same
manner for the case of fully coupled equations of spin-orbit motion.  
So we will restrict ourselves to presenting the Hamiltonian
which will give us the correct form of the triangular system in
new variables. This will not be applicable to the study of the
effect of the spin on the orbit motion if we admit
the complete equations of motion (\ref{f31})-(\ref{f33}).

In the variables introduced above the spin-orbit
Hamiltonian becomes
\begin{eqnarray}
\breve{H}\;=\;
\breve{H}_{orbt} \:+\:
\breve{H}_{spin} 
\nonumber
\end{eqnarray}
\begin{eqnarray}
\breve{H}_{orbt}\; =\; \frac{\hat{H}_{orbt}}{\pi_0} \;=\; 
- \mbox{\ae}\, x\, \breve{\pi}_{\vec{B}} 
\:+\: \mbox{\ae}\, y \,\breve{\pi}_{\vec{N}}
\:-\: (1 + h x + \alpha y)\, \breve{\pi}_{\vec{T}}\: -
\nonumber
\end{eqnarray}
\begin{eqnarray}
- \:\frac{e}{\pi_0 c} A_z
\:+\: \varepsilon 
\:+\: \frac{\sigma}{\pi_0 \beta_0 c} \frac{d E_0}{d z}
\:+\: \frac{e}{\pi_0 c}
\left(
  x\, {d \tilde{A}^0_{\vec{N}} \over d z}
\:+\: y \,{d \tilde{A}^0_{\vec{B}} \over d z} 
\right) 
\nonumber
\end{eqnarray}
\begin{eqnarray}
\breve{H}_{spin} \;=\; -\alpha\, s_x \:+\: h \,s_y \:-\:
 \mbox{\ae}\, s_z \:+\:
\frac{(1 + hx +\alpha y)} {\breve{\pi}_{\vec T}}
\: \vec{\breve{W}} \cdot \vec{s}
\nonumber
\end{eqnarray}
where
\begin{eqnarray}
\vec{\breve{W}} \;=\; -\frac{e}{\pi_0 c} \left(
\left( 1 + \gamma G \right) \vec{{\cal B}}
\:-\: \frac{\pi_0^2 G 
\left(\,\vec{\breve{\pi}} \cdot \vec{\cal B}\,\right) 
\vec{\breve{\pi}}}  
{m_0^2 c^2 (1 + \gamma)} 
\:-\: {\pi_0 \over {m_0 c}} \left( G + {1 \over {1 + \gamma}} \right)
\left[\,\vec{\breve{\pi}} \times \vec{\cal E}\,\right]
\right)
\nonumber
\end{eqnarray}
\begin{eqnarray}
\vec{\breve{\pi}} = \left(
\breve{\pi}_{\vec N} , \;
\breve{\pi}_{\vec B} , \;
\breve{\pi}_{\vec T} 
\right)
\nonumber
\end{eqnarray}
\begin{eqnarray}
\breve{\pi}_{\vec{N}} \:=\: \frac{\pi_{\vec{N}}}{\pi_0} \:=\:  
P_x \:-\: \frac{e}{\pi_0 c}\, \Delta A_{\vec{N}} ,
\hspace{1.0cm}
\breve{\pi}_{\vec{B}} \:=\: \frac{\pi_{\vec{B}}}{\pi_0} \:=\: 
P_y \:-\: \frac{e}{\pi_0 c} \,\Delta A_{\vec{B}} 
\nonumber
\end{eqnarray}
\begin{eqnarray}
\breve{\pi}_{\vec{T}} \:=\: \frac{\pi_{\vec{T}}}{\pi_0} \:=\:
\sqrt{ \frac{\gamma^2 - 1}{\beta_0^2 \gamma_0^2}
\:-\: \breve{\pi}_{\vec{N}}^2 \:-\: \breve{\pi}_{\vec B}^2 }
\nonumber
\end{eqnarray}
\begin{eqnarray}
\gamma \;=\; \gamma_0 \:+\: \gamma_0 \beta_0^2\,
\left(
\varepsilon \:-\: \frac{e}{\pi_0 \beta_0 c}\, \Delta \Phi
\right)
\nonumber
\end{eqnarray}

For such scaling the conversion formulae between
variables in the curvilinear
coordinate system connected with the closed orbit 
$\;z',\, x',\, P'_x,\, y',\, P'_y,\, E', \,t'\;$  and 
our final coordinates 
$\;z,\, x,\, P_x,\, y,\, P_y,\, \sigma,\, \varepsilon\;$
have the form
\begin{eqnarray}
\begin{array}{llllll}
x' &=& x , 
&\hspace{1.0cm}
P'_x &=& \pi_0 P_x \:+\: (e/c) \tilde{A}_{\vec N}^0 \\
\\
y' &=& y , 
&\hspace{1.0cm}
P'_y &=& \pi_0 P_y\: +\: (e/c) \tilde{A}_{\vec B}^0 \\
\\
t'  &=& t_0 \:-\: \sigma/(\beta_0 c) ,
&\hspace{1.0cm}
E' &=& E_0 \:+\: \left(\pi_0 \beta_0 c\right) \varepsilon \\
\\
z' &=& z 
\end{array}
\nonumber 
\end{eqnarray}

The rule for the substitution of the arguments reads now as 
\begin{eqnarray}
F(t',\: x',\: y',\: z' ) \;\rightarrow\; 
F \left( t_0 - \frac{\sigma}{\beta_0 c},
\hspace{0.2cm}
x, 
\hspace{0.2cm}
y, 
\hspace{0.2cm} z
\right)  
\nonumber 
\end{eqnarray}

The expressions for the projections of the magnetic field and for $A_z$
keep
the same form as in subsection 4.6, and the connection between
the electric field and the potentials becomes
\begin{eqnarray}
{\cal E}_{\vec{N}} \;=\; -\frac{\partial \Phi}{\partial x}
\:+\: \beta_0\, \frac{\partial A_{\vec{N}}}{\partial \sigma}
\nonumber
\end{eqnarray}
\begin{eqnarray}
{\cal E}_{\vec{B}} \;=\; -\frac{\partial \Phi}{\partial y}
\:+\: \beta_0 \,\frac{\partial A_{\vec{B}}}{\partial \sigma}
\nonumber
\end{eqnarray}
\begin{eqnarray}
{\cal E}_{\vec{T}} \;=\; 
- \frac{1}{1 + h x + \alpha y}
\left(
\frac{\partial \Phi}{\partial z}
\:+\: \mbox{\ae} \left(
  y \,\frac{\partial \Phi}{\partial x}
\:-\: x \,\frac{\partial \Phi}{\partial y}
\right) \right)
\:+\: \beta_0 \,\frac{\partial A_{\vec{T}}}{\partial \sigma} 
\nonumber
\end{eqnarray}

\section{FORGET-ME-NOT, a Computer Code
for the Study of Polarized Beam Dynamics}

\hspace*{0.5cm}
The computer code {\bf FORGET-ME-NOT} has been written 
for the study of unpolarized and polarized beam dynamics 
and among other things includes the following important options:

{\bf 1.} Calculation of the strengths of the imperfection 
spin resonances and  first order intrinsic spin resonances
with betatron oscillations with the help of 
an averaging method.

{\bf 2.} Calculation of one-turn Taylor maps for orbit and spin motion
up to arbitrarily high order with respect to the amplitudes
of the betatron and synchrotron oscillations and determination of

{\bf 2.1.} Invariant functions of the orbit motion.

{\bf 2.2.} Equilibrium polarization direction.

{\bf 2.3.} Dependence of orbit and spin tunes on the
invariants of orbit motion (spread of orbit and spin tunes).

{\bf 3.} Numerical tracking of particles with spin
in accelerators and storage rings preserving:

{\bf 3.1.} Symplecticity with respect to the 6-D orbit
motion.

{\bf 3.2.} Orthogonality with respect to the 3-D spin motion.

All options use the same physical model. The use of various approaches
allows us to understand the computed results from various
points of view.
FORGET-ME-NOT, for example, has been applied to the investigation of
schemes for preserving the
polarization in the TRIUMF KAON Booster \cite{tri1,tri2}, to
the investigation of spin motion at high energies in the HERA proton ring 
\cite{wash2_1} and to the study of the possibility to accelerate the
polarized proton beam in the Nuclotron ring in Dubna \cite{dubna}.

\end{document}